**Perfectly reproducible, stable, long plasma columns can be achieved using the electric field component of radiofrequency and microwave waves. These plasma columns boast an unrivalled operating range in terms of gas pressure, wave frequency, and discharge tube diameter, spanning 18 orders of magnitude of electron density. The interface between the outer surface of the dielectric discharge tube and the ambient air guides these waves. In contrast to all existing articles, this model perfectly matches the experiments, describing travelling wave discharges (TWDs)**

**Michel Moisan**

*Groupe de physique des plasmas, Université de Montréal, Montréal H3C 3J7, Québec*

## Abstract

A new category of plasma appeared in the late 1970s. It consists of a plasma column contained in a cylindrical dielectric discharge tube and maintained by the electric field component of radio frequency (RF) and microwave (MW) waves. These waves propagate at the interface between the outer surface of the tube and the ambient air (or vacuum). These plasmas are also perfectly stable and reproducible. This plasma column is known as a travelling wave discharge (TWD) and has the unique property of increasing in length with the RF and MW power absorbed. Its electron density decreases linearly along its axis until it stops abruptly at a non-zero electron density value, marking the end of wave propagation. The slope of its distribution depends solely on externally defined operating parameters, namely the carrier gas pressure $p$, the wave frequency $f$ and the inner radius of the discharge tube $R$. The model presented in this article is the only one to date capable of accurately reproducing all experimental data, particularly those relating to the end of the column, a feat that no other published model has been able to achieve. Interest in TWDs began with the arrival of efficient RF and MW field applicators (wave launchers), which occupy only a few centimetres of the resulting plasma column, which can eventually extend to several metres. Devices such as the Surfatron, Surfaguide, Waveguide-Surfatron, Ro-Box and TIAGO (the latter of which achieves plasma in ambient gas) are already in widespread use. The fact that all these devices have been patented testifies to the interest in the potential applications of TWDs and, of course, their originality. Another outstanding feature of TWDs is their unrivalled wide range of operating parameters $p$ (from a few mTorr (Pa) tested up to twice atmospheric pressure), $f$ (from a few MHz to at least 10 GHz); and $R$ (from 0.5 mm to at least 150 mm). As a result of these exceptional features, today, most publications on TWDs are focused on applications, and this sector is expanding constantly.

Keywords: Plasma physics, RF and microwave discharges, discharges sustained by electromagnetic surface waves/travelling waves.

[+] This article is dedicated to the memory of Professor Ivan Zhelyazkov (1938-2021), a plasma physics theorist (St. Clement of Ohrid University, Sofia, Bulgaria).

**Part I: Experiments show that the axial distribution of the electron density of TWDs is solely determined by the externally set parameters *p*, *f* and *R*, and not by the properties of the plasma column as is currently erroneously accepted (Secs. 1 to 3).**

## 1. Introduction

In practical terms, plasma columns sustained by travelling waves are characterized by the way in which the electron density is distributed along the axis of the column. It decreases linearly from the start to the end of the column, abruptly stopping at a non-zero electron density; this marks the end of wave propagation and, consequently, the end of the plasma column. All published papers to date explain this distribution by invoking an electromagnetic (EM) *surface wave* that propagates either along the plasma column itself or along the interface between its outer edge and the inner wall of the dielectric discharge tube, but this is erroneous. This paper is the first to demonstrate that, in travelling wave discharges (TWDs) held in a dielectric tube, the travelling wave propagates along the interface between the outer wall of the tube and the surrounding medium.[1] According to the *discharge stability criterion* (Sec. 1.5), the wave attenuation coefficient $\alpha(z)$ must be proportional to the electron density $n_e(z)$, i.e. $\alpha(z) = \frac{b}{\bar{n}_e(z)}$, where *b* is a constant. This ensures the linearity of the electron density axial distribution under all operating conditions, including at the end of the plasma column — a phenomenon that none of the published models can reproduce correctly. In addition, it makes that ions do not affect the power transfer from the wave to the plasma, whether in the RF or MW range.

It is the electric field of the wave that heats up the electrons, thus ionizing the discharge gas. This establishes a relationship between the energy dissipated by the wave at a given axial position and the energy acquired by the electrons at that position. Consequently, the axial distribution of electron density can also be considered to correspond to an axial distribution of the energy supplied to the electrons by the wave.

### *1.1. The first identification of plasmas supported by microwaves as TWDs*

MW gas discharges, the length of which increases with the applied power, were first identified in an experimental article by Tuma (1970) as being generated by an EM surface wave. In support of this, the author referenced the work of Trivelpiece (1958, 1967) and Gould and Trivelpiece (1959) concerning non-ionising surface waves that propagate along the positive column of a direct current (DC) discharge.

However, plasma columns had been observed to respond in this way to increasing microwave power long before, but this had not been correctly interpreted. They were often simply described in the literature as 'plasma columns emerging from an MW resonant cavity'. For instance,

[1] An easy way to check this is to close your hand around the discharge tube when it is switched on and observe how the light is partially blocked along its length. The degree of disturbance to the plasma column then depends on how much your hand overlaps the wall of the tube. At the very least, we can conclude that a significant proportion of the electromagnetic power of the wave is transmitted on the outside of the discharge tube.

Fehsenfeld *et al.* (1965) wrote that 'in all sources except 2A and 2B, a discharge may extend several centimetres in the tube outside the cavity', while Epstein *et al.* (2014) stated that 'the plasma column, which spreads beyond the resonator, is spatially uniform... it represents the afterglow of the microwave discharge inside the cavity'. At the time, excepting Tuma's paper, there was no indication that they would be maintained by an EM travelling wave as opposed to direct EM radiation or radiation from an antenna. This confusion remains for most of them to this day.

### *1.2. Various types of TWDs*

Reported TWDs most commonly occur in a cylindrical dielectric tube. However, plasma columns can emanate from the outlet of an open-ended dielectric tube and extend freely for several centimetres into the surrounding environment; these plasmas are sometimes called filaments as their diameter is around one millimetre or less. They can also be generated at the tip of a conductive nozzle that introduces gas into the environment (TIAGO system: see Appendix D in Part II). Furthermore, plasmas can be sustained along the dielectric plate that closes the top of a cylindrical metal container, producing planar TWDs, which are of interest in many microelectronic processes (Sugai, Ghanashev and Nagatsu, 1998).

### *1.3. The rapid early spread of TWDs in the literature is due to the availability of efficient wave launchers (EM field applicators), such as the Surfatron, Surfaguide and Ro-Box*

Interest in these plasmas has grown considerably since the introduction of the EM field applicators proposed and patented by Moisan and his team (see Moisan *et al.*, 1977; Moisan and Zakrzewski, 1989; Appendix A). These applicators facilitate straightforward impedance tuning to produce highly efficient, stable and perfectly reproducible plasma columns over an unequaled range of operating conditions. As mentioned, these include gas pressures $p$ ranging from a few mTorr (a few Pa) to at least twice atmospheric pressure, EM field frequencies $f$ ranging from radio frequencies (RF) as low as a few MHz to microwaves showing up above 100 MHz (mainly 2.45 GHz), and an inner radius $R$ of the dielectric tube ranging from 0.5 mm to at least 150 mm. Passing through all possible values of $p$, $f$ and $R$ results in an electron density spanning 18 orders of magnitude ($2.2\times10^{8}$ cm$^{-3}$ (Fig. 11)–$4.1\times10^{26}$ cm$^{-3}$ (Fig. 16)!, which is unprecedented compared to any kind of gas discharges.

These technical advances have provided theorists with a vast and compelling body of experimental data, with over 600 papers and several books having been published to date. However, it is a common misconception that these discharges are caused by an EM wave travelling along the plasma column itself (e.g. Gordiets *et al.*, 2000), along the (theoretical) interface between the plasma column and the vacuum (e.g. Margot and Moisan, 1993), or along the interface between the plasma column and the inner wall of the discharge tube (e.g. Navratil *et al.*, 2013).

### *1.4. The TWD model developed in this paper[2] is based on the verified fact that electron density decreases linearly along its entire length, whatever the values of parameters p, f and R.*

[2] The current model has developed progressively over the years. The first published version appeared in the Arxiv repository (Moisan, 2021), and the model has since evolved through 16 iterations. Until that point, Moisan contributed to the current misconception.

Theorists question this linearity when they compare their calculations of the axial distribution of electron density with experimental results. Regardless of their assumptions or calculation methods, they are unable to reproduce the linearity that we claim to observe throughout the plasma column. To establish the validity of this assertion, we will use a recognised statistical method, applying it to the data in all the figures in this document that describe the axial distribution of electron density.

Section 3 uses least-squares regression to statistically determine the axial distribution of electron density data points either from our laboratories and those from other countries. This demonstrates that, when considering more than five data points, the regression's *coefficient of determination ($r^2$)* is consistently close to one. In fact, our study not only confirms the linearity of the axial electron density distribution but shows that the slope of this distribution depends solely on the operating parameters $p$, $f$ and $R$. This means that the dispersion properties of the EM wave that sustains the discharge and the kinetic properties of the plasma do not affect the axial distribution of electron density. As stated for the first time in ArXiv Moisan v1 (2021), "Plasma columns supported by the propagation of an electromagnetic surface wave have no influence on the properties of the wave, which depend only on the operating conditions". As demonstrated further in the current paper, these discharges are supported by an EM wave guided along the interface between the outer wall of the discharge tube and the surrounding vacuum (or ambient air). Since the interface along which the guided wave propagates depends exclusively on the dielectric permittivity of the discharge tube and the surrounding vacuum, the general belief that plasma properties influence wave propagation is necessarily incorrect.

Section 3 uses least-squares regression to statistically determine the axial distribution of electron density data points from our laboratories and those of other countries. This demonstrates that when more than five data points are considered, the *regression's coefficient of determination* ($r^2$) is consistently close to one. Our study therefore confirms the linearity of the axial electron density distribution and also shows that the slope of this distribution depends solely on the operating parameters $p$, $f$ and $R$, meaning that the dispersion properties of the EM wave that sustains the discharge and the kinetic properties of the plasma do not affect the axial distribution of electron density. As first stated in ArXiv Moisan v1 (2021), 'Plasma columns supported by the propagation of an electromagnetic surface wave have no influence on the properties of the wave, which depend only on the operating conditions'. As demonstrated further in the current paper, these discharges are supported by an EM wave that is guided along the interface between the outer wall of the discharge tube and the surrounding vacuum (or ambient air). Since the interface along which the guided wave propagates depends exclusively on the dielectric permittivity of the discharge tube and the surrounding vacuum, the general belief that plasma properties influence wave propagation is necessarily incorrect.

To maintain continuity in the text, it is necessary to make a brief remark at this point before fully addressing the question: the first part of the plasma column, extending from the EM field applicator output for some distance, is generated by the antenna-like radiation of the wave launcher (see Section 3.2). As detailed in Section 3.1, a TWD only forms beyond this section.

### *1.5. The discharge stability criterion determines the linearly decreasing axial electron density distribution*

As will be clearly demonstrated in Sec. 4.1 of Part II, the slope of the axial electron density distribution of TWDs is governed by the discharge stability criterion. This is the process by which the wave's power is transferred to the electrons, maintaining a stable and reproducible plasma column. This implies that the electron density must decrease axially in a monotonic manner, regardless of the guided wave's specific dispersion characteristics. As previously mentioned, the slope of the axial electron density distribution is ultimately determined only by the operating parameters *p*, *f* and *R*, as detailed experimentation in Sec. 3 confirmed.

The fact that six different diagnostic techniques support the linearity of the observed axial electron density distribution (see Appendix B, Part I) provides an additional level of confidence in this assertion.

### *1.6. Organizing the study as Part I*

Firstly, Section 2 provides an overview of the different devices that have been used to produce TWDs. Historically and currently, many of these devices have not been recognised as wave launchers. A next step is to discuss the design of EM field applicators that can produce stable and reproducible long columns of plasma from travelling EM waves, focusing on their key features. A critical issue in this respect is ensuring the full transfer of RF and MW power to the field applicator. Initially, this functions as an antenna that radiates into space, without being in an EM field configuration that ensures TWDs. Section 3 shows, as already indicated, that the axial distribution of electron density remains linear throughout the column, regardless of the TWD's operating conditions (*p*, *f* and *R*). As this had been challenged by theorists, we had to perform a least-squares regression of the experimental data on all the reported axial distributions of electron density, as already mentioned. The validity of these data is increased by the fact that they originate from five different national research groups. This is followed by Appendices A (initial patent applications) and B (six different methods of electron density diagnostic) in this Part I.

## 2. Design of EM field applicators intended for producing plasma columns with travelling waves and their essential characteristics

### *2.1. Plasma columns supported by MWs were originally obtained empirically, without adequate theoretical support, and under rather inefficient operating conditions*

As mentioned in Section 1.1, MW discharges originating from 'resonant cavities' were initially reported without a suitable explanation, and under suboptimal operating conditions. The ability to vary the length of the plasma column at will, as well as to achieve high levels of excitation and ionisation, fascinated the scientific community. A resonant cavity is a closed, conductive enclosure in which a standing wave pattern can be established, resulting in a high-intensity electromagnetic field due to wave interference. The cavity may be circular, with the cylindrical discharge tube typically passing through its geometric centre. Alternatively, it may be a region within a closed segment of a waveguide, such as a rectangular one, from which the discharge tube emerges. The literature proposes numerous types of resonant cavity, typically ensuring that their dimensions correspond to an integer or half of the full wavelength of the EM field. This maximises the intensity of the electric field by establishing a standing wave (see Appendix C in Part II). It was also found that the best results for maximising the extent of the discharges outside the cavity were achieved by delimiting an EM field zone as narrowly as possible within the cavity, thereby possibly maximising the local intensity of the electric field component of the wave.

The question of how to improve impedance matching quickly arose. For instance, it was discovered that the impedance adjustment on most of these early microwave plasma devices had to be reset each time the discharge was restarted. Furthermore, this tuning had to be achieved through a lengthy and often frustrating process of trial and error, resulting in unreliable discharge reproduction. Ideally, the field applicator would incorporate two impedance adjustment mechanisms to optimise the coupling of RF and MW power generators to a given load: one to minimise the imaginary part of the impedance and the other to minimise the difference between the real part of the input impedance of the field applicator and the characteristic impedance $Z_p$ of the plasma column. In this case, the plasma column embedded in the dielectric discharge tube is considered a transmission line along which the EM wave propagates. The closer the net impedance of the field applicator is to that of the power generator, the longer the plasma column is observed for a given power. This results in lower energy losses that would otherwise heat and damage equipment components. These 'feats' have resulted in several short papers, mainly focusing on technical improvements in plasma column generation.

These considerations have guided the design of what should be more accurately called EM field applicators. As we shall see in Section 2.2, examining the characteristics of the Surfatron closely helps to clarify the correct operating conditions for generating long and efficient plasma columns.

Some of the practical problems encountered in this area warrant reporting. Citing van Dalen *et al.* (1978) concerning the Beenakker cavity (1976): “(a) Sometimes we could not get the plasma started at all. (b) Although the reflected power could be tuned to a minimum, it always remained high (10–50 W). (c) Sometimes, the reflected power did not change gradually but jumped erratically with a slight variation in the metal tuning screws. (d) The coaxial connector with the antenna loop became very hot. (e) The tuning range of the metal screws in our cavity was too small for proper tuning, especially when conditions were different to those of a helium plasma in a thick-walled quartz tube. Retaining the metal screws thus implied adapting the inner diameter of the cavity.” Other types of resonant cavity have also been developed to improve impedance matching. Examples include variable-size MW cavities (Fredericks and Asmussen Jr., 1971) and parallelepiped cavities mounted on a waveguide. The latter are larger than cylindrical cavities, reported as allowing for better MW coupling (Dessaux *et al.*, 1983).

Finally, it was recognised that plasmas obtained from resonant cavities were maintained by EM surface waves (Mallavarpu *et al.*, 1978). This means that these plasmas were not produced by resonant cavities in the strict sense, but rather by a concentration of the EM field within a *wave-launching gap*. This concept was introduced at the Université de Montréal (UdeM) and is discussed in more detail in Sections 2.2 and 2.3 below.

The emergence of various types of microwave-induced tubular plasma has stimulated research into more effective EM field applicators. This has been achieved by comparing applicators under various operating conditions. The EM field applicator known as the Surfatron was ultimately developed to overcome all the aforementioned issues (see Selby and Hieftje, 1987, for a review of the Surfatron's advantages).

*2.2. The Surfatron and Surfaguide are EM field applicators optimally designed to support long, reproducible TWD plasma columns and perfectly suited to microwave power generators.*

As mentioned, these devices have been patented by Moisan and his colleagues (see Appendix A). Not only are they much more efficient at transferring MW power to gas discharges, they also solve most, if not all, of the limiting issues previously encountered with other MW-supported discharges. These problems include ambiguous impedance matching and unreliable restart performance. Notably, the Surfatron exhibits the same impedance tuning characteristics in the RF domain (tested between 27 and 100 MHz) as in its MW range (100–2450 MHz). Perfect impedance matching in the RF domain is a distinctive attribute of the Surfatron. Unlike other RF plasma devices (e.g. ICPs), it does not require a separate impedance matching box: using these often results in high power loss as the matching box heats up considerably. In contrast, the Surfatron provides TWDs with the best possible operating conditions across an unrivalled range of operating parameters $p$ and $R$, with $f$ extending from 27 MHz (for Surfatrons of a very long length) to 2450 MHz.

In this context, the original concept of a wave-launching interstice (gap) confining the EM field is of major interest, since it enables TWDs to be generated with great efficiency and ease. This concept is described in Section 2.2.2 below.

*2.2.1. The Surfatron is an effective and efficient example of a MW field applicator for generating stable and reproducible plasma columns*

Fig. 1 shows typical photographs of the TWDs obtained using a 915 MHz Surfatron in argon gas at atmospheric pressure within a 6/8 mm inner diameter/outer diameter (id/od) fused silica discharge tube. The absorbed power is 300 W in each photo:

a) As there is no Faraday cage surrounding the discharge tube, high-frequency (HF) radiation (i.e. RF and MW radiation) from the antenna-like radiation region is not contained.

b) The discharge tube is enclosed in a Faraday Cage (FC) with a radius of 22.5 mm, which is well below the wave cut-off radius for a circular waveguide at 915 MHz.[3] Here, the FC length is limited to 30 mm — the minimum length required to prevent HF radiation in the room, as evidenced by measuring instruments then remaining unaffected by this stray radiation.

c) The discharge tube is again enclosed in a 22.5 mm radius FC which is now 305 mm long and extends beyond the plasma column length. The axial slot in the FC enables ***E***-field intensity and spectroscopic measurements to be made along the plasma column without the MW field leaking into the room through it. Note that when the same 300 W MW power level is used to feed the Surfatron as in a) and b), the plasma column length fully enclosed in the FC at cut-off is the longest. This is because no MW space-wave power escapes from the antenna-like domain into the room.

---

[3] When the radius of a circular waveguide surrounding a plasma tube is small enough to reach its cutoff frequency, no high-frequency waves can propagate freely inside it at this frequency or any lower one. The fundamental mode of a circular waveguide (i.e. the lowest frequency at which a wave can propagate through it) is the $TE_{11}$ mode. The corresponding wavelength is given by $\lambda_c = 2\pi R_{FC(co)}/1.841$. At 915 MHz, a Faraday cage with an $R_{FC}$ radius of less than 96.1 mm reaches its cutoff frequency. At 2450 MHz, this radius is 35.9 mm (Wendt *et al.*, 1958). However, there is no propagation restriction for a travelling wave guided by the discharge tube.

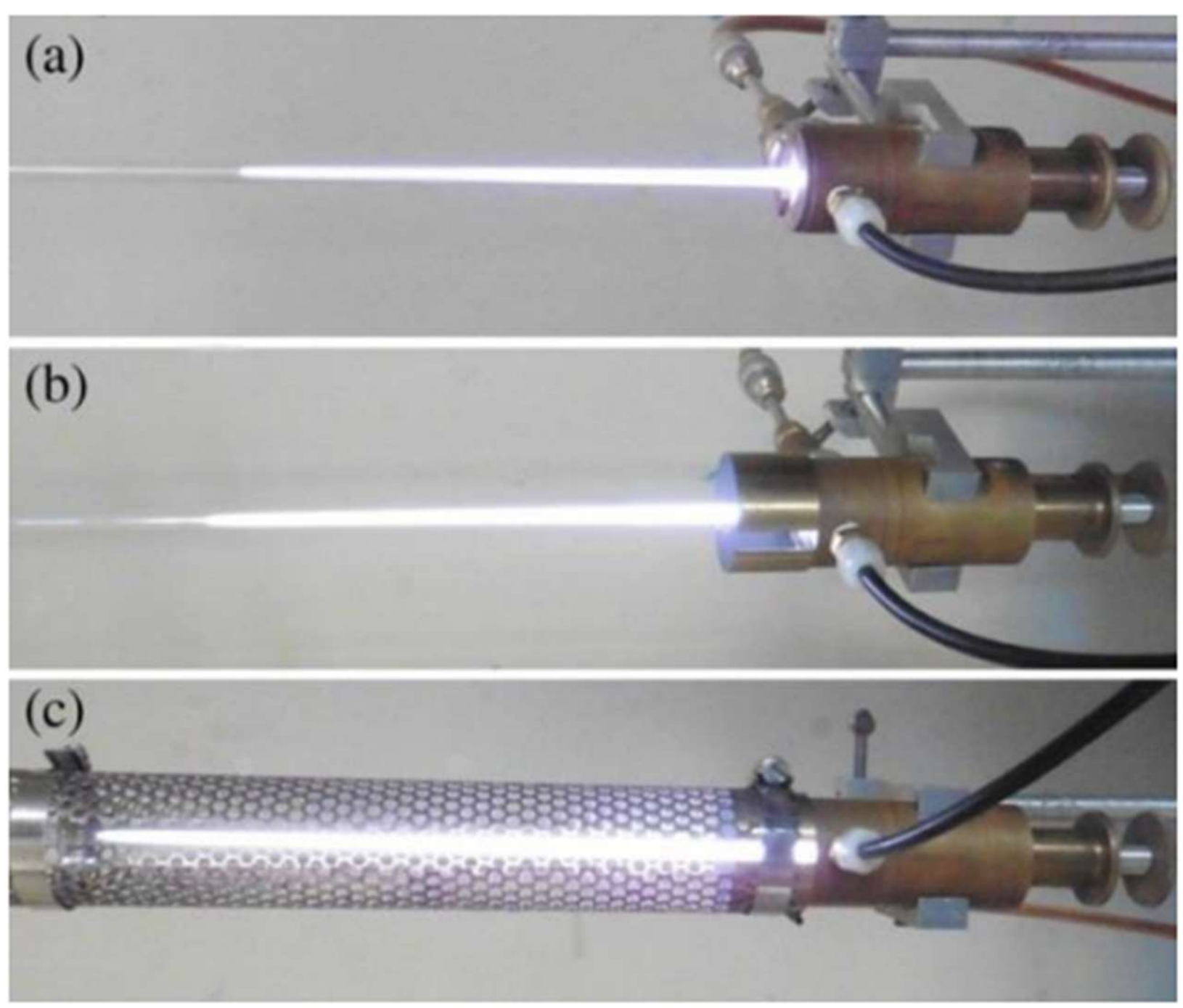

FIG. 1. Photographs showing the plasma column obtained with a 915 MHz Surfatron in a fused silica tube 6/8 (mm) id/od: a) no surrounding FC at all; b) with a radius of 22.5 mm, the FC is already well below the cutoff radius for a circular waveguide, which is at 96.1 mm at 915 MHz (footnote 3); the cage length is here 30 mm, which is the minimum FC length preventing MW field radiation in the room from affecting much the measurements; c) enclosed in a 22.5 mm radius FC, 305 mm long, which goes beyond the plasma column length. Absorbed power is 300 W in each photo. The FC's axial slit allows EM ***E***-field intensity and spectroscopic measurements to be taken along the plasma column without any leakage of MW radiation in the room (Moisan, Levif and Nowakowska, 2019).

*2.2.2. The wave field initially concentrated in the EM field interstice (the gap) radiates through the opening of the field applicator toward the discharge tube in the same way as an antenna*

In our study of TWDs, we found experimentally that that the MW radiation that interfered with the laboratory's measuring instruments was coming from the surfatron gap. The presence of a Faraday cage of minimal length and under cut-off prevented the antenna-like radiation emitted from the gap to flow in space. It reduced the power dissipated by this radiation by up to 30% of the total incident power of the surfatron (Moisan, Levif and Nowakowska, 2019).

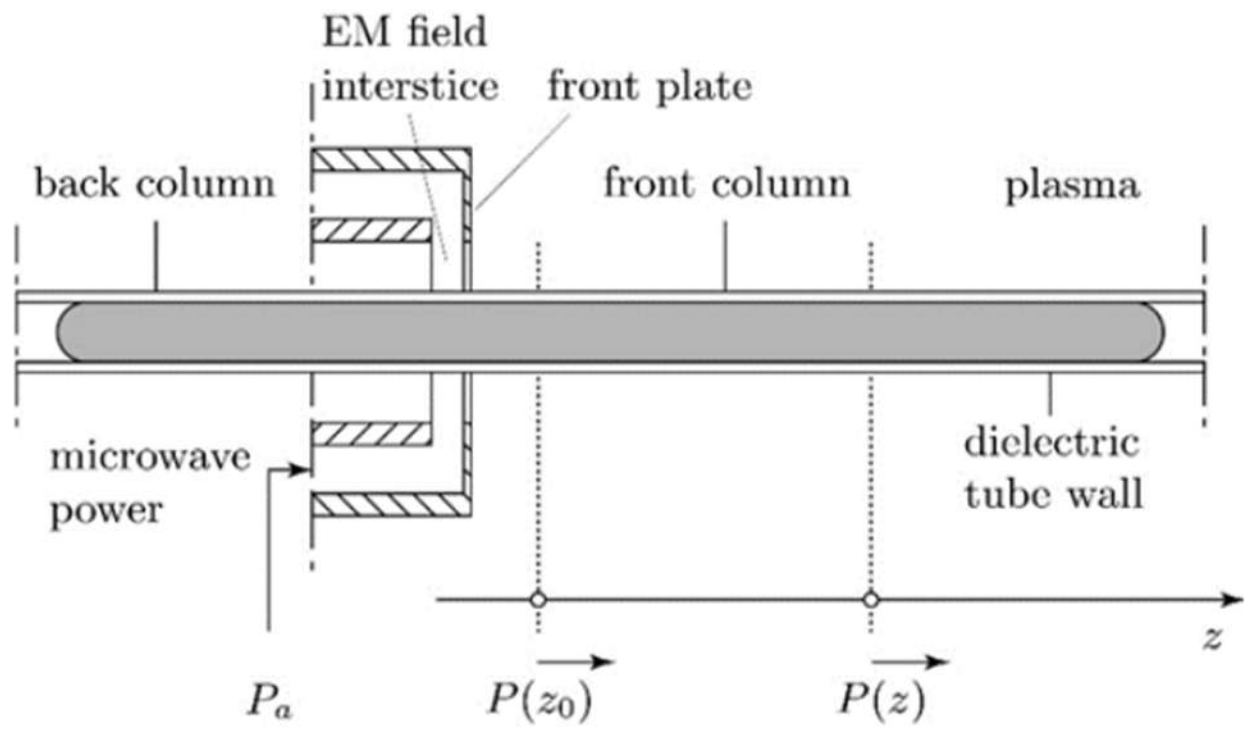

FIG. 2. Schematic representation of the configuration of the wave launching gap (EM field interstice), which is found in all UdeM EM field applicators and achieves reproducible, energy-efficient TWDs. The typical gap width is 2 mm. $P(z_0)$ marks the point at which the flow of travelling-wave power begins after the antenna-like radiation contribution (Moisan, Levif and Nowakowska, 2019).

Without going into antenna theory in detail, let's take a closer look at the type of radiation emitted by an antenna. The space surrounding an antenna can be divided into three regions: the reactive near-field, the radiating near-field, and the radiating far-field (Johnson, 1993). These regions are designated to identify the field structure in each. i) The reactive near-field region extends only a short distance from the antenna and can be neglected; ii) In the radiating near-field region, the ***E*** and ***H*** fields of the wave are neither proportional in intensity nor perpendicular to each other or in phase. These vary with position, and the wave is evanescent rather than propagating. This region extends from the antenna for a distance equal to a fraction of the travelling wave's wavelength $\lambda_0$, depending on the configuration of the electromagnetic field applicator (see Table I below); iii) The far-field radiating region begins at the end of this region. In this region, the wave propagates as an EM wave until the end of the plasma column (Balanis, 2016). We believe that, as soon as the antenna radiation turns electromagnetic in structure, it can excite the guided (and thus slower) EM wave that generates the TWD plasma column.

Unlike conventional antennas, which are described by spherical geometry, the cylindrical configuration is better suited to mapping the fields of TWDs. In the far-field region, the spatial distribution of wave power can be calculated as a function of the ***E*** field intensity to determine the radiation pattern. For TWDs, this reveals a main lobe and secondary lobes, with the majority of the radiated energy concentrated in the main lobe (Nowakowska, Lackowski and Moisan, 2020).

*Considerations on the EM field interstice* (Fig. 2). The EM field interstice (gap) plays an essential role in generating efficient TWDs, whether using Surfatron, Surfaguide, Ro-Box (Moisan and Zakrzewski, 1991) or other devices with a similar objective. The width of the gap on a Ro-Box is also 2 mm, as on a Surfatron. However, the width (height *h*) of a Surfaguide is at its optimal level when its impedance is matched to that of the plasma column then considered a transmission line (see Sec. 2.2.3).

Some of the characteristics of the Surfatron faceplate determining the EM field interstice are shown in Fig. 3. This indicates that perfect impedance matching (full power absorption) can be achieved by adjusting the axial position of the capacitive coupler of the Surfatron faceplate with a thickness of 2 mm (Moisan and Zakrzewski, 1991). Good power coupling can be obtained over a greater axial distance of the capacitive coupler with a faceplate thickness of only 0.5 mm relatively to 2 mm.

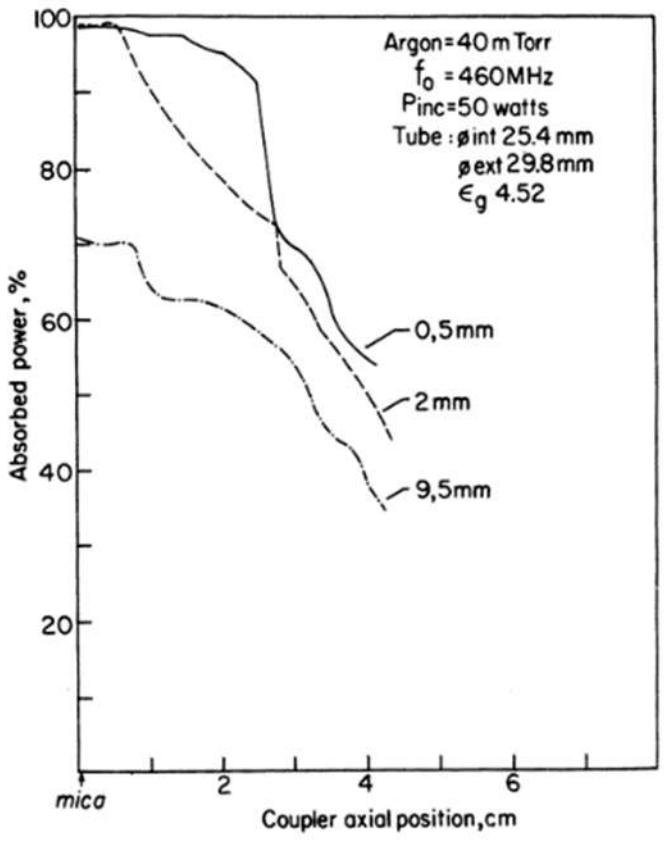

FIG. 3. Percentage of MW power absorbed for a Surfatron fitted with a capacitive coupler with axial displacement along its length (see Fig. A1 in Appendix A, also Moisan and Zakrzewski, 1991). For a given microwave power and a faceplate thickness of 2 mm or less, very little power can be reflected at the input of the field applicator, demonstrating the possibility of achieving near-perfect impedance matching. The thinner the Surfatron faceplate (e.g. 0.5 mm), the greater the axial range of impedance matching (Moisan, Beaudry and Leprince, 1975).

Experimentation has also revealed that the easier the TWD is to turn on, the sharper the inner edge of the faceplate is in the area where the discharge tube passes through. This is particularly evident at high operating frequencies, for example with a 2450 MHz Surfatron and Surfaguide. The higher the field frequency, the greater the minimum electron density required for ignition, necessitating a higher field strength.

As stated above, $L_{space}$, the length of the plasma column resulting from near-field radiation (antenna-like radiation) is a fraction of the wavelength $\lambda_o$ of the EM wave in free space. Table I confirms this for different frequencies and EM field applicators, including for the MW plasma produced using the Beenakker cavity (Lebedev, 1997), as discussed by Moisan, Levif and Nowakowska (2019). In this Table the two narrowest EM-field applicators in the direction perpendicular to the discharge tube, the Surfaguide and the Beenakker cavity, lead to the lowest $L_{space}$ / $\lambda_0$ ratio.

TABLE I Length of the plasma column in the space-wave radiation region, $L_{space}$, and its ratio to $\lambda_0$

| Frequency (MHz) | Wave launcher | $L_{\text{space}}$ (mm) | $L_{\text{space}}$ / $\lambda_0$ |
|---|---|---|---|
| 27 | Surfatron | 2800 ± 100 | 0.25 |
| 360 | Surfatron | 330 ± 10 | 0.40 |
| 915 | Surfatron | 43 ± 1 | 0.13 |
| 2450 | Surfatron | 28 ± 2 | 0.23 |
| 2450 | Surfaguide | 7 ± 1 | 0.06 |
| 2450 | Beenakker cavity | 17 | 0.14 |

*Note on the Beenakker cavity from the paper by Lebedev (1997).* The radial component of the ***E***-field radiation along the outer wall of the discharge tube was detected using a radially oriented antenna made from a tiny semi-rigid coaxial cable, the end of which had been stripped of 1 to 2 mm of its conductive wall. It first increases axially to reach a maximum intensity value at approximately $z$ = 17 mm from the cavity, after which it decreases about linearly till $z$ = 90 mm (Moisan, Levif and Nowakowska, 2019). The ***E***-field antenna was also axially set for reading at 8 mm (near-field radiating region) and then at 90 mm from the launcher (far-field radiating region). At 8 mm, the ***E***-field intensity decreases slowly in a radial direction away from the discharge tube, whereas it radially decreases exponentially from $z$ = 17 to $z$ = 90 mm (the end of the plasma column).[4]

*2.2.3. The Surfaguide: the simplest EM field applicator to build for achieving TWDs at 2450 MHz, and capable of reaching operating powers of several kWs*

Various Surfatron-like devices were first built in the early years to obtain long columns of plasma over a wide frequency range (100-2450 MHz), a frequency agility that proved decisive in revealing the properties of TWDs. Subsequently, at least 125 small axial length Surfatrons operating at 2450 MHz were marketed by the French company Sairem, and widely used for the study of TWD plasmas mainly at atmospheric pressure, but at powers not exceeding 350 W.

However, the development of applications requiring plasmas with higher electron densities led to the creation of discharges of several kWs with, for example, a Surfaguide at 915 MHz (diamond deposition: Schelz, Campillo and Moisan, 1998) and Surfaguides at 2450 MHz (purification of rare gases: Rostaing *et al*., 2000, elimination of greenhouse gases in fabs: Rostaing *et al.*, 1996 and 1999). The Surfaguide has now become the most popular EM field applicator for TWDs because of its simplicity of construction and ease of use, as well as its higher operating power than the Surfatron. Its dimensions have even been tailored so that it is no longer necessary to readjust its impedance matching under variable operating conditions (but for a limited set of operating conditions as shown in Sec. 2.2.3.2), making it ideal for industrial applications. In fact, the Surfaguide is currently the EM field applicator whose properties are best known in terms of TWD production and whose near-field radiation power loss is the lowest (Table I).

*2.2.3.1. Specific dimensions of the Surfaguide and its equivalent electrical circuit.* The Surfaguide is most commonly made from a standard rectangular WR-340 (R26) waveguide and consists of the three distinct parts shown in Fig. 4a. These are: a central section with a reduced height $h$ of the narrow waveguide wall and length $l_2$, and two sections that increase linearly in elevation from height $h$ to full height $b$ (the width of the narrow waveguide wall) at their ends. Each of these sections has a length $l_1$.

This scheme allows for a continuous and gradual transition between the central section and the two ends of the waveguide, both of which have standard dimensions. This prevents standing waves from forming. One end connects to the microwave power input, and the other to the moving short-circuit plane (tuning plunger) of length $l_s$ (Fleisch *et al*., 2007).

[4] The rather slow decrease of the ***E***-field radial intensity along the plasma column axis up to $z$ = 17 mm is ascribed to space-wave radiation (near-field radiation region) while the exponential radial decay observed along the 17 to 90 mm segment (far-field region) is clearly an attribute of TWDs (alias SWDs). This is confirmed by Fig. A12 (see Appendix E in Part II) plotted radially along a Surfatron TWD.

The central section of the Surfaguide has a circular hole in each broad wall, through which the discharge tube passes. A segment of the tube is sketched in Fig. 4a. The space between the two corresponding holes inside the Surfaguide forms the launching gap (see Fig. 2). Fig. 4b shows the equivalent electrical circuit of the Surfaguide, divided into three sections as shown in Fig. 4a, with the connections between them represented by a transformer turn ratio $k_T$. The equivalent circuit of the various TWD field applicators designed at UdeM can be found in Moisan, and Zakrzewski (1991).

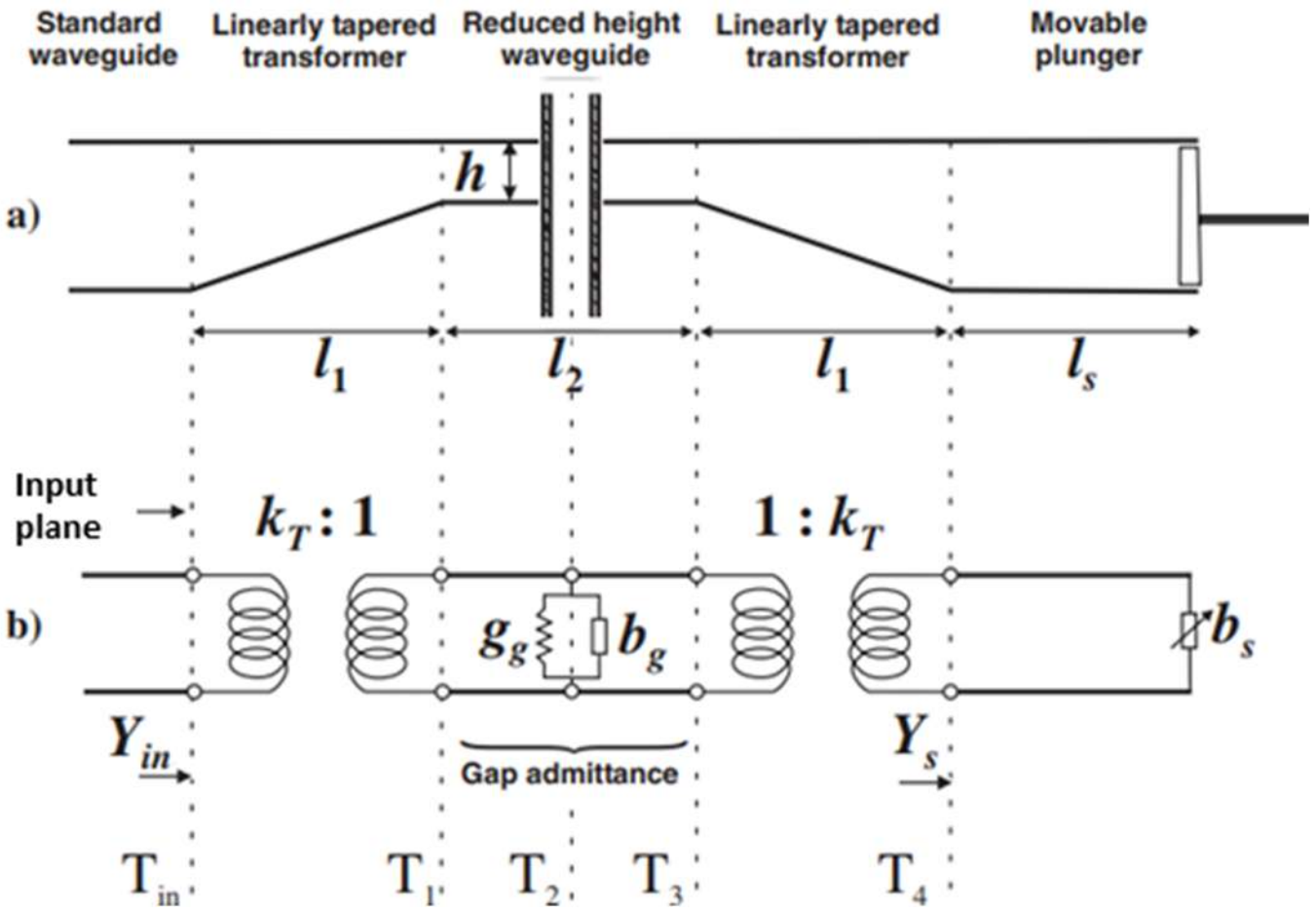

FIG. 4. (a) Schematic representation of the EM field applicator known as the Surfaguide. The vertical solid lines show the location of the discharge tube (sketched by a segment of it) and delimit the applicator's different dimensions; (b) the equivalent electrical circuit of the Surfaguide. The transformation ratio $k_T$ is used to connect two contiguous sections, the corresponding admittance of which is given by $y = g + \mathrm{j}\, b$ where $g$ is the conductance and $b$ (here) the susceptance in each section of the applicator field. The admittance is the inverse of the impedance $Z$ (Fleisch *et al*., 2007).

The reduced height $h$ of the field applicator should not be made as narrow as possible in the belief that this will increase the intensity of the $\boldsymbol{E}$ field in the launch aperture: in fact, the value of $h$, together with the position of the movable plunger, provides a means of acting on the impedance of the Surfaguide so that the characteristic impedance of the TW plasma column (considered as a transmission line) corresponds approximately to that of the Surfaguide at its launching gap (Moisan, and Nowakowska, 2018). In the case of a rectangular waveguide with broad and narrow wall values $a$ and $b$, respectively, the EM wavelength $\lambda_g$ in the waveguide, corresponding to the wavelength $\lambda_0$ in free space, is given for $TE_{mn}$ modes by (Lorrain and Corson, 1972):

$$\lambda_g = \frac{\lambda_0}{\sqrt{1-\frac{\lambda_0^2}{4}\left(\frac{m^2}{a^2}+\frac{n^2}{b^2}\right)}}, \tag{1}$$

which in the case of the $TE_{10}$ fundamental mode ($m$=1, $n$=0) comes to:

$$\lambda_g(TE_{10}) = \frac{\lambda_0}{\sqrt{1-\frac{\lambda_0^2}{4}\left(\frac{1}{a^2}\right)}} \tag{2}$$

with the corresponding waveguide impedance:

$$Z_g = \sqrt{\frac{\mu_0}{\varepsilon_0}} \frac{\lambda_g}{\lambda_0} \frac{b}{a} \quad (3)$$

where $b = h$ in the central section of the Surfaguide, implying here that $Z_g$ decreases as $h$ decreases. The optimum $h$ dimension was determined by experimentally examining the tuning characteristics of the Surfaguide under a range of specific operating conditions, as detailed in the next section.

*2.2.3.2. Determination of the range of discharge operating parameters for which no impedance retuning is necessary when using a Surfaguide with a correctly dimensioned narrow wall width h.*

Fig. 5 illustrates the experimental configuration employed to generate a TWD plasma column with a Surfaguide, as well as to diagnose the level of reflected power in relation to the position of the movable short (tuning plunger).

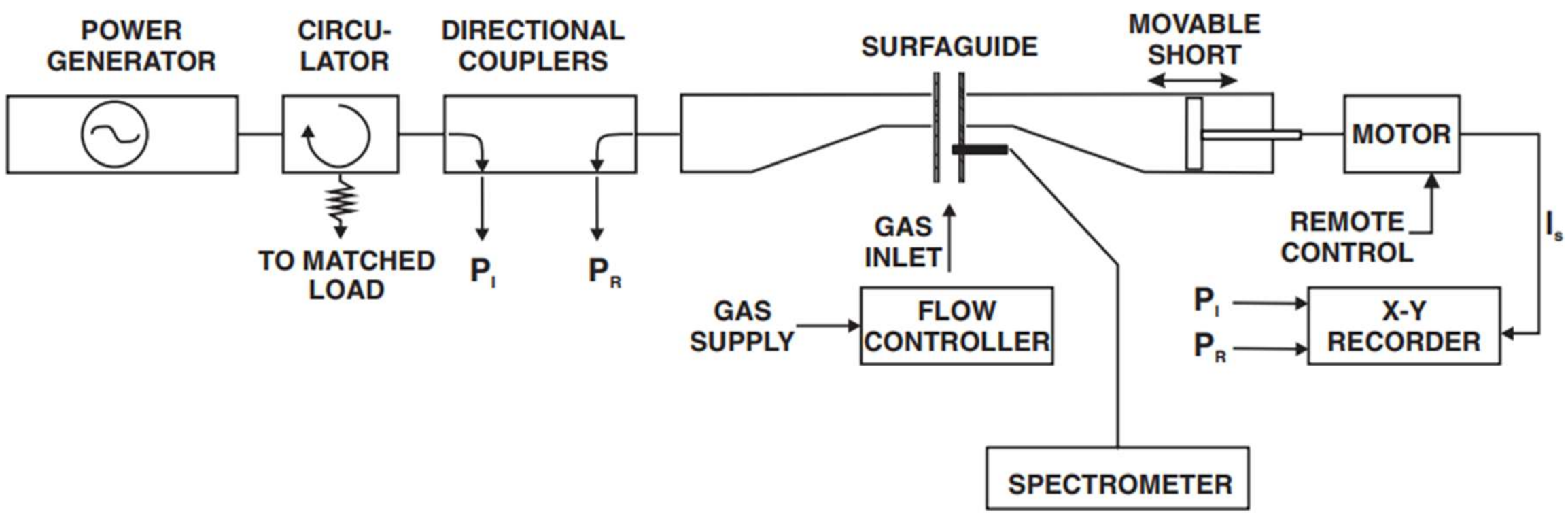

FIG. 5. Schematic diagram of the experimental arrangement used to generate a TWD with a Surfaguide, as well as analyse the reflected power $P_R$ as a function of the position of the movable plunger, for a given narrow wall width $h$ (Fleisch *et al.*, 2007).

The reflected power $P_R$ is tracked as a function of the position of the moving short, looking for the widest range of operating conditions that do not require readjustment. This scenario is examined in Fig. 6 for three reduced heights $h$ =10, 15 and 25 mm: it is $h$ = 15 mm that provides the broadest range of operating conditions for which no readjustment of the plunger position is needed. In Figs. 6b, 6e and 6h specifically, the tuning characteristics (reflected power $P_R$ over incident power $P_I$ vs. plunger position $l_s$ over waveguide wavelength $\lambda_g$) remains close to perfect impedance matching under a large range of $N_2$ gas flow (30-120 slm), MW input powers (2-5 kWs) and $N_2$ and $Ar$ mixture compositions (Fleisch *et al*., 2007). The ability to implement plasma processes across a wide range of operating parameters without constantly readjusting the Surfaguide's impedance is a definite advantage in industrial processes: this example demonstrates how efficiently $SF_6$ molecules, which are used in plasma reactors for chip manufacturing, can be removed without plunger retuning when $h$ = 15 mm (Kabouzi *et al.,* 2003).

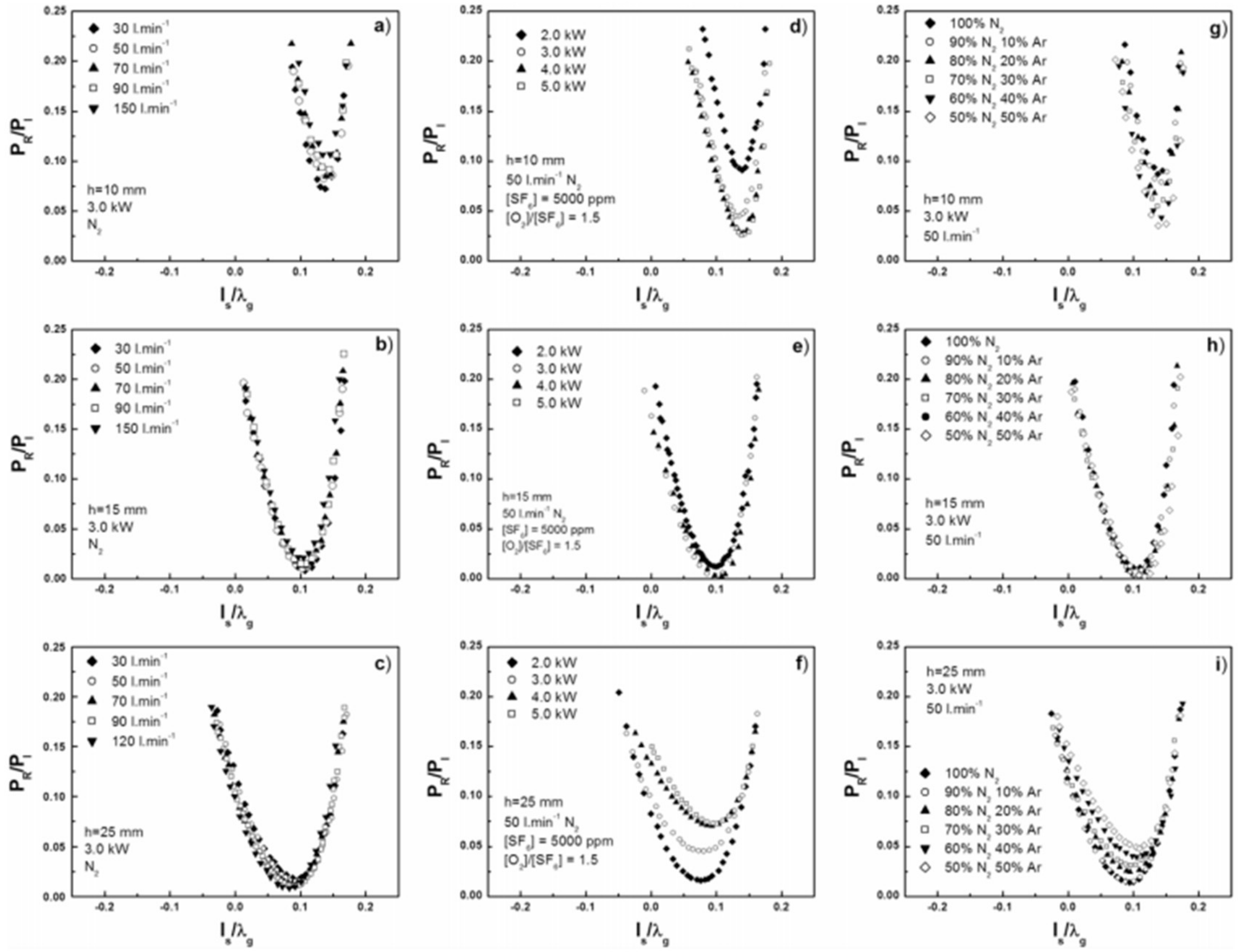

FIG. 6. Surfaguide tuning characteristics (reflected power over incident power as a function of plunger position over waveguide wavelength) considering three different heights *h*, for various discharge operating conditions: (a)–(c) influence of the $N_2$ gas flow rate for a given incident microwave power, (d)–(f ) influence of the incident microwave power for a given gas mixture comprising $O_2$ and $SF_6$ and a fixed gas flow rate, (g)–(i) influence of the concentration of admixed *Ar* gas in $N_2$ at a constant incident microwave power and given flow rate (Fleisch *et al.*, 2007).

## 3.3. The linearity of the axial distribution of the electron density of TWDs depends exclusively on the operating parameters *p*, *f* and *R*

### *3.1. The linearity of the axial distributions of electron density obtained from experimental data is statistically reliable*

In Glaude *et al.* (1980), the data points of the experimentally determined axial distributions of electron density along a TW-maintained plasma column were approximately linearly related to each other, as judged by the operator. The same was true in Fig. 3a of Chaker and Moisan (1985), for example, which is Fig. 7a. It was not until the early 2000s at UdeM that the experimental points were subjected to linear least-squares regression using specialised software, which had not been readily available previously. This recognised statistical method was used to ascertain the linearity of the axial electron density distributions, which were often disputed in the literature. Fig. 7b, obtained using this statistical technique, confirms the validity of the approximate fit of the data in Fig. 7a. This also shows that the measured points naturally align with linearity. The very high statistical determination coefficients, $r^2$, further confirm this: 0.996 for the antenna-like radiation zone and 0.994 for the TWD in Fig. 7b.

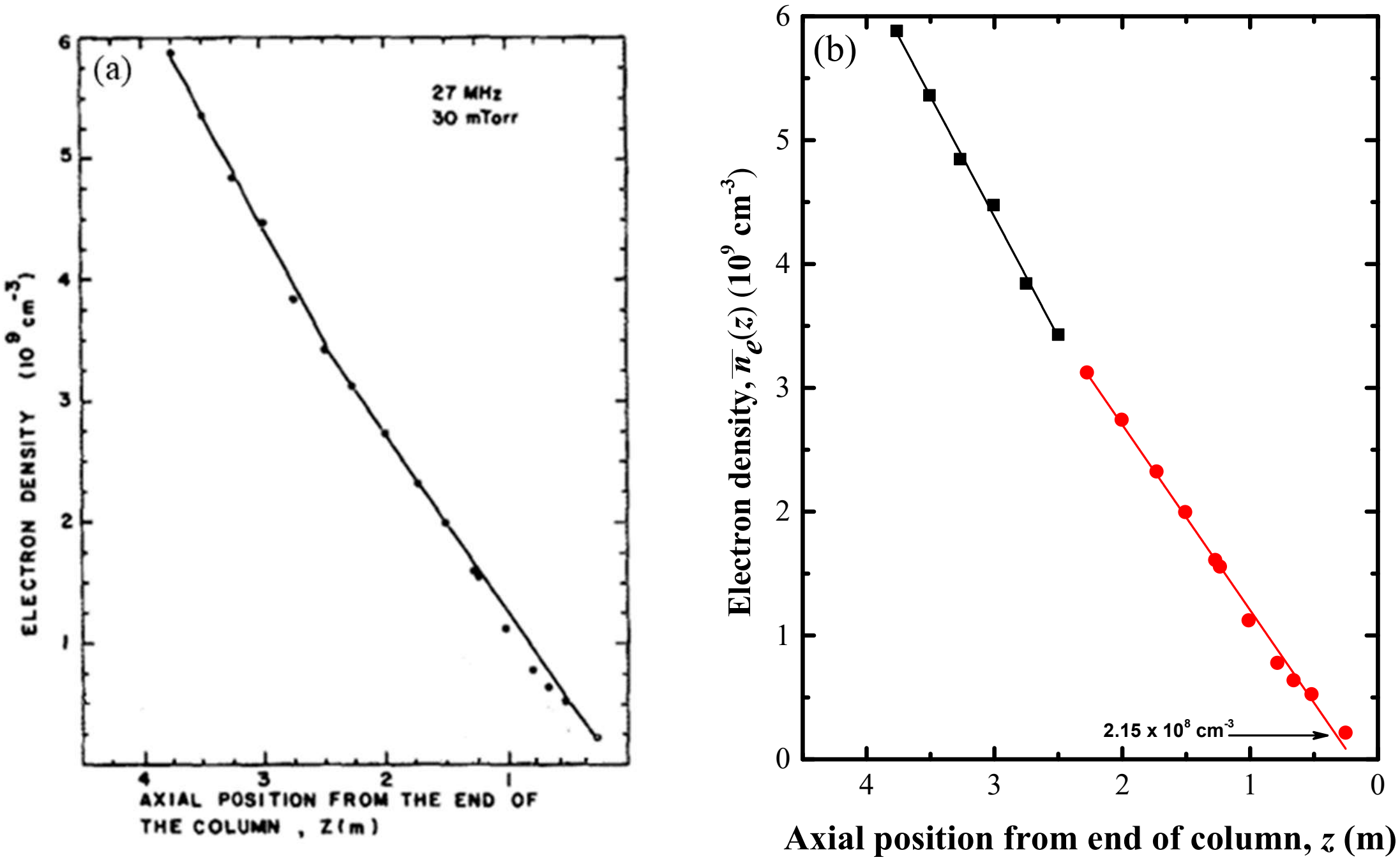

FIG. 7**:** a) Electron density determined by the TW phase-variation method along the plasma column (see Appendix B) at an argon pressure of 30 mTorrs (7.5 Pa) in a discharge tube with an inner radius of 32 mm and at a wave frequency of 27 MHz. This curve was first published in 1985 as part of Fig. 3a in Chaker and Moisan (1985): the data points were connected approximately to the eye by two successive straight lines; b) same data as in Fig. 7a, but processed by a least-squares regression, confirming the existence of two independent straight lines, with $r^2$ = 0.996 for the antenna-like radiation zone and $r^2$ = 0.994 for the TWD.

Remark: Interpreting experimental data with $r^2$ close to 1 as evidence of a perfectly linear axial profile of electron density is not justified a priori: it merely indicates a strong linear correlation. In fact, deviations above or below the mean value of the data points are masked when considering the quadratic value of the $r^2$ coefficient. In practice, this means that slight non-linearity in the data cannot be ruled out at this stage, even though $r^2$ is close to unity. It is therefore possible that the axial electron density distribution of TWDs be slightly non-linear, but this would not be detectable unless the range of electron density in a given experiment considered extends over three or four orders of magnitude. The largest observed range of electron density in a single measurement is less than two orders of magnitude for given discharge conditions. This article will focus on the observed linearity in the axial distribution of electron density, gradually clarifying its nature/origin and ultimately reaching a definitive physical conclusion: this distribution is intrinsically linear and decreases axially, ending abruptly when the propagation of the travelling wave stops.

Assuming that the straight line closest to the field applicator in Fig. 7 results from a TWD is a mistake because, as mentioned earlier, it ignores the fact that this zone, which starts at the outlet of the EM field applicator's interstice (Fig. 2), is necessarily determined by its antenna-like behaviour (designated the 'space-wave radiation zone'). This generates an unguided radiation wave that spreads through space with some intensity lobes. This intrinsic radiation zone precedes any TWD plasma column axially (see Sec. 3.2.1 below). The axial distribution of electron density in this zone can take many shapes[5], and its length depends on frequency as well as on the structural characteristics of the field applicator (Nowakowska, Lackowski and Moisan, 2020). As shown in Table I, the axial extension of this radiation zone, starting at the gap of a Surfatron, is between 0.13 and 0.4 $\lambda_0$ (EM wavelength in free space). At 27 MHz, this equates to a range of 1.4–4.4 metres. If we consider the segment close to the launcher in Fig. 7b, the radiation zone extends approximately 1.3 m long, which is consistent with the findings of Moisan, Levif and Nowakowska (2019).

Fig. 7b already shows a general feature of the axial distribution of electron density: the electron density in Figs. 7a and 7b does not decrease continuously to zero but instead stops abruptly at a non-zero value. In the present case, the observed electron density at the end of the column is approximately five times higher than the minimum required for TW propagation, namely $\bar{n}_{e(\mathrm{re})} = 4.8\times10^{7}$ cm$^{-3}$ (equation (6) below). This is because, although the gas pressure is low enough to ensure a small collision frequency $\nu$, the wave (angular) frequency $\omega$ is here too low to correspond to the conditions $\nu/\omega \ll 1$ of equation (6) (further on for details).

*3.2. The axial gradient of the electron density depends solely on the operating parameters p, f and R*

As stated in the footnote [6], the discharge gas should be characterized by its density $N$ (and its nature) rather than its pressure $p$. So far, efforts to model the axial gradient of electron density along surface wave discharges (SWDs) have highlighted its dependence on the similarity-law factor $\nu f/R$ (Aliev, Boev and Shivarova, 1982). Here $f$ and $R$ are true operating parameters, meaning they are fully set externally while the electron-neutral collision frequency for momentum transfer, $\nu$, is not since it can vary axially. We therefore suggest replacing $\nu$ with $N$, although this remains to be determined analytically. This would give the following expression for the axial gradient of the electron density:

$$\frac{d\bar{n}_e}{dz} = -\,C_0\frac{fN}{R} \tag{4}$$

where $f$, $N$ and $R$ are now all true externally set parameters, and $C_0$ is a constant. In any case, for practical reasons, the influence of the discharge gas is most often expressed in terms of $p$ instead of $N$:

[5] Fig. 8a below illustrates that the axial electron density distribution in the antenna-like region is linear at low argon pressures (20–40 mTorr) but becomes bumpy at higher pressures (80 mTorr and above).

[6] The role and importance of the gas in a discharge is usually represented by its pressure, $p$, an easily accessible and adjustable quantity. However, a more appropriate parameter is the gas density, $N$, since $p$ depends on the gas temperature, $N = p/k_BT$ ($k_B$ is Boltzmann's constant). This indicates that $p$ is not a distinctive physical parameter of the discharge, as it can correspond to different $N$ and $T$ pairs. Furthermore, the value of $N$ does not vary axially within the discharge tube. This makes it a condition that is fixed by the operator at the outset, unlike ν which develops with plasma kinetics (Aliev, Schlüter and Shivarova, 2000).

$$\frac{d\bar{n}_e}{dz} = -\,C'_0 \frac{fp}{R}\,. \tag{5}$$

*3.2.1. Minimum electron density (at the end of plasma column) as a function of wave frequency and the role of the collisional parameter ν/ω when exploring the entire TWD pressure range*

This section focuses on the minimum electron density required for travelling wave propagation. The cessation of this propagation marks the end of the plasma column. Throughout this paper, we will verify experimentally that, in our configuration, the minimum electron density allowing travelling wave propagation at a frequency *f* and in the low collision regime ($\nu/\omega \ll 1$) is obtained from the following relation (Aliev, Schlüter and Shivarova, 2000):

$$\bar{n}_{e(\mathrm{re})}(\mathrm{cm}^{-3}) \simeq 1.2 \times 10^4 (1+\varepsilon_g) f^2(\mathrm{MHz}) \tag{6}$$

where $\varepsilon_g$ is the relative dielectric permittivity of the discharge tube material (3.78 for fused silica and 4.52 for many types of Pyrex glass). When $\nu/\omega$ is no longer much less than unity, the minimum electron density found is higher than that given by equation (6). The higher the gas pressure, the higher the electron density at the end of the column, a situation not always recognized as such in, for example, Palomares *et al*. (2010b).

It should be noted that, although relationship (6) was initially demonstrated for a plasma column assumed supported by a surface wave propagating along the interface between the plasma column and the inner wall of the discharge tube, it nonetheless experimentally accurately provides $\bar{n}_{e(\mathrm{re})}$ in the case of an EM wave travelling along the interface between the outer wall of the dielectric tube and the surrounding vacuum (see the discussion concerning Fig. A3, right panel). The intensity of the wave's $\boldsymbol{E}$ field (and therefore the electron density) is reduced by the permittivity of the discharge tube as the wave passes through it to heat the electrons in the gas.

*3.2.2. Linearity of the axial distribution of electron density as a function of gas pressure*

Fig. 8a shows $\bar{n}_e(z)$, the mean radial electron density, as a function of axial position, starting from the tip of the plasma column, for five argon pressures (with each subsequent pressure being approximately double the previous one). The electron density was determined using a $TM_{010}$ mode resonant cavity method (see Appendix B). Beyond a certain axial position relative to the field applicator (here, a Surfatron), the axial distribution of the electron density becomes linear towards the end of the column. The curve of the electron density at axial positions above the straight line shown in the figure corresponds to the antenna-type radiation region. Recall, from Table I, that this region extends over a distance of between 0.13 and 0.4 $\lambda_0$ from the exit of the field applicator (Moisan, Levif and Nowakowska, 2019).

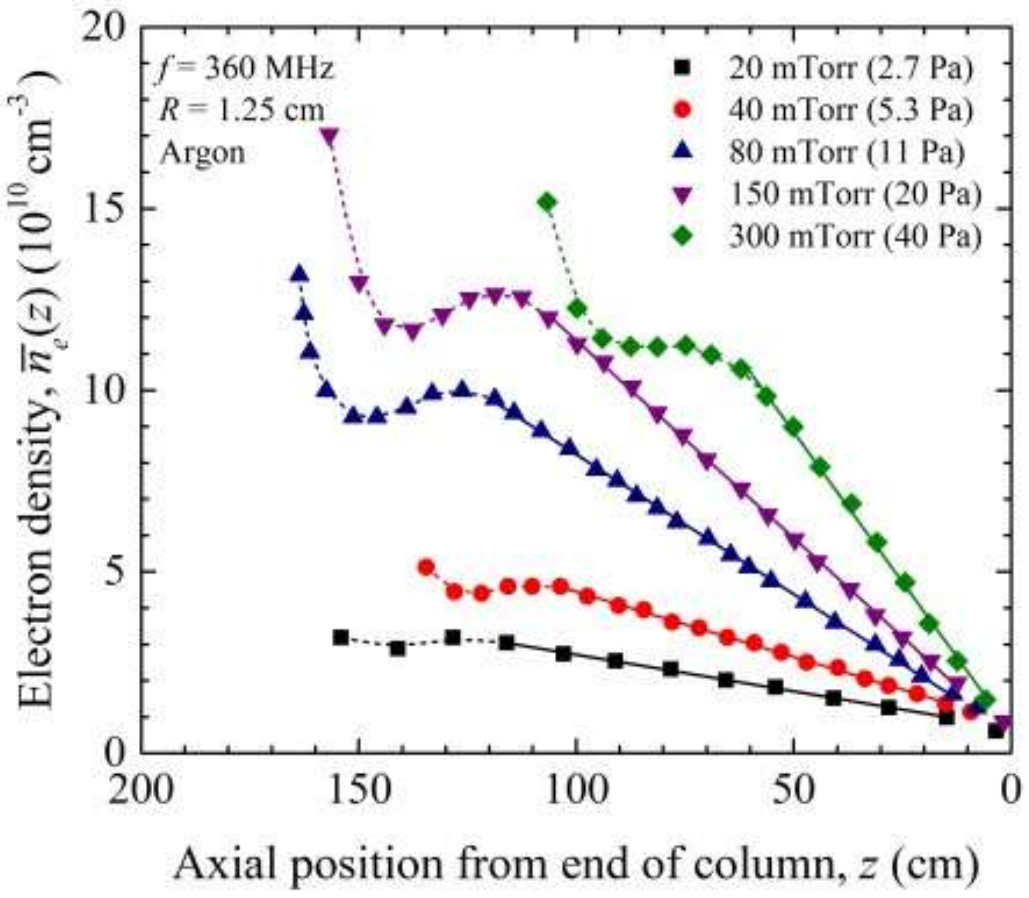

FIG. 8a. Measured axial distribution of the radially averaged electron density through the resonant $TM_{010}$ cavity method, plotted from the end of the plasma column, along the discharge sustained by the propagation of a 360 MHz TW, at five argon gas pressures. This was conducted in a Pyrex-brand glass discharge tube (Dow Corning; relative dielectric permittivity $\varepsilon_g = 4.52$) of an inner radius of 12.5 mm and external radius of 15 mm (Glaude *et al.*, 1980).

Fig. 8b corresponds to the 0-50 cm axial section of Fig. 8a. These curves were plotted using linear least-squares regressions on (many) data points, yielding coefficients of determination $r^2$ very close to unity (see the figure caption for their values) for gas pressures between 20 and 300 mTorrs. The observed axial electron density distributions are clearly straight lines that extend to the end of the plasma column. the slope of their axial electron density distribution steepens with increasing pressure, as predicted by relationship (5).

Fig. 8b also shows that the minimum electron density $\bar{n}_{e(\mathrm{re})}$ (6) that allows the wave to propagate under low-collision regime ($\nu/\omega \ll 1$) is reached at 20 mTorrs for 360 MHz. This clearly indicates that the travelling wave supporting the plasma column obeys the relation (6) at its end. Glaude *et al.* (1980) actually calculated that under current operating conditions, $\nu$ is about $1 \times 10^8$ rad s$^{-1}$, which, at 360 MHz, gives a value of about $\nu/\omega = 0.04$: this verifies that, in the present case, $\nu/\omega \ll 1$.

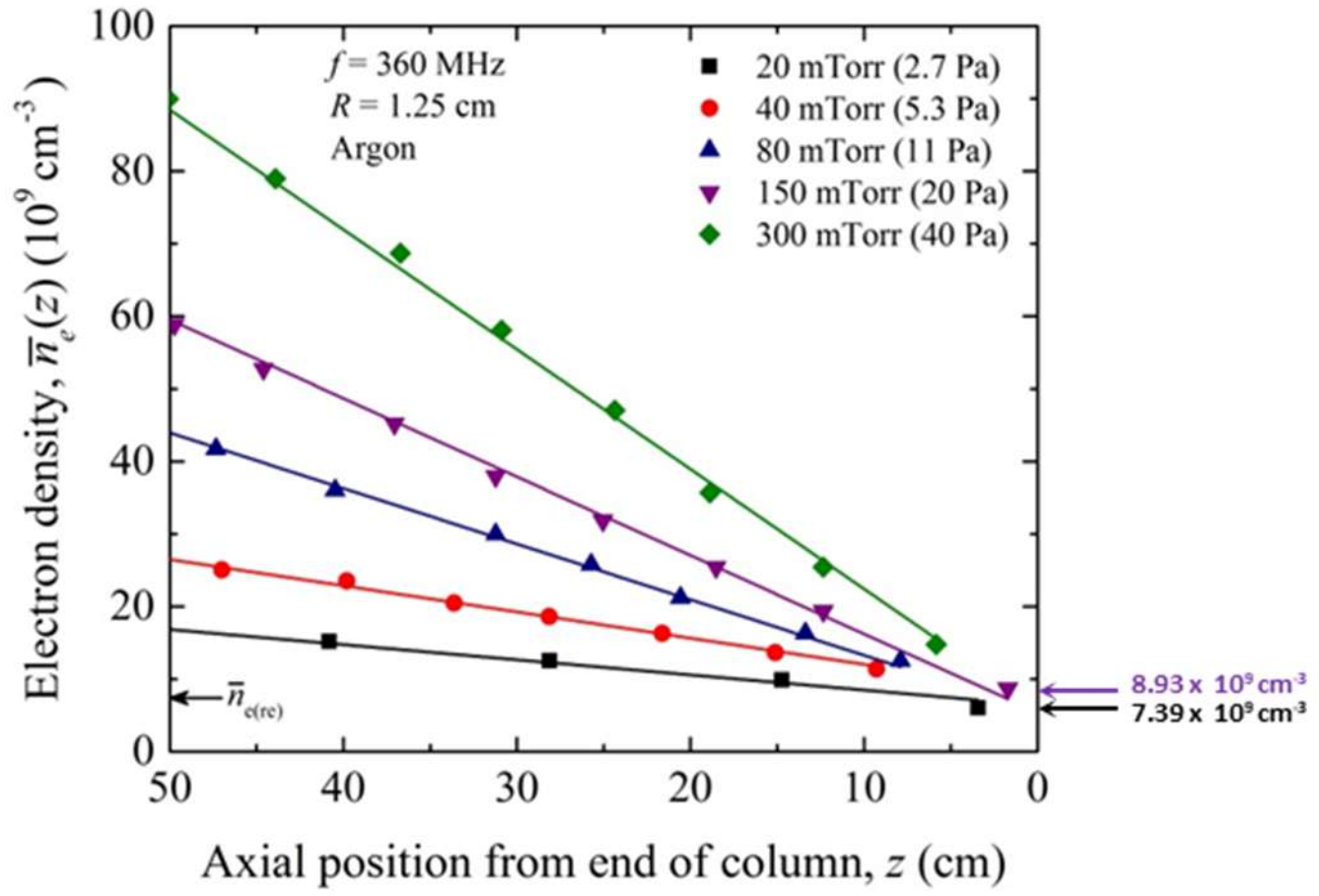

FIG. 8b. An enlargement of the 0-50 cm axial segment in Fig. 8a. The experimental data points for a given gas pressure correspond to a straight line to the end of the plasma column, with $r^2$ = 0.984, 0.995, 0.995,

0.999, and 0.999 at 20, 40, 80, 150, and 300 mTorrs, respectively. Assuming that $\nu/\omega \ll 1$, the corresponding minimum theoretical electron density (6) is $\bar{n}_{e(\mathrm{re})} = 7.4\mathrm{x}10^9$ cm$^{-3}$, as indicated by the arrow inside the figure frame. The minimum measured electron density values at 20 and 40 mTorr are designated by arrows outside, showing that the that the $\bar{n}_{e(\mathrm{re})}$ value is reached at 20 mTorr.

*3.2.2.2. Intermediate pressure range case (0.12-7.2 Torr, 16-960 Pa).* This situation corresponds, given the gas pressure and TW frequency, to the collisional regime varying from $\nu/\omega \ll 1$ to $\nu/\omega < 1$. Linearity of the TWD axial distribution of electron density is again verified with $r^2$ = 0.995 in both Figs. 9a and 9b.

According to relation (5), the axial gradient of electron density $\frac{d\bar{n}_e}{dz}$ depends on the radius of the discharge tube as *1/R*. This means that the ratio of the slopes of the electron axial density in Figs. 9a and 9b should be $R_b/R_a = 1.7$. However, the ratio of the measured slope values, 6.35/4.77 = 1.33, is lower. One possible explanation is that there is some radial contraction of the plasma column (the plasma does not entirely fill the discharge tube radially) in the larger tube radius (Fig. 9b): the radius of the plasma column is smaller on average axially than $R_b$. The axial profile of the electron density $\bar{n}_e(z)$ has been obtained by recording the axial phase variation of the EM travelling wave, then determining the electron density using the wave phase diagram (see Appendix B).

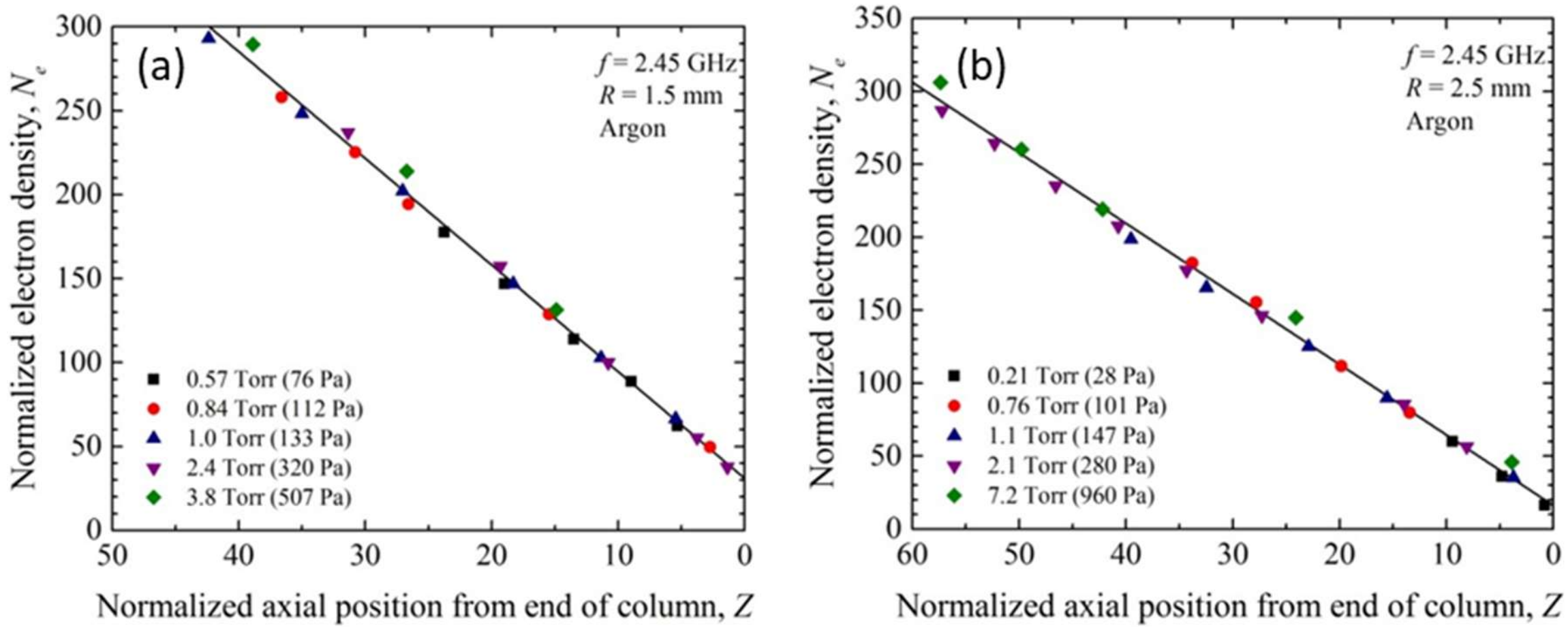

FIG. 9 (foreign laboratory). Axial distribution of the electron density of TWDs maintained at 2450 MHz at different argon gas pressures in the collisional regime ranging from low to medium pressures in discharge tubes ($\varepsilon_g = 4.8$) of two different radii: a) $R_a = 1.5$ mm and b) $R_b = 2.5$ mm; the figures are adapted from Sola, Cotrino and Colomer (1998) where *N* is here the normalized electron density and *Z* the normalized axial position. In both tubes $r^2$ is found to be 0.995.

*3.2.2.3. Atmospheric pressure domain and tube radius dependence.* Fig. 10 shows the observed axial distribution of the radially averaged electron density along a TWD-supported plasma column at twice atmospheric pressure in argon gas in three discharge tubes with inner radii smaller than one mm (Moisan, Pantel and Hubert, 1990). The electron density was determined using the broadening of the $H_\beta$ line (486.1 nm) (see Appendix B) in an argon-hydrogen gas mixture containing 0.5 % hydrogen.

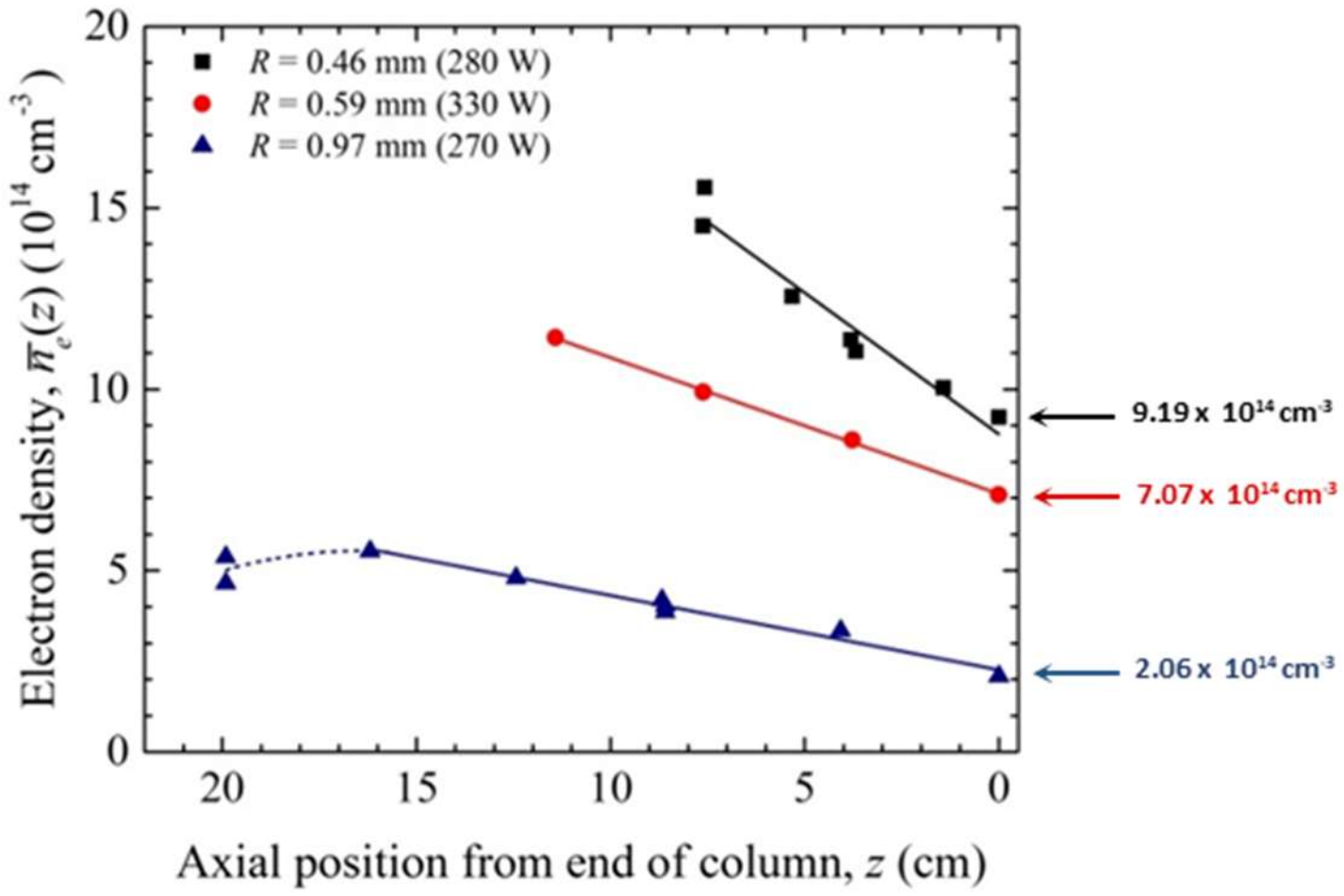

FIG. 10. Measured axial distribution of the radial mean electron density displayed from the end of the plasma column supported by a 915 MHz TW in fused silica discharge tubes. The tubes have three different inner radii $R$, but they are all thick-walled with an external radius of 8 mm. The experiment was conducted in argon gas at twice atmospheric pressure (Moisan, Pantel and Hubert, 1990). The measured electron density was by $H_\beta$ Stark broadening with 0.5 % added hydrogen (see Appendix B). The calculated minimum electron density at 915 MHz (6) is $\bar{n}_{e(\mathrm{re})} = 4.8\text{x}10^{10}$ cm$^{-3}$, which is four orders of magnitude lower than the measured electron density, as expected for such a high gas pressure. Coefficient of determination $r^2 = 0.935$, 0.999, 0.977 following the increase of the tube radius.

Although the ratio $\nu/\omega$ is much larger than unity at (twice) atmospheric pressure, the data nevertheless follow a straight line, except at the $z = 20$ cm point for the $R = 0.97$ mm tube, which belongs to the antenna-like radiation region of the field applicator (Surfatron). The slope in Fig. 10 follows the $1/R$ dependence of equation (5) at least qualitatively, since the gas pressure in the tube is not necessarily equal to the input pressure $p$. This is because the restriction to gas circulation increases as the radius of the tubes decreases sufficiently.

As mentioned previously, the discharge tubes shown in Fig. 10 all have thick walls. In contrast, all the other tubes presented in this document have relatively thin walls. For instance, Fig. 8 illustrates an inner radius of 12.5–13 mm and an outer radius of 15 mm. Therefore, as in the study by Kovačević *et al.* (2021), one might hypothesise that when the dielectric tube is sufficiently thick, the EM wave will lose power as it travels through this medium (see Section 5.2.5 for corresponding calculations). However, the experiment contradicts this hypothesis. Fig. 10 shows that the linearity of the axial electron density remains unaffected by a loss of wave power, which would otherwise increase with axial position and affect linearity as the wave propagates inside a thick dielectric tube. The absence of wave power loss in the dielectric tube is consistent with our model, in which the guided wave primarily propagates along the interface between the outer wall of the discharge tube and the surrounding vacuum.

For the reasons already mentioned in 3.1.2, the value of the electron density at the end of the column in high-pressure gas discharges should exceed the minimum electron density (6) for TW propagation, which at 915 MHz and with $\varepsilon_g = 3.78$ is $\bar{n}_{e(\mathrm{re})} = 4.8\text{x}10^{10}$ cm$^{-3}$, whereas the electron density measured at the end of the column (as shown in the figure) is $2\text{x}10^{14}$ cm$^{-3}$ in the 0.97 mm radius tube and even higher in the smaller tubes: it is four orders of magnitude higher than in the low-collision regime: nonetheless, linearity is unaffected.

The pressure range we have just examined in this section 3.2 is extremely wide, ranging from a few mTorr to twice the atmospheric pressure. It therefore corresponds to major changes in the recombination mechanism of charged particles, moving from the diffusion regime (free and then ambipolar) to volume recombination (atomic and molecular): the kinetics of the discharge are therefore strongly modified by its passage through these different regimes. However, even with such extreme variations, the linearity of the axial distribution of the electron density remains, as expected from our model.

### *3.3. Linearity of the axial distribution of the electron density in TWDs as a function of wave frequency*

Figs. 8 at 360 MHz, 10 at 915 MHz and 9 at 2450 MHz have already shown that a large variation in the wave frequency of TWDs does not affect the linearity of their axial distribution of electron density. This section extends the study to frequencies as low as 27 and 50 MHz, which belong to the RF domain. As ions can absorb energy from the ***E*** field of the wave below 100 MHz (Moisan, Ganachev and Nowakowska, 2022), we investigate whether this has any effect on the linearity of the axial distribution of electron density.

#### *3.3.1. Covering part of the RF domain up to the beginning of the microwave range (27-200 MHz) of the TWDs.*

Fig. 11 shows the value of $\bar{n}_e(z)$, the radially averaged measured electron density of these TWDs (using the $TM_{010}$ resonant cavity method), at four frequencies of the traveling wave ranging from a portion of the RF domain to the approximate beginning of the MW range, with axial position starting from the end of the plasma column (Chaker, Moisan and Zakrzewski, 1986). Due to the low gas pressure of these TWDs (30 mTorr, 4 Pa) and the fact that $\nu/\omega \ll 1$, the minimum measured electron density values (at the end of the plasma columns) should correspond to the calculated value $\bar{n}_{e(\mathrm{re})}$ (6). We observe that at each of these frequencies, the axial distribution of electron density decreases linearly, as confirmed by the $r^2$ values of the least squares regressions approaching unity (see figure caption for these values). Additionally, according to relation (5), the slope of these lines increases with increasing TW frequency.

Since the beginning of studies on SWDs, it has been observed that increasing the EM power transmitted to the field applicator under fixed operating conditions extends the length of the plasma column without affecting the linearity of the axial distribution of electron density or its slope. The 100 MHz curve in Fig. 11 illustrates this well: the arrow pointing towards 36 W shows the axial position of the start of the TW plasma column relative to its end at this power value:, increasing the MW power, for example, to 58 W does not alter the slope of the existing plasma column segment, only its length.

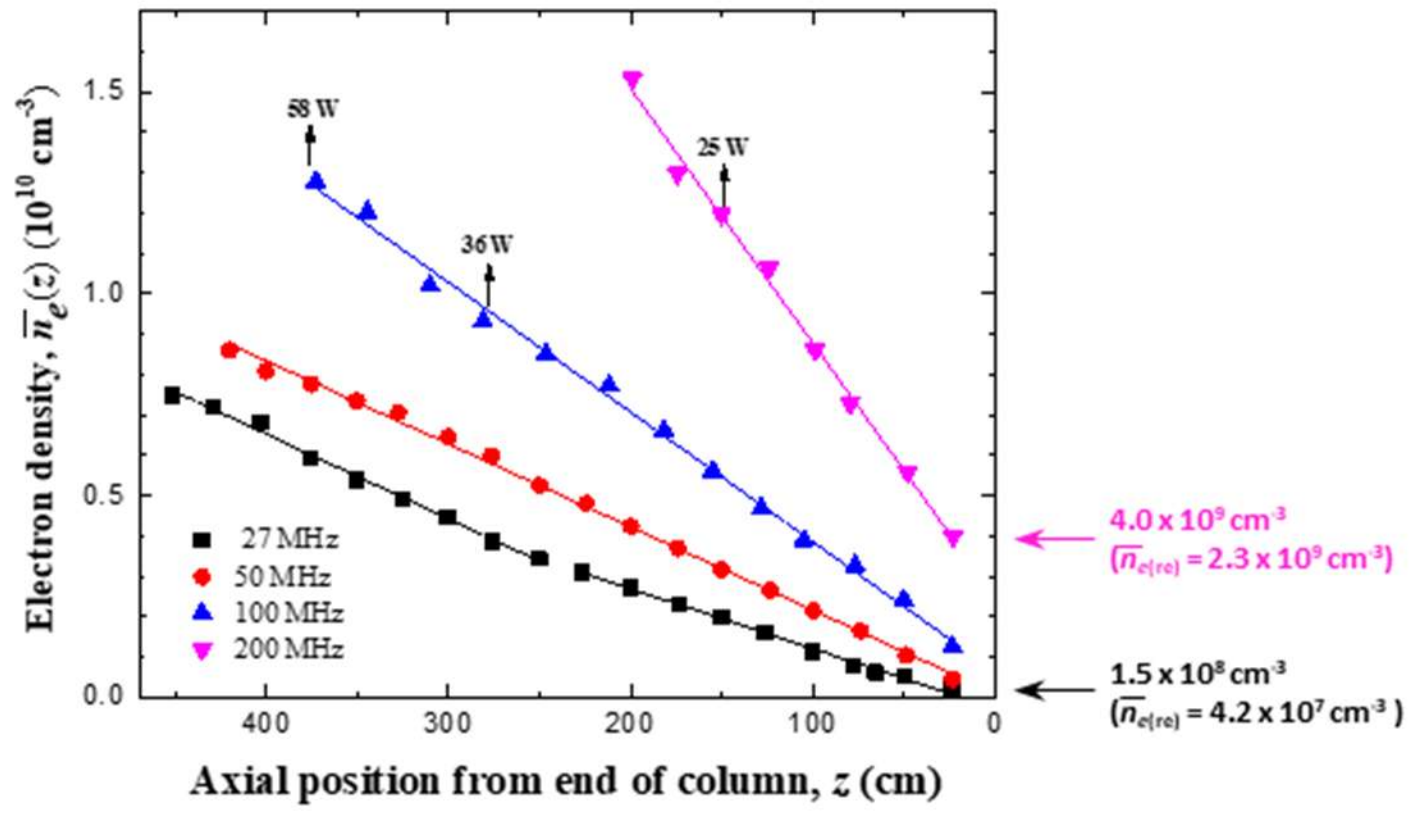

FIG. 11. Measured radial mean value of the electron density axial distribution from the column end using the $TM_{010}$ resonant cavity method. The discharge was obtained by launching an EM travelling wave excited at four different frequencies in argon gas at a pressure of 30 mTorr (4 Pa) in a fused silica tube with an inner radius of 32 mm (Chaker, Moisan and Zakrzewski, 1986). The straight lines are the result of least-squares regression of the data, providing coefficients of determination of $r^2$ = 0.990, 0.999, 0.998 and 0.998 respectively as the frequency increases. For the 27 MHz case, two different slope values can be identified for the curve: $1.5\times10^7$ cm$^{-4}$ between $z$ = 0 and $z$ = 250 cm, and $2.1\times10^7$ cm$^{-4}$ between $z$ = 250 and $z$ = 350 cm (see Fig. 7b at 27 MHz for details on the antenna-like radiation region).

Unlike in the microwave range, in the RF domain of 27 and 50 MHz, the $\boldsymbol{E}$ field of the wave heats both the electrons and the ions in the discharge: nonetheless, the linearity of the axial electron density distribution remains unaffected. One possible reason for this is the (coming) relationship (25) where $\alpha(z) = \frac{b}{\bar{n}_e(z)}$ ($b$ is a specific constant), which means that the attenuation coefficient of the wave power is related only to the electron density. This means that the ions do not come into play in the power transfer from the wave to the charged particles (see coming Sec. 3.3.4 for more information).

*3.3.2. The 200-2450 MHz MW range*

Fig. 12 shows the measured axial distribution of the average radial electron density at 210 MHz and 2450 MHz, with an argon gas pressure of 200 mTorr (27 Pa). Linear least squares regressions give $r^2$ = 0.991 for the antenna radiation segment and $r^2$ = 0.984 (based on four points only) for the TWD segment at 210 MHz. At 2450 MHz, $r^2$ = 0.997 for the TWD segment; however, the antenna-like segment is too small to be visible. Conversely, at 210 MHz, the antenna zone is longer than the TWD zone itself (Moisan, Levif and Nowakowska, 2019).

According to equation (5), the slope of the distribution at 2450 MHz should be much steeper than at 210 MHz a priori, since the operating frequency being approximately ten times higher. In this case, however, the slope at 2450 MHz is barely steeper than that at 210 MHz (TWD zone). This is because the TWD plasma column in a 4.5 mm radius tube is partially radially contracted at 2450 MHz but no at all at 210 MHz. It is generally found that, for given operating conditions, the slope of the axial electron density distribution decreases as radial contraction increases. Substituting in (5) the value of the inner radius of the discharge tube, $R$, for that of the plasma column, $r_p$, does not solve the problem. However, qualitatively speaking, it seems that the greater the difference between $R$ and $r_p$ with increasing radial contraction, the gentler the slope.

The measured minimum electron density at the end of the plasma column, indicated by an arrow in Fig. 12, is found to be 3.23 x $10^{11}$ $cm^{-3}$ and 8.67 x $10^{11}$ $cm^{-3}$ at 210 MHz and 2450 MHz, respectively. By contrast, the minimum value of $\bar{n}_{e(\mathrm{re})}$ calculated from equation (6) (assuming $\nu/\omega \ll 1$ and with $\varepsilon_g$ = 4.52) is 2.9 x $10^9$ and 3.9 x $10^{11}$ $cm^{-3}$ at 210 MHz and 2450 MHz, respectively: the observed electron density at the end of the column is close to $\bar{n}_{e(\mathrm{re})}$ only at 2450 MHz because then the ν/ω ratio is much less than one (low-collision regime).

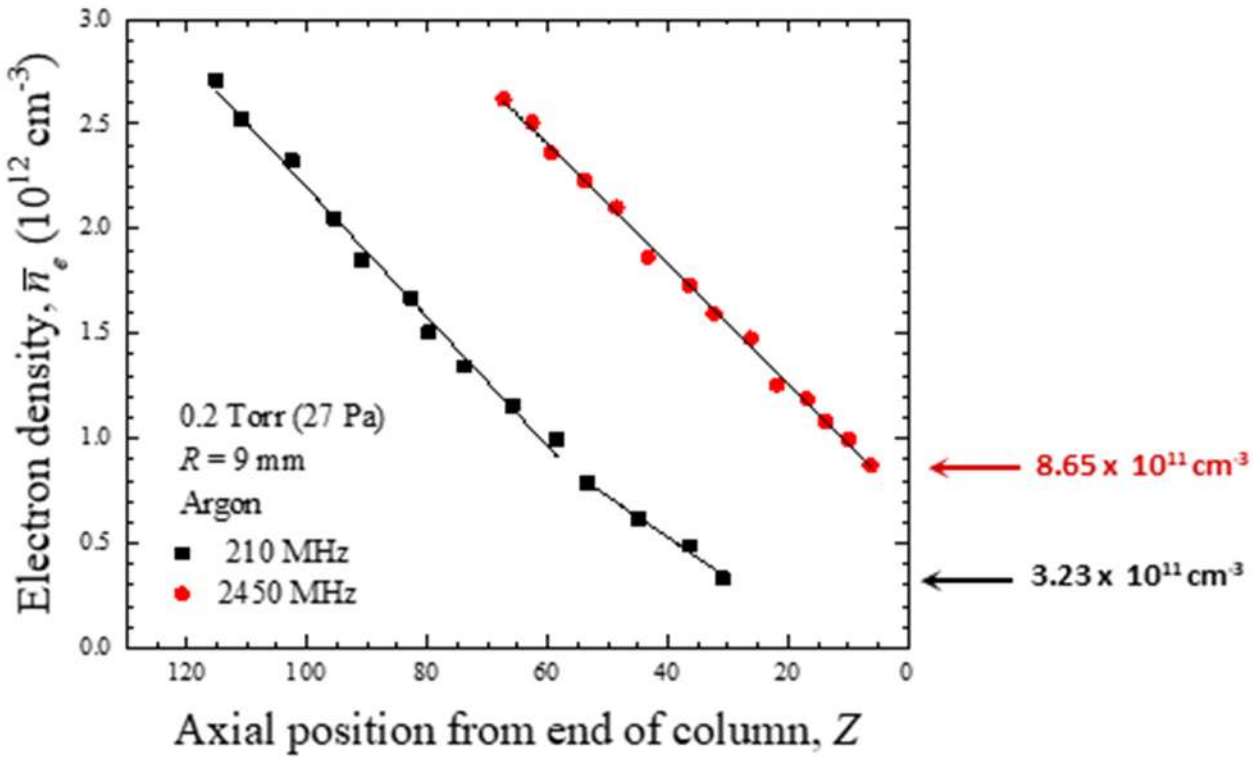

FIG. 12. Measured axial distribution of the electron density (using the $TM_{010}$ cavity method) of a TWD held in argon at 27 Pa inside a fused silica tube with an inner radius of $R$ = 4.5 mm, at frequencies of 210 and 2450 MHz, both curves obtained with a Surfatron. Least-squares linear regression yields at 210 MHz, $r^2$ = 0.991 for the antenna-like driven plasma segment and $r^2$ = 0.981 (only 4 points) for the travelling wave-sustained one. At 2450 MHz, however, there is no sign of an antenna-like zone, with $r^2$= 0.997 for the TWD segment (adapted from Chaker, Moisan and Zakrzewski,1986). The arrow outside the figure frame indicates the lowest observed electron density.

Figs. 11 and 12 confirm that the linearity of the axial electron density distribution is maintained across the entire 27–2450 MHz frequency range, although there are corresponding significant changes in the plasma properties: it does not depend on them, as erroneously currently accepted. Our opposition to the current incorrect belief is further strengthened by the finding that the corresponding operating conditions domain encompasses electron density values spanning 18 orders of magnitude (Fig. 11: 2.2×$10^8$ $cm^{-3}$ – Fig. 16: 4.1×$10^{26}$ $cm^{-3}$). Such a large variation in electron density would normally be expected to alter the electron energy distribution function (EEDF): indeed, the EEDF is most probably not Maxwellian in the low-density electron plasma at 27 MHz but should gradually tend toward a Maxwellian shape as the electron density increases with the wave frequency coming up to 2450 MHz. Nonetheless, this extremely large variation in electron density does not affect the linearity of the axial electron density distribution. Clearly, TWD behaviour is independent of EEDF characteristics, contrary to the beliefs of theorists (see Sec. 5 in Part II).

### *3.3.3. TWDs at 915 and 2450 MHz in neon*

Fig. 13 compares the axial distribution of the radial mean electron density observed at 915 MHz and 2450 MHz in a 3 mm radius tube at atmospheric pressure, this time, in neon gas. Analysis of the data for the axial distribution of electron density using least squares regression again reveals a linear relationship. However, the $r^2$ value is lower than those quoted above due to the smaller number of experimental points (only four and five data points at 915 MHz and 2450 MHz, respectively). As expected from relationship (5), the slope at 2450 MHz is steeper than at 915 MHz (Kabouzi *et al.*, 2002).

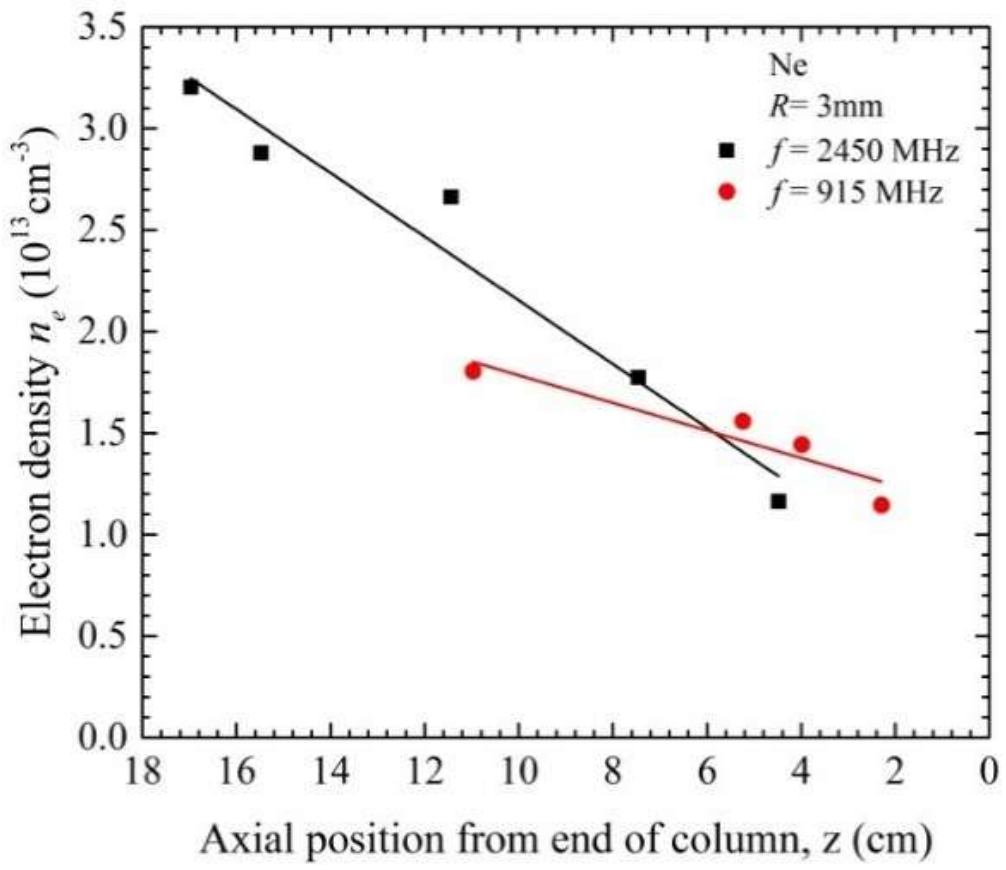

FIG. 13. Measured axial distribution of the radially averaged electron density of a TWD sustained at 915 MHz and 2450 MHz as displayed from the end of the plasma columns. The experiment was conducted in a fused silica discharge tube with an inner radius of 3 mm containing neon gas at atmospheric pressure (Castaños-Martinez, 2004). The electron density was determined from the broadening of the $H_\beta$ line (486.1 nm) caused by hydrogen atoms supplied by a minimal amount of water vapour in the discharge gas. The coefficients of determination of the least-squares regression were $r^2 = 0.80$ and 0.95 for 915 MHz and 2450 MHz, respectively. It should be noted that assigning $r^2$ values to a small number of data points is debatable.

*3.3.4. The ion density increases at the expense of the electron density in TWDs when the frequency of the travelling wave is lowered from MW towards the RF levels. However, the linearity of the axial distribution of electron density remains unaffected*

Our model assumes that the energy supplied to the electrons by the ***E***-field component of the wave causes them to heat up, resulting in the subsequent ionisation of the discharge gas. However, at frequencies below 100 MHz (i.e. in the RF range), the ions also heat up, absorbing some of the energy from the wave's ***E***-field. Nevertheless, Fig. 11 clearly showed that operating at frequencies below 100 MHz does not affect the linear behavior of the axial distribution of electron density, which additionally always obeys relation (5).

However, the slope of the axial distribution of the electron density becomes smaller than predicted by (5) as the frequency decreases from 100 to 50, more to 27 MHz, due to a decrease in the energy given to electrons in favour of ions. Table III confirms, through the measurements from Table II, our assertion that the slope of the axial distribution of the electron density decreases more rapidly than expected when the value of the frequency is lowered in the RF range. In fact, in the MW range, the quotient of the slopes agrees with the exact ratio of their frequencies (relation (5)), as shown in Table III with the ratio 200/100, whereas this is not the case for 200/27: at 27 MHz, the lowest frequency in the table, a great part of the energy of the ***E*** field is taken away from the electrons, making the ions relatively more numerous as the frequency of the wave is lowered.

| TABLE II | Measured slope of electron density from Fig. 11 (relative value) |
|---|---|
| 27 MHz | 1.7 |
| 50 MHz | 2.1 |
| 100 MHz | 3.2 |
| 200 MHz | 6.3 |

| TABLE III | Frequency ratio | Measured slope ratio | Measured slope ratio/ frequency ratio |
|---|---|---|---|
| 50/27 | 1.85 | 1.23 | 0.66 |
| 100/27 | 3.70 | 1.88 | 0.51 |
| 100/50 | 2.00 | 1.52 | 0.76 |
| 200/27 | 7.41 | 3.71 | 0.50 |
| 200/50 | 4.00 | 3.0 | 0.75 |
| 200/100 | 2.00 | 1.97 | 0.98 |

### *3.4. Linearity of the axial distribution of the electron density of TWDs as a function of the inner radius of the discharge tube*

Since $\frac{d\bar{n}_e}{dz} = -C'_0 \frac{fp}{R}$ (5), the smaller the discharge tube radius, the steeper the slope of the plasma column density gradient.

#### *3.4.1. Low-collision regime*

Figure 14 shows the measured mean radial electron density along the plasma column supported by a 100 MHz travelling wave in a fused silica discharge tube with an inner radius of either 3.2 cm or 6.2 cm, in argon gas at a low pressure of 1.8 Pa (10 mTorr) (Chaker, Moisan and Zakrzewski, 1986). As shown in the figure and caption, a straight line with a high degree of confidence is obtained for both $R$ values when a least-squares regression is performed on the axial electron density data points. More precisely, the measured slopes (expressed in relative units) for $R$ = 3.2 cm and $R$ = 6.2 cm are 13 and 5.9, respectively. It should be noted that the inverse ratio of the slopes measured from the radius of the small tube to that of the large tube is 5.9/13 = 0.45, whereas the inverse ratio of the corresponding radii (6.2/3.2) is higher, but slightly, at 0.52.

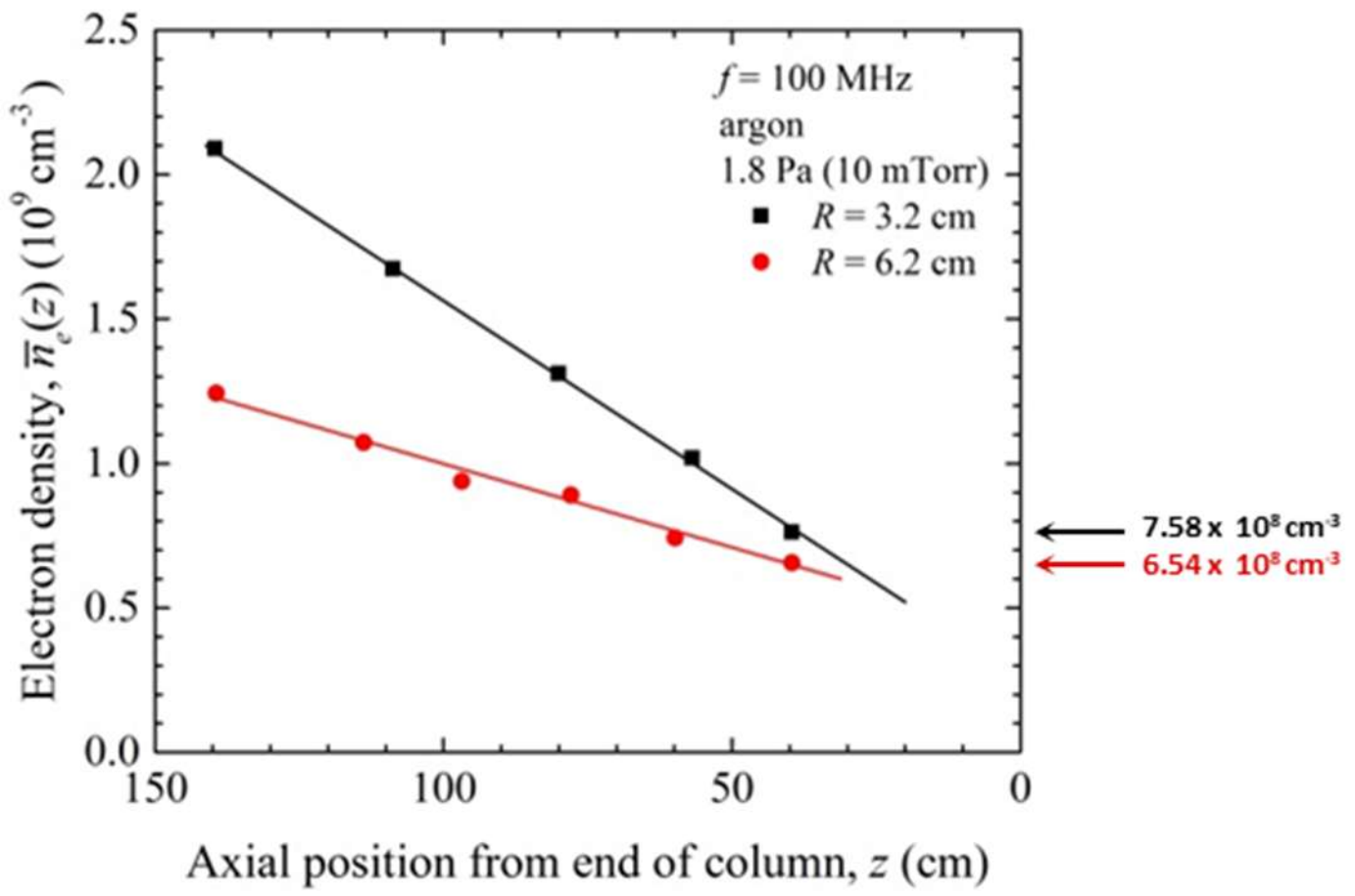

FIG. 14**.** Measured axial distribution of the mean radial value of the electron density with respect to the end of the plasma column supported by a 100 MHz travelling wave. The experiment was conducted for two (large) values of the internal radius *R* of the discharge tube, in argon gas at the (very) low pressure of 1.8 Pa (10 mTorr) (Chaker, Moisan and Zakrzewski, 1986). The electron density in the *R* = 32 mm tube was determined using a $TM_{010}$ resonant cavity method, while the axial phase variation of the TW was used for the larger *R* = 62 mm tube (Margot-Chaker *et al*., 1988). The values of $r^2$ are 0.993 and 0.986 for *R* = 32 mm and 62 mm, respectively. According to equation (6), the calculated electron density at the end of the TWD plasma column at 100 MHz and assuming $\nu/\omega \ll 1$, is $\bar{n}_{e(\mathrm{re})}$ = 6.62×10⁸ cm⁻³, which corresponds, within the experimental error, to the value measured in the tube of larger radius.

*3.4.2. High-collision regime*

Figure 10 above showed the measured axial distribution of the radially averaged electron density in a TWD supported at 915 MHz in argon gas at (twice) atmospheric pressure. Recall that it specifically concerned discharge tubes with three different internal radii and the same thick outer wall radius. The $r^2$ value of their axial electron density distribution confirmed its linearity. At the same time, this indicates that, even at pressures as high as atmospheric pressure, the lower the value of *R*, the steeper the corresponding slope of the axial electron density distribution, in agreement with relationship (5).

*3.4.3. Radial contraction of the plasma column does not affect the linearity of the axial distribution of the electron density of TWDs*

In the low-collision regime, the TWD plasma typically fills the discharge tube radially. However, radial plasma contraction does occur at pressures approaching one Torr or more if the value of one or more of the following parameters reaches a certain relative magnitude: the inner radius of the discharge tube; the frequency of the wave; and the discharge gas mass (Kabouzi *et al.*, 2002, Castaños-Martínez and Moisan, 2011).[7] In the case of Fig. 15, radial contraction of the plasma is bound to happen at atmospheric pressure, because the tube radius is large enough and the wave frequency high. However, in this case, the lower mass of neon compared to argon means that the plasma column is likely to heat less and thus contract less than an argon TWD under the same operating conditions. As shown in the figure, radial contraction causes the plasma column's radius to decrease continuously towards its end, resulting in an increasingly nonuniform plasma column.

[7] The radial contraction of TWD plasma columns is a characteristic of high-pressure electrical discharges in sufficiently large tubes. Plasma contraction in rare gas discharges is related to the influence of the gas temperature $T_g$ upon the concentration of molecular ions $X_2^+$. In contracted discharges, $T_g$ decreases strongly toward the discharge tube wall, ensuring the increase of $X_2^+$. Since the density of molecular ions is greater toward the wall, the loss of electrons by volume dissociative recombination is higher, ionization is less important, resulting in a contracted plasma (Ridenti, Spyrou and Amorim, 2014).

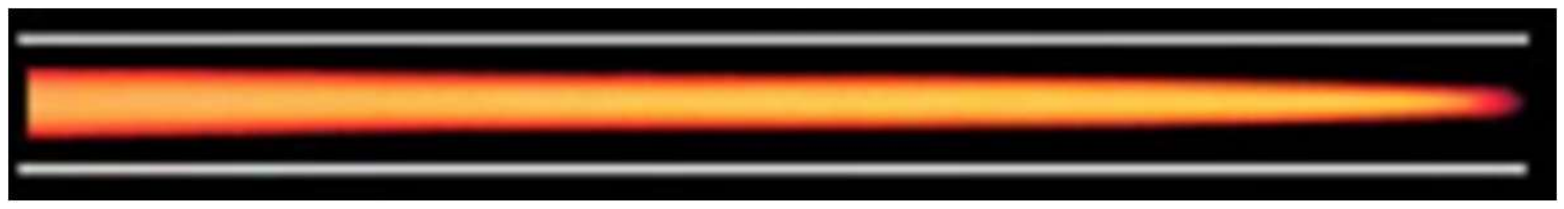

FIG. 15. Photograph showing a radially contracted TWD sustained at atmospheric pressure in neon gas in a tube with an inner radius of 6 mm and operating at 915 MHz. As can be seen, the radius of the plasma column decreases continuously towards its end, making it structurally increasingly nonuniform in the axial direction (Castaños-Martínez and Moisan, 2011).

Assuming an already contracted TWD plasma column, increasing the internal radius of the discharge tube causes the plasma column to move further and further radially away from the inner wall of the discharge tube at a given axial position. At the same time, the slope of its axial electron density distribution becomes gentler (Castaños-Martínez, 2004; Castaños-Martínez and Moisan, 2011). The degree of contraction increases as the thermal conductivity of the gas increases: for given conditions, this means successively helium, neon, argon, krypton and xenon in the case of noble gases. For instance, at 915 MHz, plasma contraction is evident in a 3 mm radius argon tube at atmospheric pressure but absent in a neon and helium tube (Fig. 4.26 in Moisan and Pelletier, 2012).

A plasma column subject to radial contraction, as shown in Fig. 15, is structurally an axially inhomogeneous medium. Nevertheless, its axial distribution of electron density remains linear (not shown), as in the case of an uncontracted plasma column. The reason for this is that the traveling wave that maintains the plasma column does not propagate on it, but along the outer wall of the discharge tube surrounded by vacuum (or ambient air): therefore, we should not expect a dependence on the properties of the plasma column, but rather on the permittivity of the dielectric tube, which is constant axially.

### *3.5. Linearity of the axial distribution of electron density as a function of the discharge gas flow rate*

It is interesting to consider how variations in gas flow within a discharge tube affect the axial distribution of electron density in a TWD. According to equation (4), the linearity of the axial distribution of electron density in a discharge tube with a constant cross-section should not be affected by the gas flow rate, since the gas density $N$ is constant along the tube's length. However, increasing the flow velocity of a gas in a tube decreases its inlet pressure (Bernoulli's law of gases). This results in a lower $N$ value and consequently a lower slope of the axial electron density distribution, as observed.

The experiment was carried out in argon at atmospheric pressure in a fused silica tube with an internal radius of 0.75 mm to prevent radial contraction of the plasma and filamentation (Martínez-Aguilar *et al.,* 2019). Gas flow rates of 0.25 and 1 standard litre per minute (slm) were used in an open-ended tube. The discharge was at 2450 MHz using a surfaguide, — an ***E***-field applicator with a launching gap on either side of the waveguide structure (see Fig. 4a) — producing two plasma columns. Depending on the direction of the gas flow, these columns are described as either 'forward' or 'backward'. The two columns initially receive an equal share of the power emitted by the surfaguide. However, the columns in the opposite direction to the gas flow are shorter, as shown in Fig. 16 and explained in the theoretical work of Kabouzi *et al.* (2007). The electron density was obtained from the broadening of the $H_\beta$ 486.1 nm Balmer line ((Palomares *et al.*, 2009)). Key points to note from this figure are that the axial distributions of electron density remain linear in all cases and that the slope corresponding to the greatest gas flow (smallest value of *N*) is the smallest in both directions, which is consistent with the role of *N* in relation (4).

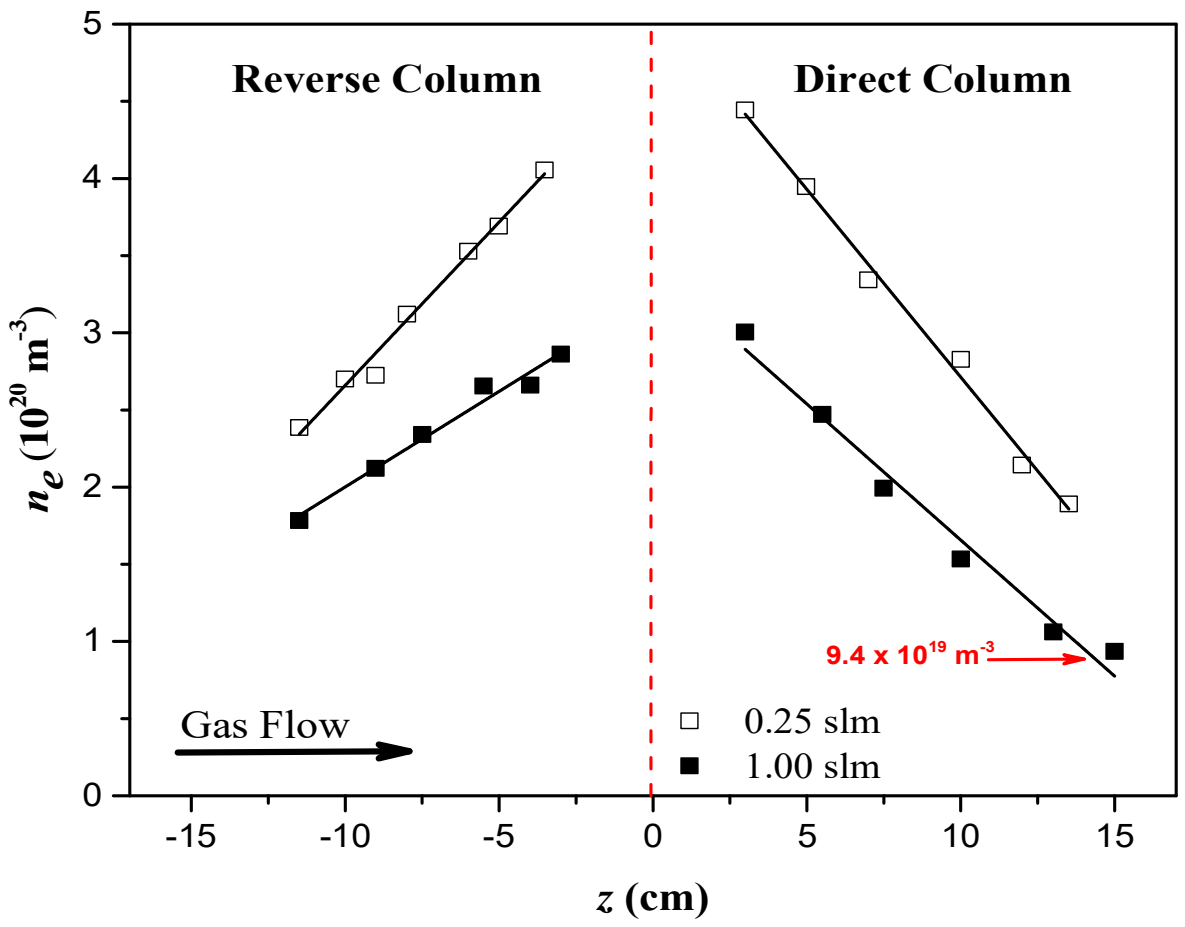

FIG. 16 (foreign laboratory). Experimental axial distribution of the TWD electron density accounting for the argon gas flow direction (forward and backward) and two gas flow rates at atmospheric pressure with $R$ = 0.75 mm and $f$ = 2450 MHz. The electron density was determined by the broadening of the $H_\beta$ Balmer line (486.1 nm). The data points from Martínez-Aguilar *et al.* (2019) were processed using least-squares regressions, yielding $r^2$ values of 0.985 and 0.992 for the 0.25 slm backward and forward columns, respectively, and 0.971 and 0.974 for the 1 slm backward and forward columns, respectively. This confirms the linearity of these axial distributions. The predicted minimum electron density for travelling wave propagation $\bar{n}_{e(\mathrm{re})}$ (6) at 2450 MHz is 1.7x10$^{16}$ m$^{-3}$ whereas the TW is observed to stop propagation at 9.4x10$^{19}$ cm$^{-3}$, meaning cessation of wave propagation occurs more than three orders of magnitude above $\bar{n}_{e(\mathrm{re})}$.

*3.6 Linearity of the axial distribution of electron density in TWDs of molecular gases*

As all previous figures concerned TWDs in atomic gases such as argon and neon, it is worth examining what happens to TWDs generated in molecular gases. We know that these discharges are mainly composed of molecular ions at low electron density. However, at high electron density, the increased rate of molecular dissociation means that they comprise an ever-greater proportion of atomic ions (Kabouzi *et al.*, 2002; Kabouzi *et al.*, 2007). As we shall see, the difference in concentration of atoms and molecules in these discharges has no effect on the linearity of their axial electron density distribution.

### 3.6.1. Nitrogen

Fig. 17 Illustrates the axial electron density distribution in an $N_2$ TWD sustained within a Pyrex™ glass tube ($\varepsilon_g$ = 4.52) featuring a substantial inner radius of 22.5 mm and operating at a low MW frequency of 500 MHz, with at a gas pressure of 0.5 Torr (~67 Pa). These combined conditions correspond to a radially uncontracted discharge (Dias *et al.*, 1998).

Electron density was determined using the Langmuir probe technique (see Appendix B). According to Grosse *et al.* (1994a), the technique was initially intended to "gather information about the actual EEDF", but it has since been found to "act as a source of inhomogeneity in the plasma" due to the disturbance it causes to wave propagation. This method provides electron density values that are significantly lower than the minimum value calculated for TW propagation, $\bar{n}_{e(\mathrm{re})}$ = 1.66 x$10^{10}$ cm$^{-3}$ (indicated by the arrow outside the figure frame). We have previously confirmed in the current paper the existence of this minimum value experimentally, using both the $TM_{010}$ resonant cavity method (see Figs. 8b, 11 and 12) and the axial phase variation of the travelling wave (see Fig. 14). Therefore, a TWD cannot have an electron density lower than $\bar{n}_{e(\mathrm{re})}$. Despite this inconsistency, the axial distribution of electron density obtained using this method is as linear as that obtained using the various diagnostic methods mentioned above. This also means that the linearity of the axial distribution of electron density is maintained, even when dealing with a molecular discharge gas rather than an atomic one.

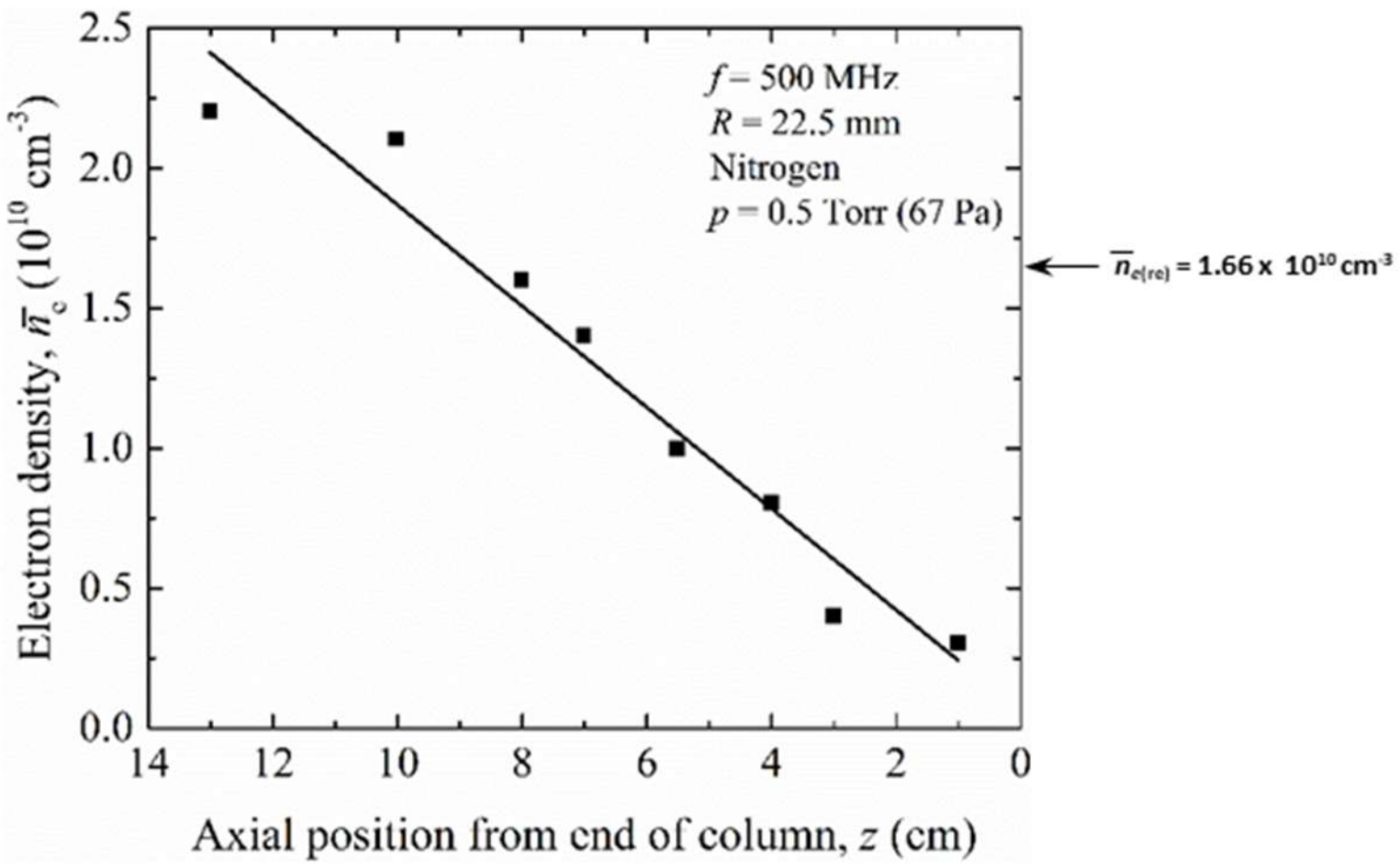

FIG. 17 (foreign laboratory). Axial electron density distribution determined by using the Langmuir probe technique for a TWD sustained in $N_2$ gas at 67 Pa within a dielectric discharge tube ($\varepsilon_g$ = 4.52) with an inner radius $R$ = 22.5 mm at 500 MHz. The corresponding minimum electron density for wave propagation along the TWD (6), $\bar{n}_{e(\mathrm{re})}$ = 1.66x$10^{10}$ cm$^{-3}$, is shown on the outside of the figure frame. Least-squares regression of the data points from Dias *et al.* (1998) gives a linear fit with $r^2$ = 0.986.

### 3.6.2. Hydrogen

Fig. 18 shows the measured axial distribution of electron density in an $H_2$ TWD operating under the same conditions as $N_2$ in Fig. 17, also using a Langmuir probe to diagnose electron density (Gordiets *et al.,* 2000). Data point $Z$ = 28 (normalised axial position) is attributed to the antenna radiation region of the field applicator (Moisan, Levif and Nowakowska*,* 2019) and is therefore not part of the TWD plasma column. As briefly mentioned in relation to $N_2$ TWD, the microwave field actually disrupts the probe measurements, as it causes it to collect a time-varying current.

This shifts the apparent floating potential towards more negative values, leading to distortion of the probe characteristic and its second derivative. This results in the minimum electron density starting well below $\bar{n}_{e(\mathrm{re})}$ = 1.66x10$^{10}$ cm$^{-3}$. Nevertheless, we observe that linearity of the axial distribution of electron density is preserved whatever the kind of molecular discharge gases. This lends further support to our model.

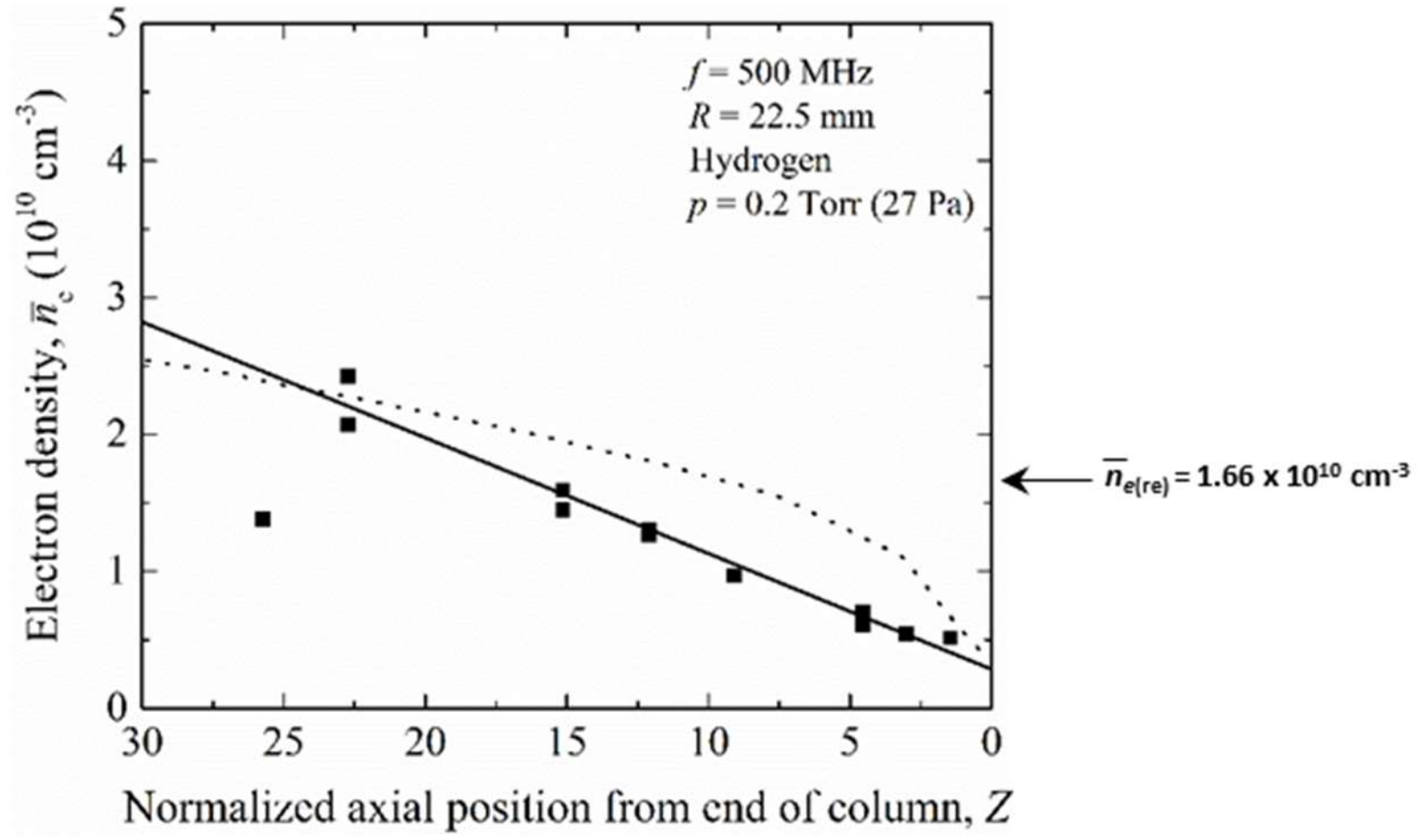

FIG. 18 (foreign laboratory). Measured axial distribution of the electron density in a TWD held at 500 MHz in $H_2$ gas at 27 Pa within a Pyrex$^{TM}$ glass discharge tube ($\varepsilon_g$ = 4.52) with an inner radius of 22.5 mm, as determined using the Langmuir probe diagnostic method. Excluding the value at the largest $Z$ (related to the antenna-like region of the field applicator), the least-squares regression yielded an $r^2$ value of 0.948. The whole figure was restructured from Gordiets *et al.* (2000) where the dotted curve is theoretical. It deviates significantly from the experimental results and will be discussed in Sec. 5.2.6.

*3.7. If the plasma column contracts radially, a diagnosis of electron density that considers only part of the radial distribution of electrons at each axial position will not produce a linear axial distribution of electron density*

*3.7.1. The total number of electrons in a radial section of a TWD plasma at a given axial position is proportional to the power dissipated by the travelling wave at that position*

Recalling the concept of power absorbed per electron (Moisan, Ganachev and Nowakowska, 2022), it is easy to agree that the total number of electrons produced in a radial section of the plasma column is proportional to the amount of power absorbed from the wave that generated them at that particular axial position.

If there is no radial contraction of the plasma column, its radial structure remains constant, i.e. does not vary along the axis. Therefore, whichever given portion of the radial electron distribution is considered in the diagnostic technique, this remains axially proportional to the total electron distribution, and consequently to the power dissipated by the travelling wave along the axis. In this case, a linear distribution of electron density should be observed.

In the $TM_{010}$ resonant cavity, the total electron density is considered because the entire radial distribution of electrons is exposed to the EM field within the cavity. Similarly, the electron density diagnostic by the axial variation of the wave phase yields an average total electron density for each radial section of the plasma column at each axial position (see Appendix B).

In both diagnostic techniques, the graph of the radially averaged electron density is therefore proportional to the power lost by the wave at that axial position. This observation, previously made just before Sec. 1.1., provides an innovative new perspective on the axial distribution of electron density, revealing an additional law of energy conservation, which links the wave and the electrons of the plasma column (first proposed by Aliev, Boev and Shivarova in 1982).

3*.7.2. This is the case of electron density diagnosis, in which only some of the electrons in the radial distribution are considered while the plasma column undergoes radial contraction*

Consider, for example, Fig. 19a, which illustrates the Thomson scattering (TS) system employed for measuring the electron density within a Surfatron plasma column. The intersection of the laser beam and the focal spot of the triple grating spectrometer (TGS) (whose slit is parallel to the laser beam axis) defines the TS detection volume. Due to the narrowness of the laser beam (75 µm), only a very limited part of the electron radial distribution is detected.

Two situations can be envisaged:

i) there is no radial contraction of the plasma column (see case 3.7.1 above). Fig. 19b shows that this occurs at sufficiently low gas pressure (below 0.65 mbar), yielding an axial electron density distribution that is linear.

ii) in Fig. 19b, at gas pressures of 20 mbar and above, linearity is lost, as the radial electron density profile has contracted, according to the author of the paper.

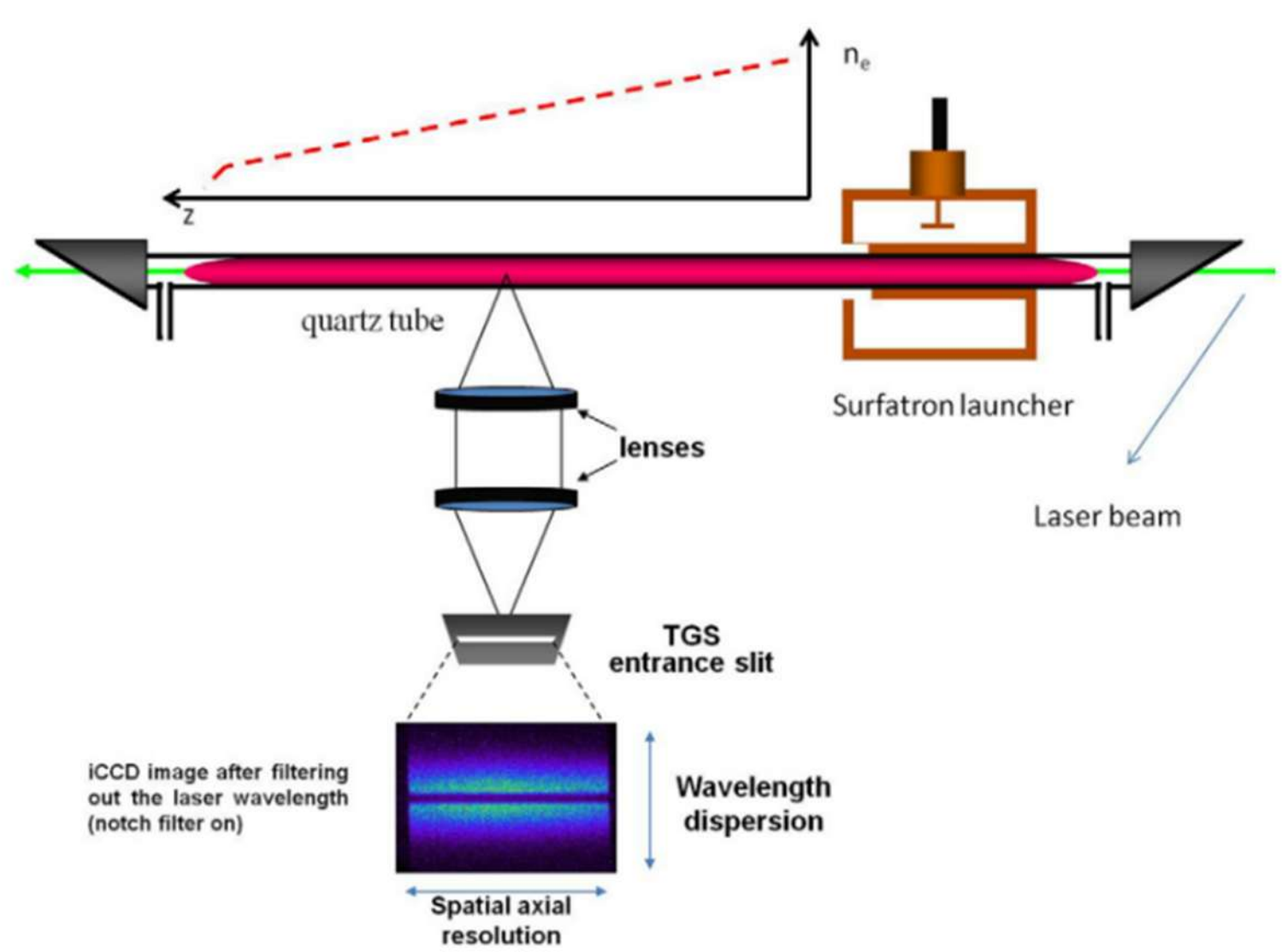

FIG. 19a (foreign laboratory). Diagram displaying the Surfatron plasma source and the TS diagnostic detection system. It illustrates how the entrance slit of the triple grating spectrometer (TGS) is focused on a small radial portion of a given axial section of the plasma column (Carbone *et al.*, 2012).

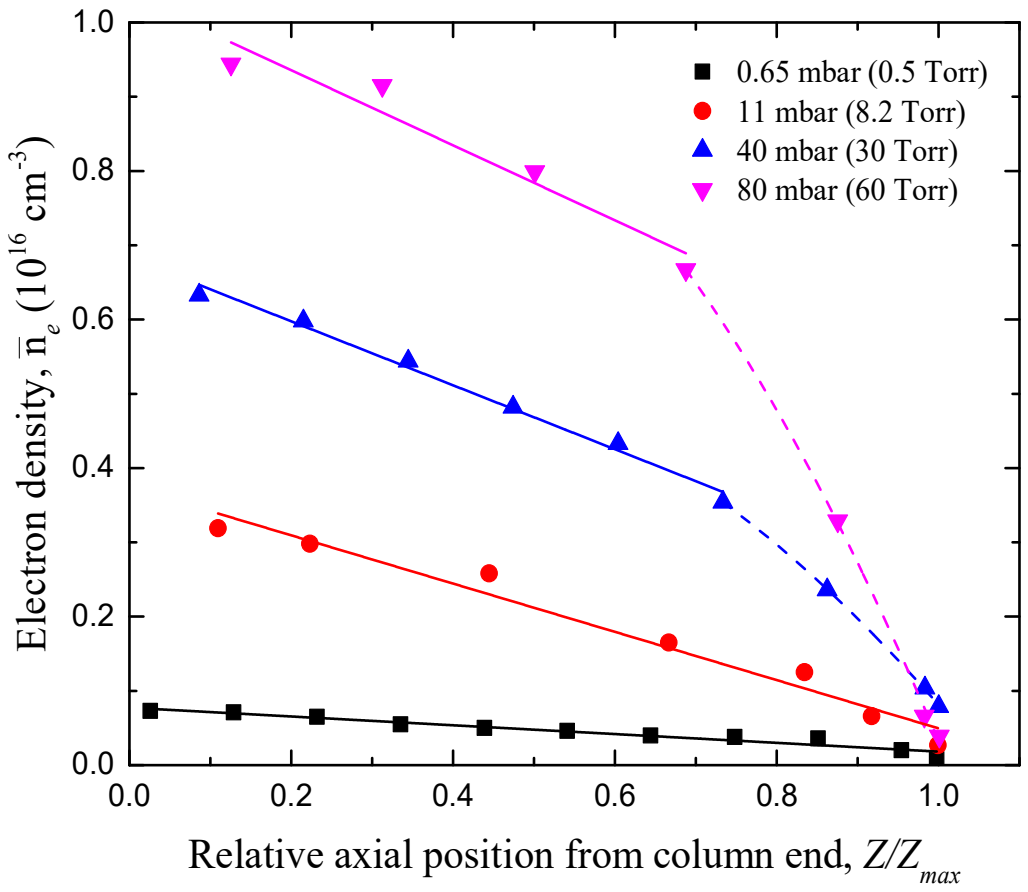

FIG. 19b (foreign laboratory). Axial electron density distribution determined by Thomson scattering along a Surfatron plasma column maintained at 2450 MHz in argon at four different gas pressures, inside a 3 mm radius fused silica tube (adapted from Hubner 2013a and 2013 b).

Note: Contrary to the linearity observed throughout Section 3, Fig. 7 at 270 W in Moisan, Pantel and Hubert (1990) shows experimentally a deviation from the linearity of the axial distribution of electron density, where the TWD is maintained at 915 MHz in a 0.97 mm radius tube at atmospheric pressure in argon. The axial electron density distribution of the plasma column at 270 W curves towards the end, which according to our model should not do so. We attribute this bending to an error in the data processing. In contrast, the 500 W and 700 W curves in the same Fig. 7 both show a linear electron density distribution along the entire length of the plasma column. As Fig. 12 in the current paper demonstrates, the axial electron density distribution of TWDs remains linear throughout its axial distribution of electron density, regardless of the total RF/MW power supplied.

This first part concludes that the axial distribution of electron density in TWDs is linear whatever the relevant diagnostics used and depends solely on the operating parameters $p$, $f$, and $R$, this over an electron density range covering 18 orders of magnitude: an exceptional and unique property of TWDs. This new conclusion was first expressed by Moisan in the initial version of his article on ArXiv (v1, June 2021). It contradicts the widely held belief in the field that it depends on the properties of the plasma column. It also suggests that TWDs could be used to optimise a given application in terms of its operating conditions.

The second part, which will follow, will show how traveling waves dissipate their energy whilst maintaining a TWD-type density distribution. It will also explain the mechanisms underlying the linear distribution of electron density observed in this first part.

## APPENDIX A: THE INITIAL PATENT APPLICATIONS THAT LED TO THE SURFATRON THE SURFAGUIDE AND RO-BOX EM FIELD APPLICATORS (TRAVELLING WAVE LAUNCHERS)

-*Surfatron.* The original patent application for the EM field applicator, later named the *Surfatron*, was filed by the Agence Nationale de la Valorisation de la Recherche (ANVAR), a French government agency, and was designed by Michel Moisan, Philippe Leprince, Claude Beaudry and Émile Bloyet. It was registered on 31 October 1974 by the Institut National de la Propriété Industrielle (INPI, National Agency for Industrial Property), it was given the number FR7436378A and the title:

*Perfectionnements apportés aux dispositifs d'excitation, par des ondes HF, d'une colonne de gaz enfermée dans une enveloppe* (Improvements to devices for exciting with HF waves a column of gas enclosed in a casing).

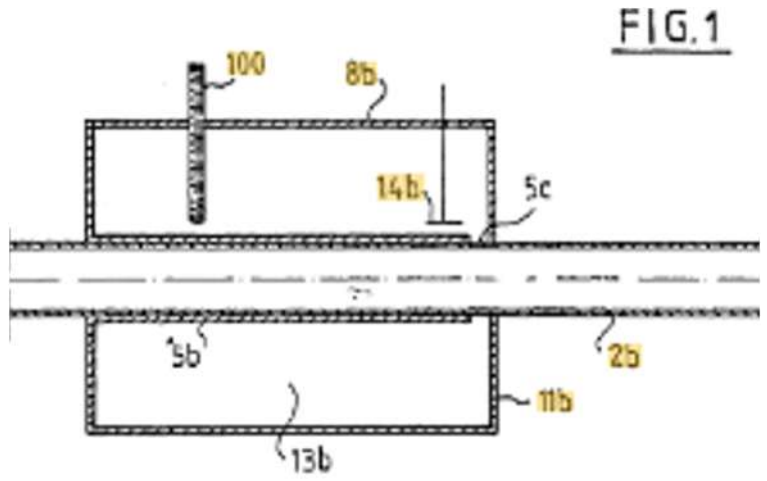

FIG. A1 (FIG. 1 in the initial patent disclosure). Schematic description of the cylindrical EM field applicator of the reentrant-type cavity, which encompasses a dielectric discharge tube coaxially (2b). It shows the radially movable high-frequency (HF) power coupler with its capacitive plate perpendicular to its end (14b). This coupler is located near the leakage interstice (gap) between the EM field and the tube (5c); this ensures impedance matching with the HF generator. A tuning screw (100) could also facilitate impedance matching of this device, which was later named the 'Surfatron'. Moisan, Zakrzewski and Pantel (1979) emphasise that the applicator enables stable, reproducible and energy-efficient generation of TWDs.

On 28 May 1975, a supplement to the first application was registered with priority over FR7436378A. This supplement was designated as the certificate of addition FR7533425 and was entitled:

-*Surfaguide. Perfectionnements aux dispositifs d'excitation, par des ondes hyperfréquences, d'une colonne de gaz dans une enveloppe allongée* (Improvements to microwave excitation devices for a column of gas in an elongated envelope).

FIGS. A2a and A2b (FIG. 2 and FIG. 2A in the original patent disclosure) show a schematic representation of a waveguide-type EM field applicator, later designated as a 'Surfaguide'. This applicator is fed by a coaxial high-frequency (HF) power connection (135) to a MW generator. At the opposite end of this connection is a movable impedance-tuning piston (136).

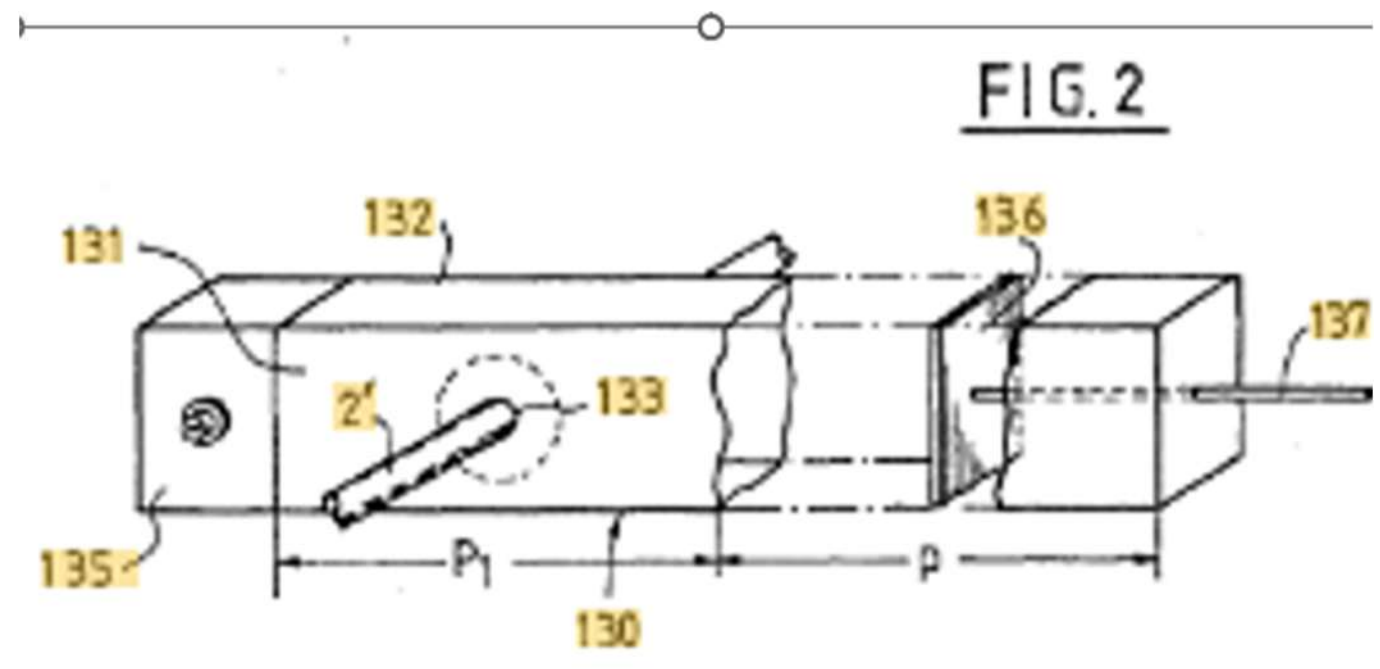

FIG. A2a: first step in the design of the future waveguide-based EM field applicator intended to become the Surfaguide.

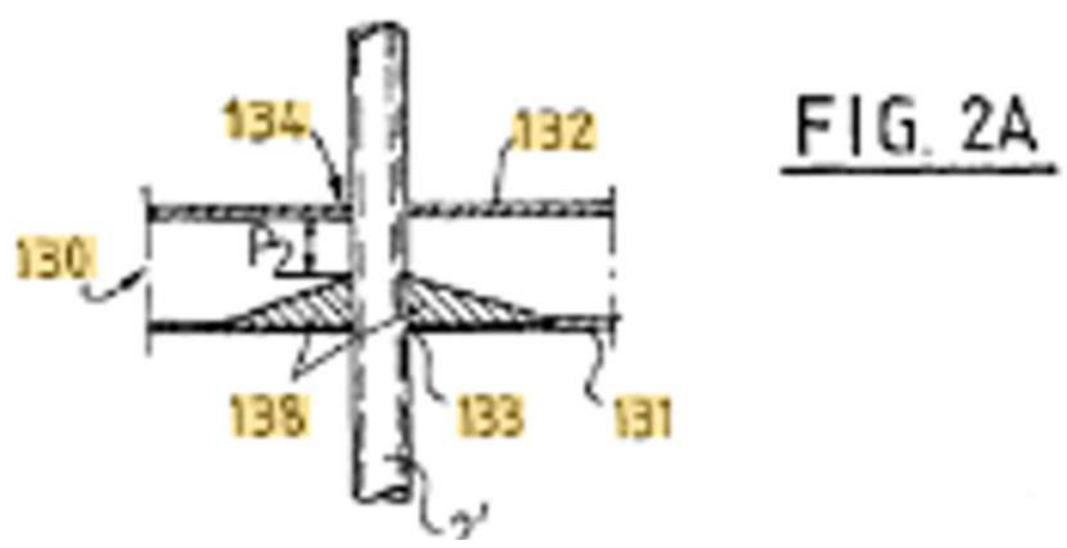

FIG. A2b (FIG. 2A in the patent disclosure). Optimising impedance matching with the MW generator by reducing accordingly the waveguide narrow-wall height ($P_2$) close to the launching gap (134).

Patent application FR2290126B1 was granted on 8 December 1978 for both the Surfatron (Fig. A1) and the Surfaguide (Figs. A2a and A2b). The patent was then extended to other countries, including the United States, where it was registered under the title 'Improvements relating to devices and methods of using HF waves to energise a column of gases enclosed in an insulating casing' (US Patent 4,049,940, 1977 covers both the Surfatron and the Surfaguide).

The term 'Surfatron' (Fig. A1) first appeared in the literature in Bertrand *et al.* (1977), while the term 'Surfaguide' (Figs. A2a and A2b) first appeared in Moisan *et al.* (1984).

-*Ro-Box.* The Ro-Box, an EM field applicator that can produce TWDs at frequencies typically below 100 MHz.

The device is designed to extend the Surfatron's frequency range below 100 MHz, as otherwise, the Surfatron's length would become unwieldy. The patent application was initially filed in Canada in December 1988 as: *'New surface wave launchers to produce plasma columns and means for producing plasmas of different shapes'*, and was granted as Canadian patents 1,246,762 (August 1990) and 1,273,440 (a division of the initial Canadian parent patent), respectively: this division distinguishes the EM field applicator device from the ability to achieve TWDs in dielectric vessel of various shapes. The corresponding US patent, number 4,906,898 (granted in March 1990), was a division of the US parent patent, number 4,810,933.

-*TIAGO*. The TIAGO[8] plasma torch: a high electron density TWD confined in an ambient gas at atmospheric pressure (Appendix D). Patented as: Moisan *et al*. (2002). First appearing in literature under the name TIA plasma torch in Moisan *et al.* (1994), the TIAGO impedance matching system, which is the one used in the Surfaguide, is significantly simpler and easier to use than that of the TIA. Many small-size TIAGO plasma torches can be assembled on a single waveguide ensuring non-interference and equal sharing of MW power among the various torches. French patent application FR 99/11422 (Pelletier *et al.*,1999) and, for example, US patent 6,727,656.

## APPENDIX B: SIX DIFFERENT MEASUREMENT TECHNIQUES WERE USED TO HIGHLIGHT THE PERFECT LINEAR SHAPE OF THE AXIAL DISTRIBUTION OF ELECTRON DENSITY ALONG THE TWD

[8] TIAGO is the French acronym for: Torche à Injection Axiale sur Guide d'Onde.

In Section 3, we demonstrated that the observed axial distribution of electron density decreases at a constant rate until it reaches a non-zero value at the axial position at which the EM wave can no longer propagate. Verifying the perfect linearity of this experimental distribution is essential, as it is the basic characteristic of TWDs. Consequently, any theoretical model that fails to accurately reproduce this distribution can be considered erroneous or at least incomplete. The various electron density diagnostics used are as follows:

-1 The discharge tube passes through the central hole of a flat, narrow, cylindrical conductive cavity operating in the $TM_{010}$ resonant eigenmode

The plasma is represented by its equivalent and relative permittivity $\varepsilon_p$, which can be calculated and/or calibrated with respect to the shift of the cavity resonance peak observed on an oscilloscope screen relative to the known permittivity of low-loss dielectric liquids (e.g., benzene). Varying the RF/MW power sets the length of the TWD plasma column relative to its end. This enables the different segments of the plasma column to pass through the cavity in succession, allowing the frequency shift of the resonance peak of each segment to be recorded relative to the resonance peak at the end of the plasma column. The axial distribution of the electron density can then be reconstructed segment by segment (Glaude *et al.,* 1980). One variant of this method involves calibrating the light emission intensity from several plasma segments (positioned at different points along the plasma column) against the corresponding electron density in the cavity. The intensity of light emitted along the column is fully recorded and compared with the light emission calibration in relation to electron density.

In all cases, the resonant cavity approach is limited by the fact that it becomes impossible to obtain a well-defined resonance peak above a certain level of electron density and gas pressure due to damping and distortion. Additionally, the resonant cavity of a large-diameter discharge tube must be much larger than the orifice for it to work properly, which makes it unwieldy and heavy. The resonant cavity method was used in Figs. 8, 11, 12 and 14.

-2 Determining the axial distribution of the electron density is achieved by recording the wavelength $\lambda$ of the travelling wave through its phase variation along the plasma column

This phase variation is detected using an ***E***-field antenna[9], which moves along the discharge tube. Its signal feeds the first leg of a double balance mixer. The second leg is connected to the MW generator and acts as a phase reference. The mixer's output signal provides the axial variation of the wave phase, shown as curve *c* of the left panel of Fig. A3. Some values of the deduced wavelengths $\lambda$ are shown on it by way of example. The corresponding electron density can then be determined by comparing it with the calculated phase diagram, as discussed.

A *phase diagram* plots $\omega/\omega_{pe}$ (where then $\omega$ is constant in contrast to a wave *dispersion diagram*) versus $\beta R$ where the wave number $\beta = 2\pi/\lambda$. Here, the phase diagram (solid line in the right panel) is calculated under the assumption of a cold plasma column with a locally uniform electron density, and surface-wave propagation along it (Zakrzewski *et al.*, 1977). The corresponding data points $\beta a$ ($a$ is the plasma radius) are plotted on the calculated phase diagram (right panel). The agreement between the experimental points and the calculated phase curve is quite good. However, the calculated curve overshoots above $\left(1+\varepsilon_g\right)^{-1/2}$ due to the authors' assumption that the travelling wave propagates in the ionised discharge gas (In contrast, our model claims that the travelling wave primarily moves along the interface between the outer wall of the discharge tube and the surrounding vacuum (see Appendix E)). Nevertheless, the factor $\left(1+\varepsilon_g\right)^{-1/2}$, as a short, dashed line on the phase diagram, correctly marks the observed end of wave propagation in terms of electron density, i.e. the minimum electron density under low collision regime (6), thus excluding the calculated values resulting from this assumed overshoot above $\left(1+\varepsilon_g\right)^{-1/2}$ in the diagram.

[9] It is made from a semi-rigid coaxial cable with a diameter of approximately 2 mm. This cable is terminated by its bare conductor over a length of a few millimetres and is directed radially towards the outer wall of the discharge tube, at a distance of less than one mm.

Margot-Chaker *et al.* (1988) developed a detailed technique using the phase diagram to determine the electron density along TWDs. This method was applied to generate Figs. 7, 9 and 14 ($R$ = 62 mm) and has the advantage that it can be used with discharge tubes of a larger diameter and at a higher gas pressure than with resonant cavities. However, beyond a certain pressure limit, the TW phase diagram ceases to vary significantly with $\beta$: it then becomes impossible to accurately determine the electron density (see, for example, Moisan, Pantel and Hubert, 1990).

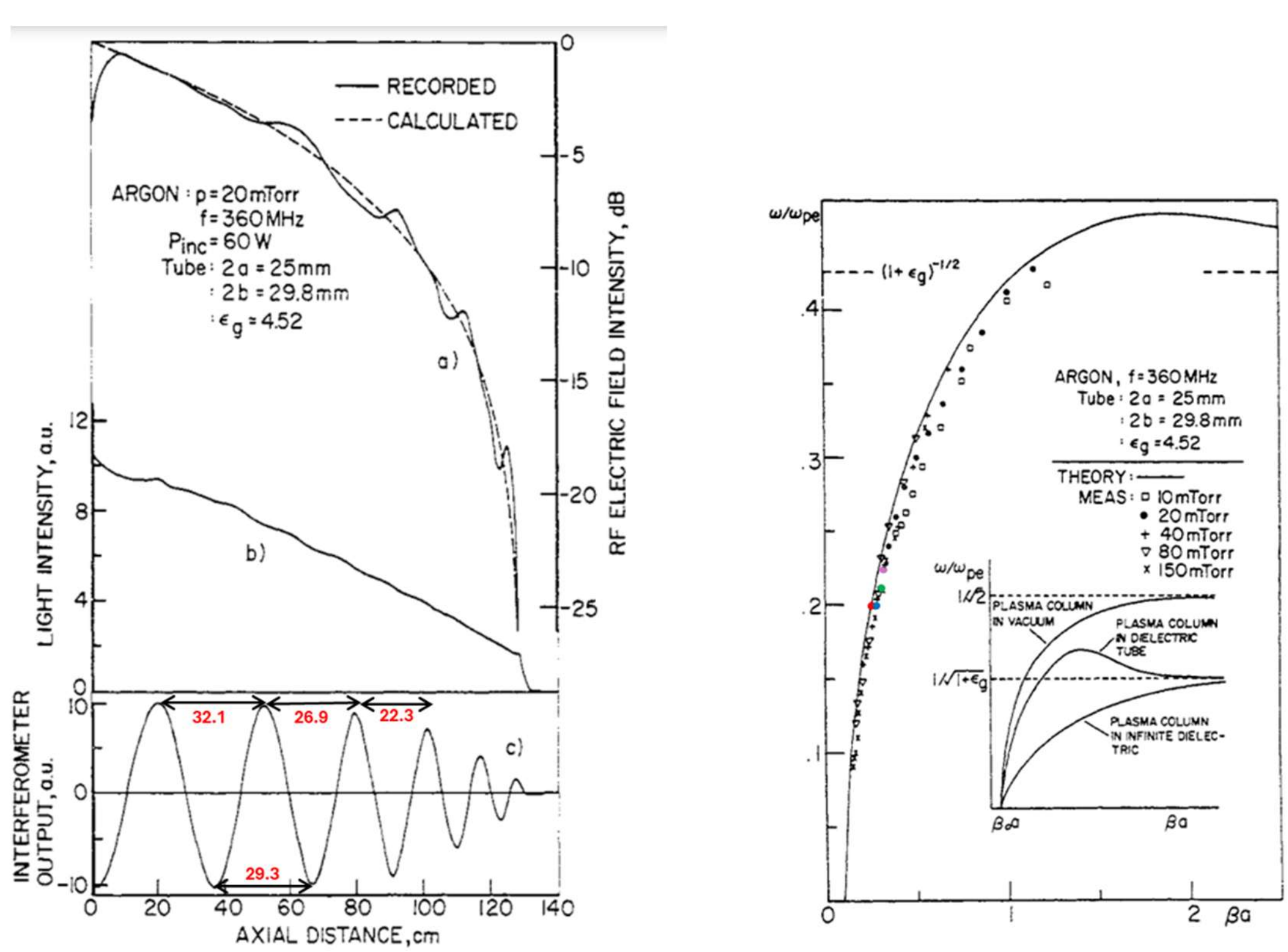

FIG. A3. Left panel: a) the electric field intensity of the travelling wave recorded axially along the external wall of the discharge tube using a semi-rigid coaxial cable acting as an antenna for the wave electric field component. The antenna is directed perpendicular to the discharge tube at less than a mm from its outer wall (see footnote 9); b) axial light intensity distribution recorded along the outside of the discharge tube; c) variation of the TW phase signal at the output of the double-balance mixer as a function of axial position from which the wavelength $\lambda$ is obtained providing the wave coefficient $\beta = 2\pi/\lambda$. The experimental $\beta a$ values (*a* is the plasma radius) are plotted in the right panel of Fig. A3 and compared to the calculated solid curve, which assumes a surface-wave extending axially along the plasma column at constant $\omega$ (*Zakrzewski et al.*,

1977)[10]. The maximum observed value of $\omega/\omega_{pe}$ is $(1+\varepsilon_g)^{-1/2}$, which corresponds to the observed minimum electron density obtained under low-collision regime, following equation (6). It really marks the end of the plasma column.

3- Electron density diagnostic from the Stark broadening of the $H_\beta$ emission line (486.1 nm)

Hydrogen atoms are made available in the discharge gas in the following ways:
i) in the argon plasma at twice atmospheric pressure in Fig. 10, 0.5% added hydrogen was used, which did not alter the length of the pure argon plasma column (Moisan, Pantel and Hubert 1990);
ii) in the neon plasma at atmospheric pressure (Fig. 13), the hydrogen atoms came from a minimal amount of added water vapour, such that its addition did not modify the length of the plasma column (Castaños-Martinez, 2004);
iii) the percentage of $H_2$ admixed with argon at atmospheric pressure in Martinez-Aguilar *et al.* *(*2019) is unspecified (Fig. 16);
iv) the gas in the TIAGO plasma torch operating at atmospheric pressure was a 9:1 argon-hydrogen mixture (see Fig. A10 in Appendix D).

-4 Electron density diagnostic through the Thomson laser scattering method

Figure 19a shows that the method of recording electron density from Thomson scattering provides a relatively high level of spatial resolution. This method can be used to diagnose a TWD plasma in a non-intrusive way. "This is an active spectroscopic method based on the scattering of laser photons by free electrons in the plasma. It does not need any model to account for the state of equilibrium departure" (Palomares *et al.*, 2010a and 2010b).

Fig. A4 refers to a Surfatron plasma at 2450 MHz and a pressure of 20 mbar in a 3 mm diameter discharge tube and with $\varepsilon_g$ = 3.84. It shows that excluding the data points closest to the Surfatron due to the antenna-like radiation from the EM field applicator, the axial distribution of the electron density is linear.

[10] The wavelength of the EM travelling wave in a vacuum, $\lambda_0$, at 360 MHz is 83.28 cm, while Fig. A3c (left panel) shows a wavelength of 32.1 cm at one given axial position along the plasma column. Since the phase velocity of an EM wave is $v_{ph} = \omega/\beta$ where $\beta = 2\pi/\lambda$, the guided wave is travelling at this point at 38% of the speed of light in a vacuum, decreasing as the electron density decreases toward the column end.

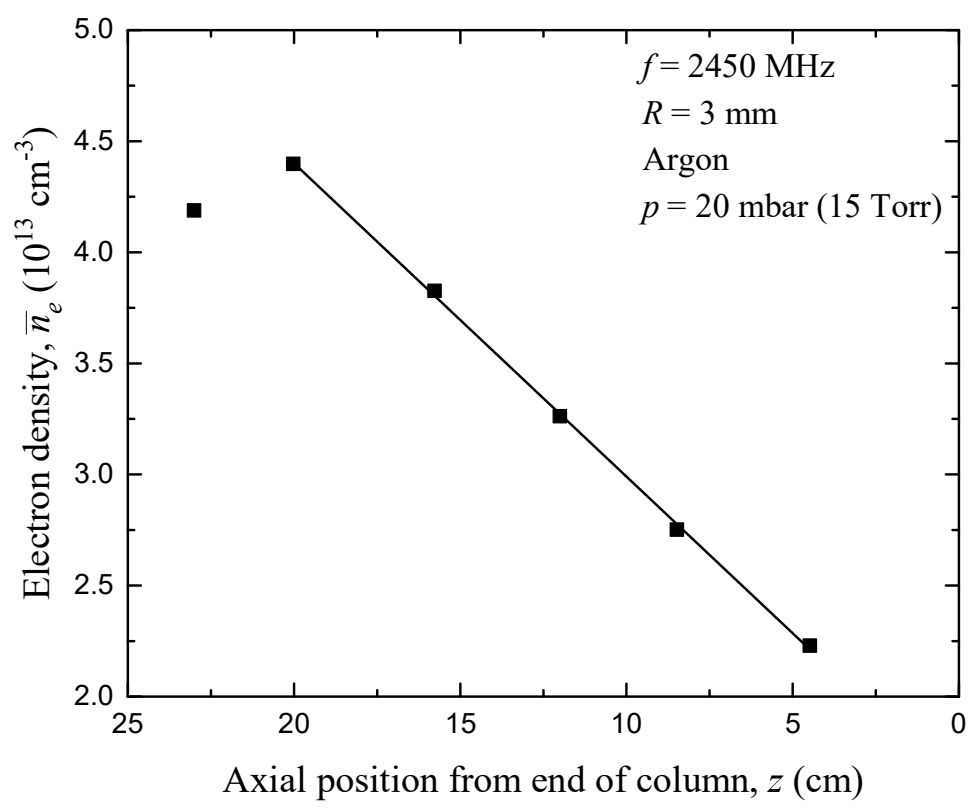

FIG. A4 (foreign laboratory). Axial electron density distribution obtained through Thomson laser scattering along a Surfatron plasma at 2450 MHz in a $R$ = 3 mm tube at 20 mbar (15 Torr), by selecting a few electrons in the radial cross-section of the plasma column in absence of radial contraction (Palomares *et al.*, 2010a and 2010b). Beyond the data point closest to the field-applicator (antenna-like radiation), the plasma column is linear and ends abruptly at a non-zero electron density, as expected from TWDs. Coefficient of determination $r^2$ = 0.9993.

However, this is not the case at higher gas pressures of 40 and 80 mbar (Fig. 19b), where the observed axial electron density distribution is no longer linear. This is primarily because the TWD plasma column undergoes radial contraction at these higher pressures. Due to the narrow width of the probing laser beam (see Sec. 3.7.2) and the electron density diagnostic method employed, only a portion of the radial distribution of the electron density is considered. However, linearity is restored when the entire radial cross-section of electron density is considered (not shown).

5- Electron density diagnosed through the continuum radiation of the argon line at 648 nm, calibrated with a standard tungsten ribbon lamp and expressed in absolute units (Iordanova *et al.*, 2008a and 2008b)

Fig. A5 shows the electron density $\bar{n}_e$ obtained using this diagnostic method in a Surfatron plasma at 2450 MHz and 15 mbar (11.2 Torr) in a $R$ = 3 mm tube with $\varepsilon_g$ = 3.84. Once again, the axial distribution of electron density is perfectly linear when the data point closest to the Surfatron — originating from the antenna-like radiation of the field applicator — is disregarded.

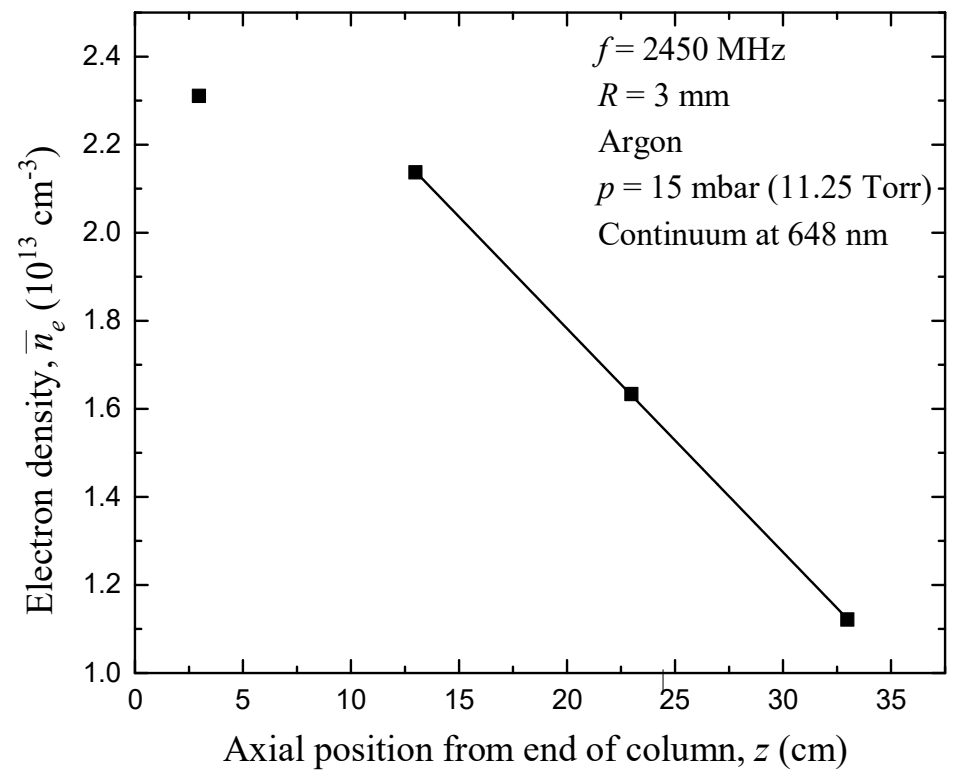

FIG. A5 (foreign laboratory)**.** Axial distribution of electron density measured using continuum radiation from the 648 nm argon line, calibrated using a standard tungsten ribbon lamp. The results are expressed in absolute units (Iordanova *et al.*, 2008a and 2008b). The coefficient of determination (with only three data points!) is $r^2 = 0.99996$.

6- Electron density diagnosed using the ion saturation current of an axially moving Langmuir probe paired with a reference probe placed inside the Surfatron (Gordiets *et al.*, 2000).

The authors indicate that installing of a Langmuir probe in a Surfatron TWD was intended to 'gather information on the actual EEDF', but it 'acts as a source of inhomogeneity in the plasma' (Grosse, Schluter and Tatarova, 1994b). This disturbance causes some of the electron density values shown in Figs. 17 and 18 to fall below the minimum electron density $\bar{n}_e$ (6), which contradicts the results obtained using all the other diagnostic methods for TWDs cited in this paper thus far.

In summary, all measurement methods produce the same axial electron density distribution for TWDs. This is characterised by a linear decrease of electron density from start to finish, followed by an abrupt end at a non-zero value. This result remains consistent regardless of the measurement technique employed, provided the antenna-like EM radiation region of the field applicator is avoided and provided, in the case of plasma radial contraction, that the whole radial distribution of electrons is considered. The coming Part II focuses on modelling.

---------------------------------------------------------------------------------------------------------------

**Part II: The origin of the linear axial distribution of electron density, the fundamental property of TWDs, is determined. Since the travelling waves that maintain TWDs are azimuthally symmetric TM waves, their field components are $E_r$, $E_z$ and $H_\varphi$, with $E_r^2 > E_z^2$ and $H_\varphi$ constant. As a result, these are shown to generate two Poynting vectors: one that describes the wave power flowing primarily along the outer surface of the dielectric discharge tube, and one that is radially directed and carries a smaller fraction of the total wave power, which is used to sustain the TWD locally.**

Part II is organised as follows: Section 4 provides a physical explanation of specific aspects of TWDs, leading to equations that accurately reproduce the characteristics of their plasma column, as obtained experimentally in Section 3. Section 5 offers a critical review of articles that deal with modelling the axial distribution of electron density along TWDs, highlighting erroneous modelling aspects. Section 6 concludes the article with a summary and discussion. Appendix C examines the axial distribution of electron density under various standing wave conditions. Appendix D considers the TIAGO plasma torch, and Appendix E uses the Poynting vector to determine how wave power flows to sustain the TWD.

## 4. Key mechanisms controlling the linearity of the axial electron density distribution in TWDs

As shown in Section 3, the axial distribution of electron density is always linear and depends only on $p$, $f$ and $R$, which are parameters set by the operator. This section will investigate the various mechanisms responsible for such a linearity.

### *4.1. For TWDs to be stable (stationary), the electron density must decrease monotonically towards the end of the column*

Early studies of SWDs (Glaude *et al.*, 1980) were based on the assumption, stemming from observation, that the electron density $\bar{n}_e(z)$ along these discharges was proportional to the power dissipated by the wave at each axial position. This made it possible to empirically establish a relationship between the wave power attenuation coefficient, $\alpha(z)$, and $\bar{n}_e(z)$[11]. Later, Zakrzewski (1983) sought to determine, in the case of a long plasma column (i.e. avoiding specifying the characteristics of the end of the column), the conditions necessary for stable discharge, whereby the production and loss of charged particles are balanced in steady state. This imposes the wave power and electron density to decrease monotonically all along the plasma column. This principle emerges from the following set of inequalities in wave power flow and electron density proposed by Zakrzewski (1983):

$$dP(z)/dz.\ d\bar{n}_e(z)/dz > 0 \text{ and } d\bar{n}_e(z)/dz < 0. \qquad (7)$$

who suggested then to express the proportionality of $\alpha(\bar{n}_e(z))$ to electron density analytically as:

[11] This empirical fact was first introduced in a theoretical article by Aliev, Boev and Shivarova (1982), but it was not physically justified (for that purpose, see Sec. 5.1.2 below).

$$\alpha(n) = An^k, \tag{8}$$

where $k$ is here an integer and $A$ is a constant, with the electron density $n$ normalised to its value $n_{cs}$ at the start of the TWD plasma column.

Recall the following expression from Glaude *et al.* (1980):

$$\frac{dn}{dz} = -2\alpha(n)n(z)\left(1 - \frac{n(z)}{\alpha(n)}\frac{d\alpha(n)}{dn}\right)^{-1}, \tag{9}$$

then from (8) and (9), there comes:

$$\frac{n(z)}{n_{cs}} = \left(1 + 2An_{cs}^k \frac{k}{1-k} z\right)^{-1/k} = \left(1 + \frac{2k}{1-k}\alpha_{cs} z\right)^{-1/k} \tag{10}$$

and $\frac{P(z)}{P_{cs}} = \left(\frac{n(z)}{n_{cs}}\right)^{1-k}$, (11)

where the normalised quantities $n_{cs}$, $a_{cs}$ and $P_{cs}$ are those at the start of the TWD plasma column. Fig. 20 is a graph of equation (10).

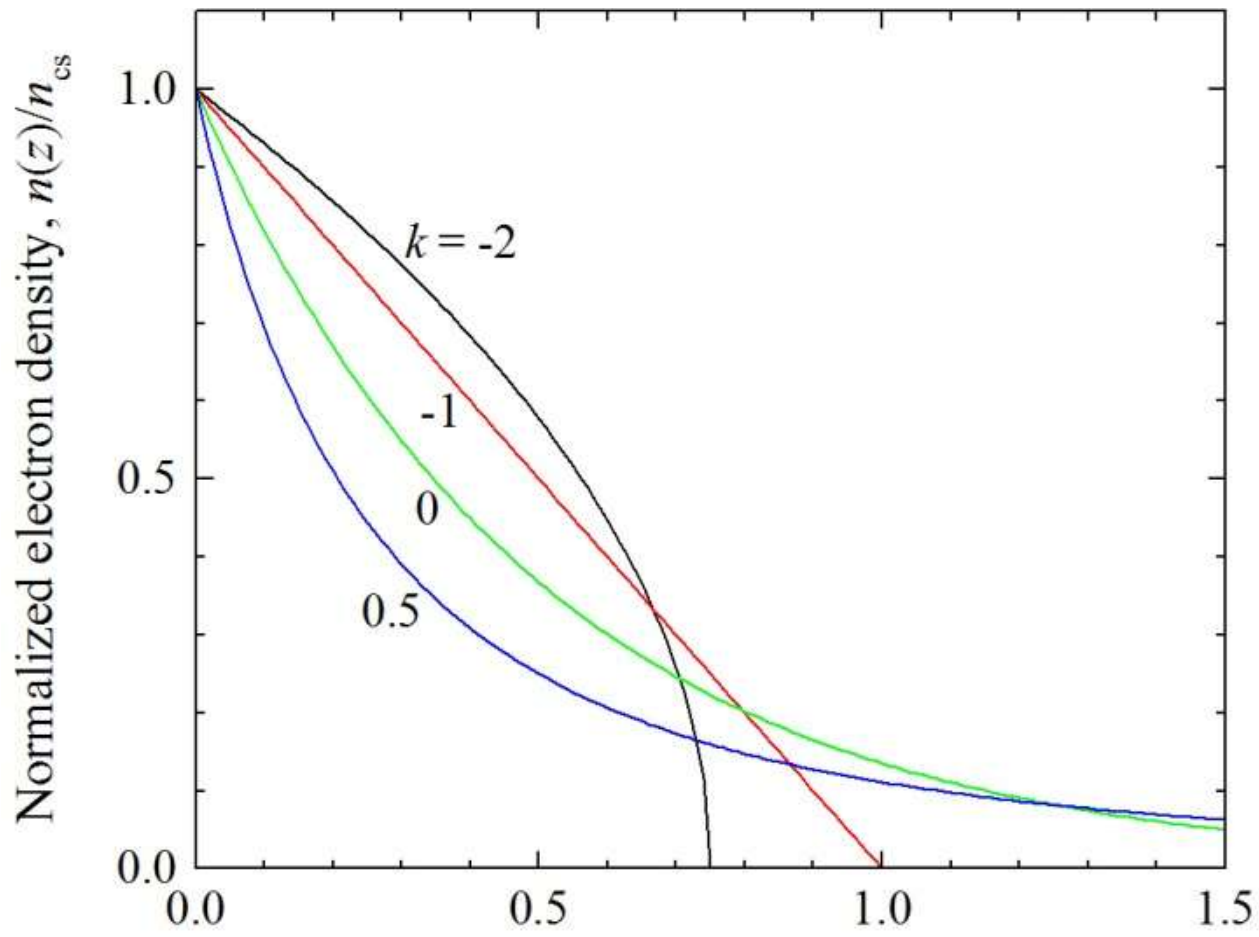

FIG. 20 (foreign laboratory)**.** Electron density calculated as a function of axial position, according to the relation (10). This relation reflects Zakrzewski's (1983) criterion for ensuring the stability of a *long* TWD (meaning ignoring the characteristics of its end). The quantities $n_{cs}$ and $a_{cs}$ are normalised with respect to their values at $z$ = 0, assuming $\alpha(n) = An^k$ (8). The $k$ = -1 curve corresponds to the linearly decreasing axial distribution of electron density, experimentally observed in Sec. 3 under an extremely large range of operator set values $p$, $f$ and $R$.

Fig. 20 shows that the axial profile of the electron density distribution, calculated from relation (10), can be concave ($k$ = 0.5 and 0), linear ($k$ = -1) or convex ($k$= -2). All of these possibilities comply strictly with the stability criterion $d\overline{n}_e(z)/dz < 0$ (7), which requires monotonical decay of the electron density along the plasma column. These electron density profiles are obtained for discharges maintained by EM travelling waves of any kind, i.e. irrespective of their specific dispersion properties or the kinetic characteristics of the generated plasma column. Of the different values of $k$ shown in Fig. 20, only $k$ = -1, which corresponds to the observed axial electron density distribution, remains once the transient period preceding the steady state is over (see below for details).

Zakrzewski (1983) put forward the argument that the SW axial field intensity $E_0$ is constant along a long plasma column, which guarantees a linear decrease in the axial distribution of electron density. Regarding the end of the plasma column, he incorrectly assumed that the intensity of $E_0$ would fall sharply. Experimentally, rather, there is a sudden increase in the intensity of $E_0$ at this point and in its immediate vicinity (Moisan *et al.*, 1991, Moisan, Ganachev and Nowakowska, 2022). In reality, the increase in $E_0$ at the end of the column is compensated by a reduction in the cross-section $S$ of the plasma column (see Sec. 4.3 below), ensuring that the linearity of the axial electron density distribution of the plasma column is maintained up to its end, as observed throughout Sec. 3.

As briefly mentioned, the fact that the unique $k$ = -1 electron density profile is reached under stationary conditions is linked to what occurs during the transient period that precedes the stationary state. More specifically, at the start of this period, the wave propagating towards the end of the column produces distinct plasma clusters on its way (Hamdan *et al.*, 2017). As the discharge evolves towards stability, the electron density along the axis decreases monotonically, leading to the merging of the initial individual plasma segments to form a single plasma column. A possible further mechanism contributing to the linearization of the axial electron density distribution would be the existence of electron density axial curvature (gradient) with profiles other than $k$ = -1: these favour the axial transport of electrons, eventually eliminating any curvature of these gradients.[12]

*4.2. Derivation of the relationship between $\alpha(z)$, the attenuation coefficient of the travelling wave power, and the axial electron density $\overline{n}_e(z)$, including the column end*[13]

The axial distribution of the electron density is determined by the rate at which power is lost by the travelling wave along the plasma column. Assuming that this wave propagates in the z-direction (with z = 0 representing the end of the plasma column in this model), the fundamental equations for power flow can be expressed as follows:

$$\alpha(z) \equiv \frac{1}{2P(z)} \frac{\mathrm{d}P(z)}{\mathrm{d}z} \tag{12}$$

and

$$L(z) = \frac{\mathrm{d}P(z)}{\mathrm{d}z} \tag{13}$$

where α($z$) defines the attenuation coefficient of the wave power along $z$, and $P$($z$) is its power flow[14]. $L$($z$) is defined as the power lost by the electrons per unit length due to collisions of all kinds at $z$, compensated by $dP/dz$, the absorbed wave power. $L$($z$) can then be expressed as (Moisan, and Zakrzewski, 1991):

[12] Assuming that the discharge tube is longer than the plasma column produced.

[13] The analytical derivation that follows (inspired from Moisan and Zakrzewski (1991)) is proposed here for the first time.

[14] Note that this demonstration does not require to specify in which medium(s) the wave propagates and loses its power.

$$L(z) = \bar{n}_e(z)\theta(z)S(z) \tag{14}$$

where $\theta$ is the power absorbed per electron and $S$, recall, is the cross-sectional area of the plasma column (Moisan, Ganachev and Nowakowska, 2022), both depending a priori on $z$ but their product being axially constant in accordance with what is imposed in (18) below. The experimentally observed axial distributions of electron density displayed in section 3 unambiguously suggest as a starting point for our model that:

$$\bar{n}_e(z) = n_0 + bz \tag{15}$$

where $n_0$ is the electron density at the plasma column end and $b \equiv \frac{\mathrm{d}\bar{n}_e}{\mathrm{d}z}$, the (constant) slope of the axial distribution of electron density. Expressing (14) fully as:

$$L(z) = (n_0 + bz)\theta(z)S(z), \tag{16}$$

then from (13) $P(z)$ can be expressed as an integral over $L(z)$ in the following form:

$$P(z) = \theta S \int_0^z (n_0 + bz)\mathrm{d}z + P_0. \tag{17}$$

As previously mentioned, the values of $\theta$ and $S$ depend on axial position $z$ but their product, $\theta S$, must remain constant. Otherwise, $P(z)$ in (17) would depend on S(z), contradicting the experimentally demonstrated independence of axial electron density (15) from plasma radius variation during radial contraction (see Sec. 4.3 below). After integration and referencing (12) and (17), the following is obtained:

$$P(z) = (n_0 z + \frac{1}{2}bz^2)\theta S + \frac{n_0 \theta S}{2\alpha_0}, \tag{18}$$

identifying the remaining power flow at the column end as:

$$P_0 = \frac{n_0 \theta S}{2\alpha_0}. \tag{19}$$

Therefore finally:

$$P(z) = (\frac{n_0}{2\alpha_0} + n_0 z + \frac{1}{2}bz^2)\theta S. \tag{20}$$

From equations (13) and (20), and bearing in mind that the product $\theta S$ is independent of $z$, we have:

$$\alpha(z) = \frac{n_0 + bz}{\frac{n_0}{\alpha_0} + 2n_0 z + bz^2}. \tag{21}$$

Relation (21) can also be written as follows:

$$\alpha(z) = \frac{b\,\bar{n}_e(z)}{\bar{n}_e(z)^2 + c} \tag{22}$$

where:

$$c = n_0\left(\frac{b}{\alpha_0} - n_0\right). \tag{23}$$

Posing c = 0 (see below):

$$b = n_0 \alpha_0, \tag{24}$$

then from (22):

$$\alpha(z) = \frac{b}{\bar{n}_e(z)} \tag{25}$$

where $b$ in (15) is $\frac{\mathrm{d}\bar{n}_e}{\mathrm{d}z}$ making that:

$$\alpha(z) = \frac{\mathrm{d}\bar{n}_e}{\mathrm{d}z} \frac{1}{\bar{n}_e(z)} \tag{26}$$

or equivalently:

$$\alpha(z)\bar{n}_e(z) = \frac{\mathrm{d}\bar{n}_e}{\mathrm{d}z} \tag{27}$$

where $\frac{\mathrm{d}\bar{n}_e}{\mathrm{d}z}$, recall, is experimentally constant till the very end of the plasma column (see Sec. 3) and, therefore, such must be the product $\alpha(z)\bar{n}_e(z)$ all along the plasma column, unlike the result of previous model calculations (Glaude *et al.,* 1980) where this product grew exponentially towards the column end. The independence of $\alpha(z)\bar{n}_e(z)$ from the characteristics of the plasma column comes from the fact that the travelling wave propagates along the interface between the outer wall of the discharge tube and the vacuum (ambient air) surrounding it, as we advocate in this paper, and not along the interface between the outer edge of the plasma column and the inner wall of the discharge tube. The present derivation involves only the electrons in the discharge (as the medium absorbing the wave power), and no other properties of the plasma. Setting $c = 0$ in equation (23) gives $\alpha(z) = \frac{\mathrm{d}\bar{n}_e}{\mathrm{d}z} \frac{1}{\bar{n}_e(z)}$ (26), which makes that $k$ = -1 when writing equation (8) as $\alpha(n) = A/n^{-k}$. This corresponds to the experimentally observed axial electron density distribution.

### *4.3. The independence of the axial position of the product q(z)S(z) ensures a linear electron density distribution up to the end of the plasma column*

The axially constant nature of the product $\theta_A(z)S(z)$ is an essential and original feature of our model (recall Sec. 4.2). To illustrate this, consider a low-pressure TWD in which the value of $\theta_A$ (absorbed EM power per electron) is experimentally constant throughout the plasma column, except at its end where it increases sharply (see Fig. 21a; for detailed physical reasons, see Moisan and Nowakowska, 2018). Simultaneously, the total radial luminosity of the plasma decreases at the end of the column (see Fig. 21b), which indicates a reduction in the plasma's cross-sectional area $S$ (Fig. 21c). In fact, the product $\theta(z)S(z)$ remains constant throughout the column because the sudden increase observed in $\theta_A$ at the end of the column (Fig. 21a) is offset by the corresponding reduction in $S$. These axial variations can be observed in both the photograph (Fig. 21b) and the plot of the luminous diameter of the plasma column (Fig. 21c).[15]

[15] Clearly, rigorously checking that $\theta(z)S(z)$ is a constant would require determining $\theta(z)$ and $S(z)$ under the same operating conditions.

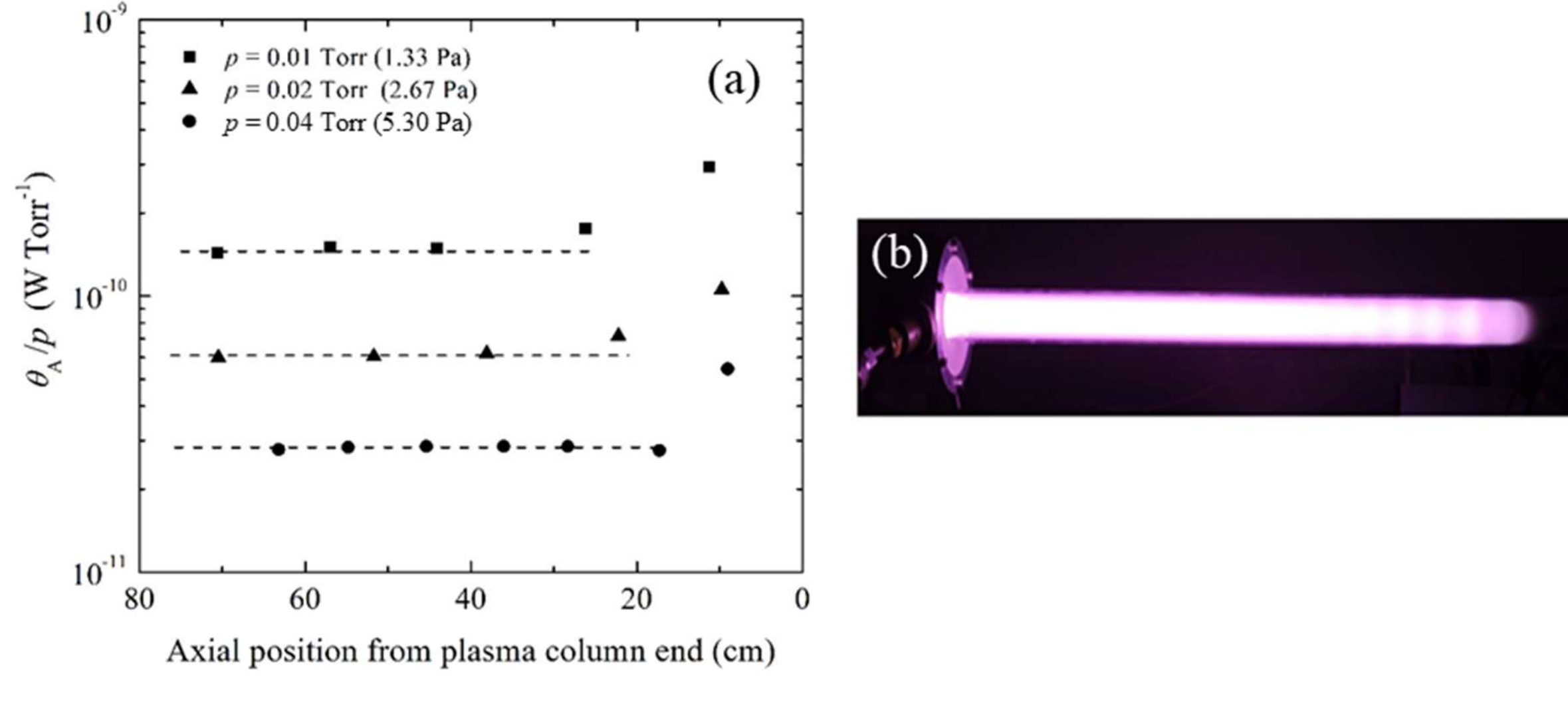

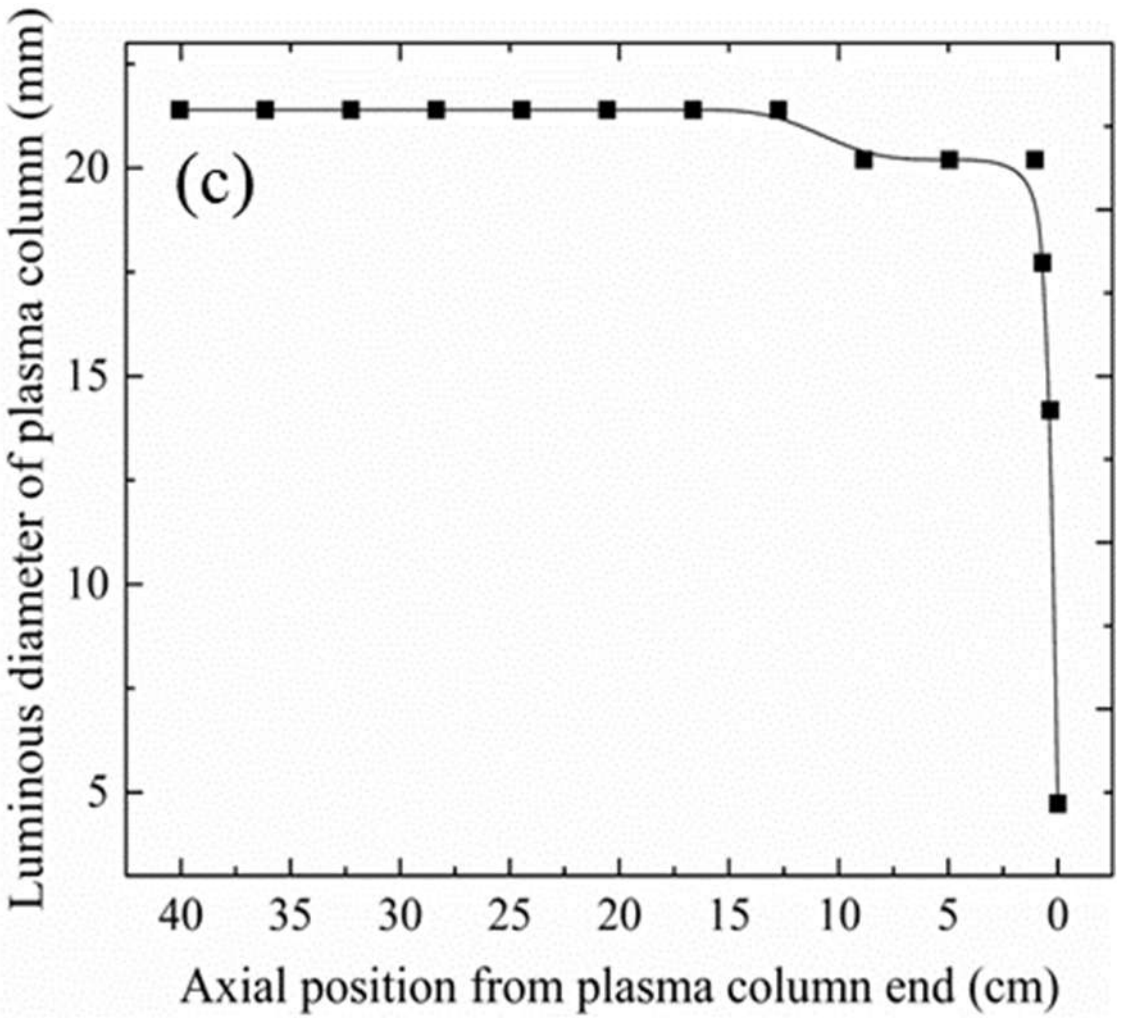

FIG. 21**:** a) Measured values of $\theta_A/p$ as a function of axial position relative to the end of the TWD plasma column sustained at 200 MHz in a tube with inner and outer radii of 13 and 15 mm, respectively, for three different gas pressures $p$ in the low-collision regime. For a given gas pressure, the power absorbed per electron ($\theta_A$) does not vary with axial position, except at the end of the column where there is a sudden increase (Moisan *et al.* 1991); b) photograph of an argon TWD at 0.05 Torr (6.7 Pa) sustained at 915 MHz in a tube with an inner radius of 10.7 mm. The wave runs from the left (Surfatron gap) to the end of the plasma column on the right. The radius of the plasma column decreases towards the end as the plasma becomes rounder. The alternating variations in luminosity at the end of the column are due to the combined reflection of an azimuthally symmetric surface wave ($m$ = 0) and a dipolar surface wave ($m$ = 1): see Appendix C, case 1; c) The measured luminous diameter of the plasma column (leading to $S$) indicates that it fills the discharge tube, except at the very end where it decreases abruptly over approximately 10 mm.

The role of the product $\theta(z)S(z)$ in this behaviour was ignored by everyone, most probably because no one considered the possibility of an accompanying reduction in $S(z)$. This made it impossible to adequately resolve the problem raised by the sudden surge in $E_0$ (and $\theta$) at the end of the column (Moisan *et al.,* 1991), This condition on *θ(z)S(z)* is thus essential for obtaining $\alpha(z)\bar{n}_e(z) = \frac{d\bar{n}_e}{dz}$, which leads to $k$ = -1, i.e. a linearly decreasing axial distribution of electron density from the beginning to the end of the plasma column.

*4.4. Even the shortest experimentally obtained TWD plasma column can be linearly extrapolated to any given length to provide the correct axial distribution of electron density. This is, as already mentioned, an essential characteristic of TWDs*

This feature of TWDs (referred to as SWDs in the literature) is well known and has been documented in experiments (see Sec. 3). Its validity has been demonstrated regardless of the operating parameters $p$, $f$ and $R$. The way in which the distribution of electron density decreases axially stems first from our representation of the wave's power distribution as an energy conservation law that links the wave axial power loss to electron heating throughout the discharge process (see Sec. 4.2). As the TW propagates in a vacuum along the surface of the outer wall of the dielectric tube, it dissipates power linearly. This monotonical decay of electron density guarantees the stability of the discharge (Sec. 4.1). Finally, it is the electric field component of the wave that heats the electrons in the discharge gas, thereby ionising the plasma column.

In conclusion, we emphasise that the phenomenon linking wave attenuation and the resulting electron density distribution is expressed by the relationship $\alpha(z) = \frac{b}{\bar{n}_e(z)}$, regardless of the kinetic characteristics of the plasma column. Furthermore, this relationship applies throughout the TWD plasma column. None of the models published to date reproduce this linearity as well in a short TWD plasma column as in a very long one. In fact, all their calculations show significant deviation from linearity at the end of the column, contrary to the linearity observed over its entire length (see Section 3).

5. A critical review of articles on modelling the axial distribution of electron density along so-called surface-wave discharges (SWDs)

While all SWD theorists concur that these MW discharges are governed by an EM surface wave, whether propagating along the plasma column itself or along the interface between the plasma and the inner surface of the dielectric tube, this is erroneous. This inaccuracy has been reiterated in more than 500 articles published on SWDs, but our study challenges this analysis. In the following discussion, the reasons why these authors did not obtain the perfectly linear axial electron density distribution required by the experiment will be explored. This distribution is, in contrast, provided by our model. A thorough analysis of their publications reveals flaws in the wave propagation mode they advocate. The publications in question, all of which are classified in the SWD category, can be divided into two groups:

i) The surface wave attenuation coefficient is intrinsically associated with the electron density of the plasma column.

ii) One of the kinetic properties of the plasma column is imposed from outside in diverse forms.

*5.1. The properties of the surface waves in these papers are determined either analytically or by numerical simulation by assuming propagation in a cold, homogeneous plasma characterised by its relative permittivity. The travelling wave power flow heats the electrons, ionising the discharge gas*

*5.1.1. Glaude et al. (1980) present experimental results supplemented by numerical calculations*

This pioneering paper has paved the way for the study of the fundamental properties of TWD-sustained plasmas. It reveal the characteristic properties of these plasma columns and shows how the electron density $\bar{n}_e(z)$ (related to $\omega_{pe}$ in the real part of $\varepsilon_p$, the relative permittivity describing the plasma) depends on the wave frequency $\omega/2\pi = f$ (present in $\varepsilon_p$), on the inner radius $R$ of the discharge tube through $S = \pi R^2$, and $p$, via the electron-neutral collision frequency for momentum transfer $\nu$ (the imaginary part of $\varepsilon_p$). A notable experimental finding is that the number of electrons $\bar{n}_e(z)\, S\, \Delta z$ in the cylindrical plasma section of small axial extension $z,\, z + \Delta z$ is proportional to the power absorbed within it, $P_a(z)$, according to the following equation:

$$\bar{n}_e(z)\, S\, \Delta z\, \theta = P_a(z) \qquad (28)$$

where $\theta$ is the power absorbed/lost per electron[16] (Moisan, Ganachev and Nowakowska, 2022). Since Glaude *et al.* (1980) assumed that $\theta$ is axially constant, the following can be written for two successive short cylindrical plasma segments at $z_1$ and $z_2$:

$$\theta = P_a(z_1)/n(z_1)\, S\Delta z_1 = P_a(z_2)/n(z_2)\, S\Delta z_2. \qquad (29)$$

The electron density and absorbed wave power values for the first cylindrical segment, as determined by the measurements, enable the iterative calculation process for the second segment to begin since the first segment initially provides the value of $\theta$ (see Glaude *et al*., 1980 for the full procedure). By keeping the three parameters $\omega$, $S$ and $\nu$, which define the operating conditions, at constant values, the calculations generate a linear axial distribution of electron density, except at the end of the column where "numerical oscillations" are observed. Furthermore, similar calculations, in which one of the parameters $\omega$, $S$ and $\nu$ is varied at a time, also lead to a linear axial distribution of electron density (again, except at the end of the column). The slope of these distributions increases with $f$, $\nu$ and decrease with the inner radius of the discharge tube $R$ (via $S$); this dependence appears as a similarity law of the form $\nu f/R$. Ultimately, the authors demonstrate that the gradient of the axial electron density distribution can be expressed as follows, given the attenuation coefficient of the travelling wave power $\alpha(z)$ assumed to be given by:

$$dn/dz = \alpha n. \qquad (30)$$

According to the numerical calculations by Glaude *et al.*, this relationship remains approximately constant, ensuring the linearity of the axially decreasing electron density distribution, except at the end of the column where α diverges: it means that the model is incorrect/incomplete.

[16] The quantity $\theta$ was initially introduced as the inverse of its current value, simply as a constant of proportionality (Glaude *et al.* (1980)) but more appropriately identified afterwards by Chaker *et al.* (1982) as the power required to generate an electron.

.

*Conclusion on Glaude et al.* (1980)*:* relation (30) is, in fact, formally identical to our relation (26) $\alpha(z) = \frac{\mathrm{d}\bar{n}_e}{\mathrm{d}z}\frac{1}{\bar{n}_e(z)}$, the difference lying in the way the value of α is assessed. In relation (30), the attenuation coefficient assumes that the travelling wave propagates through the plasma column, the dielectric tube and the surrounding vacuum. In contrast, our current model asserts that the travelling wave propagates (and attenuates) guided along the interface between the outer wall of the dielectric tube, and the surrounding vacuum. This follows from the stability criterion (Sec. 4.1) formalised by the relation $\alpha(n) = An^k$(8), with $k$ = -1 leading to a linearly decreasing distribution of electron density from the beginning to the end of the plasma column, in accordance with the experiment. Although Glaude et al.'s paper contains the key elements of TWD wave physics, it does not consider the true propagation medium of the guided wave. As previously mentioned, in our model this medium is the interface between the outer wall of the discharge tube and the vacuum. This solution had not been considered until the present study.

*5.1.2. Model of a plasma column sustained by a surface wave focussing on the radial distribution of electron density (Ferreira, 1981)*

The main point of the paper is on recovering the observed radial distributions of excited argon atoms obtained with Surfatron plasmas. 'Depending on the working conditions (gas pressure, tube radius and absorbed HF power), the radial profiles are either flat or exhibit a maximum near the wall'. The axial distribution of the electron density is however not considered.

*5.1.3. The first analytical model to reproduce the linearly decreasing axial distribution of electron density in a plasma column maintained by an EM surface wave* (*Aliev, Boev and Shivarova, 1982)*

Based on the experimental results of Glaude *et al.* (1980), the authors initially assumed an energy conservation law: the electron density at *z* is proportional to the absorbed wave power at the same axial position (as observed by Glaude *et al.,* 1980). Furthermore, the electrons are heated by the electric field component of the wave, thereby ionising the discharge gas.

They went on, noting that, in the thin-cylinder approximation and for a plasma column surrounded by a vacuum, the only significant power flow occurred in the *z*-direction and in a vacuum, which fits our model (see Appendix E). They further considered that the expression of the slope of the axial distribution of electron density should include an average (constant) axial value representing the dependence of wave dispersion on plasma properties: this is a basic error of appreciation. As shown in Sec. 3, the axial distribution of electron density depends exclusively on *p*, *f* and *R*. not on the properties of the plasma or wave dispersion. Nonetheless, these approximations produce a linear decrease in the axial distribution of electron density throughout the entire length of the generated plasma column, as observed in the experiment.

The following presentation literally quotes the main points of their article, highlighting their contribution (CGS units are used).

"Consider a weakly damping electromagnetic wave of TM type having both axial and radial electric field components $E_z$, and $E_r$, respectively, and an azimuthal magnetic field component $B_\varphi$. The slow variation along the *z*-coordinate of the seek for quantities is determined by the energy conservation relation of the surface wave:

$$\frac{d\bar{S}(z)}{dz} = -\bar{Q}(z) \quad (31)$$

where $\bar{S}$ (designing the Poynting vector) is the SW power flow averaged over a wave period:

$$\bar{S} = \frac{c}{8\pi} Re \int_0^{2\pi} d\phi \int_0^{\infty} r E_r(r) B_\phi^*(r)\, dr \tag{32}$$

and *Re* and * denote the real part and the complex conjugate of a quantity, respectively, *c* being the speed of light in vacuum. As for the value of $\bar{Q}$, the Joule collisional heating losses in the plasma column per unit axial length, it is given by:

$$\bar{Q} = \frac{1}{2} \int_0^{R} r\, dr \int_0^{2\pi} d\varphi\, (Re\, \sigma) |E|^2 \tag{33}$$

where σ is the conductivity of the low-temperature plasma. Further it is assumed that the frequency of the elastic electron-neutral collisions ν does not exceed the surface wave frequency ($\nu < \omega$), thus neglecting collisional damping. The energy conservation law is expressed, for simplicity, in the thin cylinder approximation (the plasma cylinder of radius *R*, located in vacuum, is smaller than the skin depth). $E_z$ is the axial electric field component which in the case of the thin cylinder has a constant value over the plasma column cross-section and exceeds considerably the radial electric field component. The plasma (relative) permittivity:

$$\varepsilon_{P1} \approx -(\omega_{Pe}^2 \,/\, \omega^2) \tag{34}$$

is negative with an absolute value much greater than unity (according to the case of surface waves in the thin cylinder approximation). This condition allows neglecting in the power flow the contribution of the wave propagation in the plasma (the obtained general expression for $\bar{S}_z$ shows that in the thin cylinder case only the power flow forward in the *z*-direction is considerable and this flow is in vacuum". Finally, assuming that the local value of the electron density is proportional to the absorbed wave power at that same point:

$$|\varepsilon_{p1}| = \beta E_0^2 \tag{35}$$

where $\beta$ is in their paper some constant and "$E_0$ is the axial electric field component which in this case of the thin cylinder has a constant value over the plasma column cross-section and exceeds considerably the radial electric field component".

The authors use the SW dispersion relation of a homogeneous cold plasma column surrounded by vacuum whose electron density is provided by (34) and (35). "The analytical results for the axial electron density distribution along the gas discharge axis is in the end:

$$n_e(z) \,/\, n_e(0) = 1 - z \,/\, L \tag{36}$$

where:

$$L = (R/\nu\omega) \bar{f} (12\pi e^2/m)\, n_e(0), \tag{37}$$

*L* is, here, the length of the produced plasma column, $n_e(0)$ is the electron density value at $z = 0$ where the surface wave generator is situated[17], $\bar{f} = 0.16$ is a wave dispersion average value and *m*

[17] Aliev, Boev and Shivarova (1982) were not aware at the time that the TWD plasma column does not begin immediately at the field applicator aperture, as was later documented (Moisan, Levif and Nowakowska, 2019); this results in an antenna-like radiation region.

(here only) is the electron mass. In the end, the gradient of the electron density distribution can be expressed as:

$$\frac{dn}{dz} = -(\nu\omega\ /\ R)\left(\bar{f}^{-1}m/12\pi e^{2}\right) \qquad (38)$$

where the authors assume that the electron-neutral collision frequency $\nu$ does not vary along the plasma column which allows them to preserve the observed axial linearity of the electron density all along the plasma column. Considering, in this condition, $\nu$ instead of the gas density $N$ as an operating condition (see footnote 6) amounts to the same thing from a calculation point of view. "The initial value of the wave power (at $z = 0$) determines the length $L$ of the plasma column but it does not influence the electron density axial profile".

Comparison with the experiment showed that the calculated properties of their plasma column, despite the very restrictive assumptions of the model (low-collision regime ($\nu/\omega \ll 1$), thin-cylinder approximation, plasma column in vacuum) gave a correct picture of wider operating conditions.

*5.1.4. Aliev, Boev and Shivarova (1984) use the approximation of a thin plasma cylinder as before, but with a plasma column enclosed within a dielectric tube*

As in their 1982 paper, they assumed that the contribution of wave dispersion depends on plasma properties, appearing as an average value in the slope expression (which makes that linearity is not affected, but conceptually erroneous as mentioned), although using a more complex expression for $\bar{f}_1$. Nonetheless, it leads to a similar expression as (38) (CGS units):

$$\frac{dn(z)}{dz} = -\ (\nu\omega\ /\ R)\ (\bar{f}_1^{-1}m/4\pi e^{2}). \qquad (39)$$

They indicate that 'the wave power flux in a vacuum is greater than the energy flux in the plasma', which is a distinctive feature of their modelling and is consistent with our own modelling (see Appendix E). However, their assertion that plasma parameters must be described by nonlinear equations is meaningless for two reasons: first, the axial distribution of electron density is independent of plasma properties (Sec. 3), and second, our fully linear model is consistent with the experimental results.

*5.2. Appraising the influence of the kinetic features of the plasma column on the axial distribution of electron density when incorporating some of these elements into the wave equation*

As demonstrated by our model, the travelling wave sustaining the discharge propagates along the interface between the outer wall of the dielectric discharge tube and the surrounding vacuum. Therefore, it can be inferred that the wave is independent of the properties of the plasma column. This standpoint is at odds with the approach adopted in the following papers.

*5.2.1. Mateev, Zhelyazkov and Atanassov (1983) conducted an analytical study of the axial electron density distribution of SWDs, with a particular focus on the role of Maxwell's equations*

These are developed in the electrostatic (slow wave) approximation by applying the standard boundary conditions for the field components. The plasma column is represented by the relative (complex) permittivity of a cold, homogeneous, plasma embedded in a vacuum.

As the wave propagates along the plasma column, its energy "dissipated into the medium heats the electrons and they ionize the neutral gas sustaining the plasma column. The wave energy flux decreases during the wave propagation, and it becomes zero at the end of the column".

An original part of their contribution lies in the introduction of "universal" curves, $\bar{n}_e(\zeta)$ and $E^2(\zeta)$), resorting to a dimensionless axial space coordinate $\zeta = \nu z/\omega R$, which aggregates in the form of the observed similarity law (Glaude *et al.*, 1980) depending on the operating parameters $\nu$, ω and $R$. Another original aspect examined by these authors is to consider that the influence of electron density on the properties of the discharge should differ between the low and high gas pressure regimes:

i) "at low pressure, the electron density is proportional to the wave power absorbed per unit length. This occurs when the average lifetime of charged particles is determined by their diffusion towards the inner walls of the discharge tube". In this first approximation, they provide a relationship for the axial gradient of electron density that is similar to equation (38) of Aliev, Boev and Shivarova (1982), namely:

$$\frac{dn}{dz} = 0.73 \times 10^{-8} \frac{\omega}{2\pi} \frac{\nu}{R} (cm^{-4}). \tag{40}$$

ii) at relatively high pressure, electron density is proportional to the total energy of the wave field per unit length. "This occurs when the average lifetime of the charged particles is defined by bulk (volume) recombination".

In both pressure cases, the axial distribution relating to the electron density deviates, although very little, from perfect linearity by showing a slight overall circular curvature extending from the beginning to the end of the plasma column. More dramatically, the axial distribution of electron density at the end of the column clearly shows a downward curvature, which is contrary to experiment (Sec. 4.4): the authors justify this discrepancy by referring to the corresponding increase in the intensity of the wave ***E*** field at the end of the column. In any case, their model is inadequate, as axial electron density linearity is not achieved over the full length of the SWD column.

*5.2.2. In a continuation of the theoretical approach pioneered by Mateev, Zhelyazkov and Atanassov (1983), Zhelyazkov, Benova and Atanassov (1986) proceeded to extend it by providing a comprehensive EM treatment of the SW field components*

The authors introduced the parameter $\sigma = \omega R/c$, where $\sigma = 0$ corresponds to the electrostatic approximation. They used this parameter to track the evolution of the axial distribution of electron density, moving from the electrostatic approximation to a more complete EM description. They found that, as σ increases, the axial distribution of electron density becomes more linear. Indeed, the slight overall axial curvature away from linearity observed by Mateev, Zhelyazkov and Atanassov (1983) disappears when $\sigma$ exceeds 0.2. However, the downward curvature observed by the aforementioned authors at the end of the plasma column persists in the complete EM treatment. This suggests that their calculations are inconsistent with the experimental results and must therefore be erroneous on this basis.

*5.2.3. Using the thin cylinder approximation, Aliev et al. (1993) propose an analytical derivation that extends the work of Aliev, Boev and Shivarova (1982), specifically considering volume recombination and ambipolar diffusion*

This paper makes various approximations and assumptions to obtain analytical expressions for $\frac{\mathrm{d}\bar{n}_e}{\mathrm{d}z}$ and $E(z)$. The model clearly distinguishes between the roles of the charged particle recombination regime and the influence of the wave dispersion features, as well as the operating conditions of the discharge, namely:

i) The charged particle loss mechanism acts as a numerical factor *a* determining the slope of the electron density distribution, which thus remains linear in all cases. The slope of the axial electron density distribution is smaller in the case of volume recombination than in the case of ambipolar diffusion, given the same discharge parameters;

ii) the wave dispersion still appears as an average value, <f>, taken over a limited interval of electron density;

iii) the operating parameters of the discharge are considered in the form of the observed similarity law, namely, $\nu\omega/R$.

The authors endeavoured to ensure that the axial electron density distribution remained linear, in accordance with the experiment (see Sec. 3), specifically in the end:

$$\frac{dn}{dz} = -(1/2)\, \nu\omega/R\; m_e/(4\pi e^2)\; 1/\langle f\rangle \text{ (CGS units)}, \tag{41}$$

except at the end of the plasma column where the electron density, again, falls rapidly. The authors exclude this terminal segment from the region of validity of the thin cylinder approximation. Note that the slope of the electron density distribution is related to the wave dispersion by the mean value <f>, allowing the authors to assert that the wave dispersion characteristics and the axial plasma density distribution are interdependent: this is fundamentally contrary to reality (Sec. 3). Therefore, their model is incomplete and can be considered erroneous.

*5.2.4. Aliev et al. (1995a) introduce the concept of wave power absorption by resonance and compare it with collisional absorption*

To simplify the calculations, the authors consider a plasma slab rather than a plasma column. Recall that resonance absorption requires a transition region with a longitudinal (axial) gradient of electron density, allowing the resonance condition, $\omega = \omega_{pe}$, to be met at a specific point $z$ along this gradient path: according to the authors, resonance would occur at the beginning and end of the plasma slab. However, this situation does not arise because, at both low and atmospheric pressures (see Sec. 3), the axial electron density profile of an SWD is experimentally linear along the entire plasma slab (column). Therefore, there is no specific transition region at either end of the plasma column to allow for resonance.

*5.2.5. Kovačević et al. (2021) proposed an analytical approximation of the SWD that ensures the axial electron density distribution is fully linear, as required by the experiment*

The authors employed the *square-root* analytical *approximation* of the surface wave dispersion relation introduced by Babović, Aničin and Davidović (1997). This involves limiting the calculations of the dispersion relation to the $\beta R < 1$ case, where $\beta$ is here the SW wavenumber. This approach facilitates the determination of the wave power attenuation coefficient $\alpha(z)$ by 'avoiding the more difficult-to-handle cluster of Bessel functions of different types, orders, and arguments'. Using this approximation, the authors obtained linear axial distributions of electron density from the beginning to the end of the plasma column. However, the physical significance of the somewhat restrictive $\beta R < 1$ approximation is not discussed.[18]

The authors also examine the contribution of losses arising from the assumption that the wave propagates besides the plasma column in the dielectric discharge tube material. The losses in a dielectric tube increase significantly with both the frequency of the wave field and the thickness of the tube. They therefore conclude that, besides the collisional losses in the plasma column, $\alpha(z)$ should account for dielectric losses in the discharge tube material, since these could impact the linearity of the axial distribution of electron density. Their calculations show that linearity is 'a good approximation' insofar as $R_{oi}$, the ratio between the outer and inner radius of the tube, is less than 1.5, whereas for a thicker tube, the axial distribution of electron density should become increasingly parabolic as $R_{oi}$ increases.

Such a possible contribution can be verified experimentally in two ways: by using discharge tubes with particularly thick walls, such as those with an outer radius of 8 mm and an inner radius of less than 1 mm (see Fig. 10), or by varying the frequency of the travelling wave, for example from 200 MHz to 2450 MHz (see Fig. 11). In both cases, the observed axial distributions of electron density remain linear. This result is no surprise for us since the travelling wave propagates along the surface of the outer wall of the discharge tube in the surrounding vacuum (see Appendix E), therefore without penetrating the discharge tube. However, note that the intensity of the electric field component of the travelling wave, which heats the electrons in the discharge gas, should be lower once passing through the tube dielectric wall.

*5.2.6. Gordiets et al. (2000) examined the kinetics of the plasma column and surface processes, which they claim are self-consistently linked to the electron density axial distribution*

As a matter of fact, the paper first attempts to compile all available information data on the properties of plasma, wave electrodynamics, gas dynamics, thermal balance and surface processes (those occurring on the inner wall of the discharge tube). “The volume and surface kinetics in discharges operating in molecular gases are strongly coupled, and this coupling significantly affects the whole discharge physics”. “A SW discharge is sustained by a wave which simultaneously propagates and creates its own propagation structure”, an incorrect assumption. In that sense, they claim that anything that happens in the discharge tube is reflected in the axial distribution of electron density. However, this statement is contradicted by the experimental results analysed in Sec. 3, where the electron density distribution only depends on $p$, $f$ and $R$, on nothing else. Fig. 18 illustrates the outcome of their approach, clearly showing that their model does not replicate the observed linear axial distribution of electron density. Furthermore, they get a downward curvature of $n_e(z)$ at the end of the column, a general feature when assuming that the plasma column and the wave interact. The article is based on many wrong premises.

*5.2.7. Aliev et al. (1995b) considered that the wave propagating at the outlet of the EM field applicator and in its immediate vicinity is a true EM wave that is faster than the guided wave generating the TWD. This maintains a plasma column with a higher electron density, resulting in a quadratic axial electron density profile*

---

[18] Their approximation requiring $\beta R < 1$ avoids considering the true end of the plasma column, thus skipping the problem of downward curvature appearing in the calculated axial distributions of electron density by all.

As demonstrated in Section 2.2.2 and by Moisan, Levif and Nowakowska (2019), the electric field present at the output of the EM wave launcher and extending a certain distance beyond, is that of the radiation generated by an antenna. Furthermore, it is not an EM wave in the strict sense because the ***E*** and ***H*** components of its field in this region are neither of equal intensity, nor in quadrature or phase, as discussed in Section 2.2.2. In fact, the distribution of the electron density recorded in this region cannot be attributed a specific axial profile: for example, at 27 MHz (Fig. 7), this profile is linear, whereas at 360 MHz it is erratic (Fig. 8a). Moreover, the authors postulate that "the nature of this type of discharges combines in a self-consistent nonlinear manner wave and discharge properties": this claim of non-linearity is based on incorrect premises directly with respect to our model, which excludes interaction between wave and discharge properties.

*5.2.8. The Zhelyazkov and Atanassov (1995) report on SWDs*

This comprehensive 122-page report is constructed like a textbook and focuses on the main characteristic of plasma columns maintained by EM surface waves: their axial electron density distribution. In all cases studied, Zhelyazkov and Atanassov consistently agreed, directly or indirectly, with the authors reviewed, but wrongly, that the surface wave propagates into the discharge gas or at the interface between the plasma column and the inner surface of the discharge tube, and consider this 'a basis for modelling surface-wave plasmas'. However, experiment (Sec. 3) and our model (see Appendix E) show that this is an incorrect approach.

The Zhelyazkov and Atanassov report covers SWDs up to 1995. It is divided into two themes. The first describes the surface waves, which can propagate in two ways (with azimuthal and dipolar symmetry, noted as $m = 0$ and $m = 1$, respectively). The second theme considers the structure and specific properties of the plasma column that the waves are expected to generate. It also examines the impact of a static, homogeneous external magnetic field on the propagation of EM surface waves. The development of the subject in each section follows a chronological order and is sometimes accompanied by contributions from Zhelyazkov and Atanassov themselves. Their comments shed additional light on the subject by identifying and summarising key issues more clearly:

1) The different aspects of SWD modelling are divided into three categories: (i) "approaches in which the surface wave and the plasma column are coupled; (ii) approaches that ignore the wave aspect of the problem but deal in detail with the equilibrium of charged particles in the plasma in the presence of an HF field (generally assumed to be uniform); and finally, (iii) approaches that focus on the propagation and attenuation characteristics of the wave and its power transfer to the plasma, but pay little attention to the equilibrium of the charged particles".

Quoting again the authors of the report:

2) "The axial gradient of the electron density is entirely determined by the discharge conditions (as defined by Moisan and Zakrzewski, 1991), i.e., the composition and pressure of the gas, the dimensions of the discharge tube and metal enclosure (if any), the permittivity of the tube, the strength of the applied static magnetic field $\boldsymbol{B_o}$, the mode and frequency of the wave; the electron density gradient does not depend on the power of the HF wave, which is then considered as a condition separate from the discharge conditions";

3) "a simple model for HF discharges produced by EM travelling waves is based on the assumption that the treatment of the physical processes occurring in the discharge can be divided into two separate parts: the first concerns the maintenance processes within the discharge; the second, the electromagnetic behaviour of the HF waves sustaining the plasma";

4) "the local wave dispersion relation is assumed to be linear as far as the field intensity is concerned. This validity was checked by Zakrzewski *et al.* (1977) up to high wave power levels inclusive. Increasing the surface wave power density (in diffusion regime strictly) leads to a plasma density increase, not to an increase of the wave amplitude into the plasma".

In summary, their report assumes an interaction between the wave and the plasma column throughout, whereas the experiment (Sec. 3) and our model reject this idea: the wave propagates independently of the plasma column. The calculated axial distributions of electron density presented in the report are at best described as 'mostly' linear (hence the need in Sec. 3 for a statistical check of the observed linearity). However, the calculated or simulated theoretical axial distribution of electron density reported certainly do not remain fully linear till the end since they all curve downwards (sometimes upwards). This contrasts with the experiment where the axial distribution is totally linear from start to finish. To be additionally noted, some authors have conducted nonlinear calculations to the extent that the linear axial profile of electron density is considered an approximation of a nonlinear situation. However, experiment again contradicts this and, furthermore, our model correctly reproduces experimental axial distribution of electron density without the need for linear approximations!

Their report is limited to papers published at low gas pressure. At that time, little data was known about SWDs at atmospheric pressure; in fact, the first well documented SWD publication on such SWDs was published as early as 1990, but in a 'confidential' journal (Moisan, Pantel and Hubert, 1990). As for the existence and characteristics of the antenna-like radiation zone of the field applicator, it had not yet been confirmed experimentally until 2019 (Moisan, Levif and Nowakowska, 2019).

*5.2.9. Concluding comments on the reviewed papers*

Of all the articles we have just reviewed, only that of Kovačević et al. (2021) reports a calculated axial distribution of electron density that remains linear over the entire length of the plasma column. However, this result is due to an approximation that removes calculations from the end of the column. In fact, to reiterate our point, none of the published axial distributions of electron density obtained so far by calculation or simulation end linearly, apart from that of our model. Given that the linearity of the axial distribution of electron density over the entire length of the plasma column is an essential observed characteristic of TWDs (see Sec. 4.4), it follows that all the models presented in the existing literature are incorrect: it is difficult to support such a conclusion in the face of so many authors.

The fundamental error in all these articles also appears in the conclusion that the propagation of the travelling wave is closely linked to the properties of the plasma column produced. However, if we assume instead that the wave propagates at the interface between the outer wall of the dielectric tube and the surrounding vacuum, this statement becomes irrelevant. Furthermore, the stability criterion for TWDs is another feature that clearly demonstrates that the wave sustaining the discharge is entirely independent of the properties of the plasma column, as it only requires a monotonic axial decrease in electron density and power lost by the wave. This strict common requirement on both electron density and electron energy is related to the fact that, as proposed, the axial distribution of electron density is also the axial distribution of energy lost by the travelling wave.

## 6. Summary, discussion and conclusion

### *6.1. Summary*

Our model is principally based on experimental results, which demonstrate that the axial distribution of electron density in plasma columns sustained by travelling EM waves decreases linearly from start to finish. The operating parameters $p$, $f$ and $R$ are the only factors that influence this distribution, specifically on its slope, and no other characteristic of these discharges has any effect. This experimental starting point was meticulously confirmed and developed in Sections 3 and 4, respectively. However, this inference remains to be disputed in all theoretical papers published to date. Indeed, all the papers referenced at the end of this article — including our own work up to June 2021 — are affected by the fundamental error of assuming that the propagation of travelling waves along TWDs depends on the kinetic properties of the discharge. It could be argued that it was only natural to think this way!

A key modelling element introduced in our research is the Zakrzewski stability criterion for discharges that are sustained by EM travelling waves. This criterion ensures that maintaining a plasma column in this way results in a monotonically decreasing axial electron density and electron energy distribution, regardless of the type of EM travelling wave involved. The utilisation of this property to comprehend the transient period preceding the stationary state of the plasma column is also a significant advance. This phenomenon demonstrates how the numerous reflections of the EM wave during the transient period— a consequence of the requirement for the electron density to decrease monotonically along the plasma column — gradually eliminate local gradients. Ultimately, a single linearly decreasing axial electron density distribution remains. Moreover, the absence of correlation between this stability condition and the dispersion properties of the EM travelling waves serves to confirm that the wave feeding the plasma column does not interact with it, contrary to general claims.

This article demonstrates (and does not merely assume) for the first time that the axial distribution of electron density reflects the axial dissipation of wave energy through the electric component $\boldsymbol{E}$ of the wave, which heats the electrons, in both the RF and MW range; the subsequent ionisation of the discharge gas then maintains the plasma column. In summary, the wave energy attenuation coefficient, α(z), which governs the process of energy transfer, depends solely on the electron density, ultimately expressed as $\alpha(z) = \frac{b}{\bar{n}_e(z)}$ (26). This confirms our assertion that the travelling wave propagates along the interface between the outer wall of the dielectric tube and the surrounding vacuum. This explains why the wave does not depend on the properties of the plasma column, since it has thus no connection with the plasma column, as it rather depends on the dielectric permittivity of the discharge tube.

In the case of the plasma produced by the TIAGO torch at 2450 MHz (see Appendix D), the electron density observed along the plasma filament initially corresponds to the antenna-type radiation zone, extending approximately 2 mm from the nozzle. This is followed by the dart, characterised by a linearly decreasing electron density distribution along the axis, confirming the identification of TIAGO plasma as a TWD.

*6.2. Discussion*

In order to eliminate any potential doubts regarding the linearity of the axial electron density distribution, a least-squares regression analysis was performed on all the presented diagrams to validate straight lines drawn in our graphs, some of which originated from other laboratories. Those who oppose the current model like to point out observed deviations from linearity in the axial distribution of electron density. Deviations of this nature have been reported in cases of radial contraction of the plasma column combined with an inadequate electron density diagnostic, for example when using Thomson scattering (see Sec. 3.7.2).

Several issues raised in this article warrant further investigation and documentation. For example, the minimum electron density at which travelling waves cease to propagate in a low-collision regime ($\nu/\omega \ll 1$), as defined by equation (6), should be analysed in greater depth. Moreover, it should be clarified how the increase in gas pressure beyond the low-collision regime leads to a gradual increase in the minimum electron density for wave propagation relative to $\bar{n}_{e(\mathrm{re})}$, as delineated in Section 3.

There are several other points that require clarification or confirmation:

i) it is necessary to collect experimental data on $\theta$ and $S$ under identical operating conditions at the end of the plasma column. This should clearly demonstrate that an increase in the local value of $\theta$ is strictly compensated by a reduction in the plasma cross-sectional area $S$, thereby ensuring that the product of these two values, $\theta S$, remains constant axially over the entire length of the column (see Figs. 21a and 21c);

ii) it is imperative to conduct a detailed monitoring of the time evolution of the axial electron density distribution during the transient period. This is required to follow the merging of the initial individual plasma segments, which is caused by the monotonical decrease of their axial electron density (in accordance with the Zakrzewski criterion). Eventually, all axial gradients (see Fig. 20) should disappear in favour of a linearly decreasing gradient;

iii) the radial contraction of the plasma column, including the formation of multiple filaments, could be documented more thoroughly.

iv) the substitution of gas density $N$ for $p$, as we have suggested, requires an analytical derivation leading to equation (4).

v) determining how the wave power flows when the TWD is sustained by a $m = 1$ mode would be of interest since the corresponding wave is a linear combination of both TM and TE waves (Margot-Chaker et al., 1989), making power flow more complex than for the TM wave of the $m = 0$ mode (see Appendix E).

*6.3. Conclusion*

Recall that it has been meticulously demonstrated that the axial distribution of electron density in TWDs is linear, and this depends exclusively on the operator set parameters $p$, $f$ and $R$. This is an experimental fact that cannot be contested. Consequently, any suggested interaction between the wave and the plasma column can be definitively ruled out. The model proposed challenges the previously universally accepted notion of surface waves propagating along the plasma column or at the interface between the plasma column and the inner wall of the discharge tube. This development represents a substantial paradigm shift within the field. The only conceivable scenario for EM travelling waves with transverse components is for them to propagate on the surface of the outer wall of the dielectric discharge tube, which is in a vacuum (ambient air).

The ready availability of TWDs with unrivalled operating conditions provides a unique, reproducible, energy-efficient plasma medium. Combined with the present in-depth knowledge of their mechanisms, this makes them powerful research tools for developing and easily optimising new applications in a wide range of fields. Indeed, in contrast to all other types of RF and MW discharges, it is not necessary to modify the discharge configuration when exploring a wider range of operating conditions as a function of $p$, $f$ and $R$, thanks to the flexibility and variety of their operating conditions.

## APPENDIX C: VARIOUS STANDING WAVE CONFIGURATIONS DESIGNED TO REDUCE THE NON-UNIFORMITY OF THE AXIAL DISTRIBUTION OF THE ELECTRON DENSITY OF TWDs

Maintaining a TWD plasma column using a travelling guided EM wave results in a linear decrease in electron density along the axis, yielding a highly non-uniform plasma column. To mitigate this issue, the column is, for instance, fed by two travelling waves that interfere with each other. This creates various types of standing wave, which reduces the degree of axial non-uniformity of the electron density. The interfering waves may or may not have the same azimuthal symmetry and may be in or out of phase.

1- Interference at/near the end of the plasma column of an azimuthally symmetric ($m = 0$) mode travelling wave with its corresponding dipolar ($m = 1$) wave

Fig. 21b (see Section 4.3) showed alternating light and dark circular rings along the axis of the plasma column near its end. The formation of these rings is attributed to the interference of two travelling waves: one of the symmetric ($m = 0$) type and the other of the dipolar ($m = 1$) type. These waves are reflected at the end of the plasma column. In order for this phenomenon to occur, the $fR$ value must be sufficiently high to excite the $m = 1$ mode but not so high that the $m = 0$ wave ceases to exist; the exclusive presence of the $m = 1$ mode begins at $fR = 1.8$ GHz-cm (Margot-Chaker *et al.*, 1989). As illustrated in Fig. 21b, the discharge tube has a radius of 1.07 cm and the frequency of the wave field that sustains the plasma column is 915 MHz. This results in an $fR$ product of 0.98 GHz-cm, which allows the $m = 0$ and $m = 1$ propagation modes to coexist. However, the faster the amplitude of one of the two waves decreases axially, the more limited the interference zone at the end of the column becomes.

It is noteworthy that the selection of the propagation of given $m$ modes depends on two of the three operating parameters ($p$, $f$ and $R$) acting in relation to each other as $fR$, whereas in equation (5) it appears as $f/R$. This phenomenon again confirms the exclusive control exerted by these operating parameters over the TWDs (see Section 3).

.

2- Interference between two independently launched azimuthally symmetric ($m = 0$) travelling waves

*a)* Figure A6a shows the experimental setup consisting of a Surfaguide coupled to a moderate-radius discharge tube. This configuration generates a pure azimuthal-mode ($m = 0$) travelling wave from the field applicator, propagating in two opposite directions, to the left and right of the Surfaguide (Rakem, Leprince and Marec, 1990). The wave launched from the left slot propagates towards 'short-circuit 1' (the fixed reflector plane) and is then reflected back, passing through the left and then right launch apertures of the Surfaguide towards 'short-circuit 2' (the movable reflector plane). At this stage, the wave initially coming from the rear propagates in the same direction as the forward wave emerging directly from the right launch slot of the Surfaguide. These two waves interfere due to their phase difference.

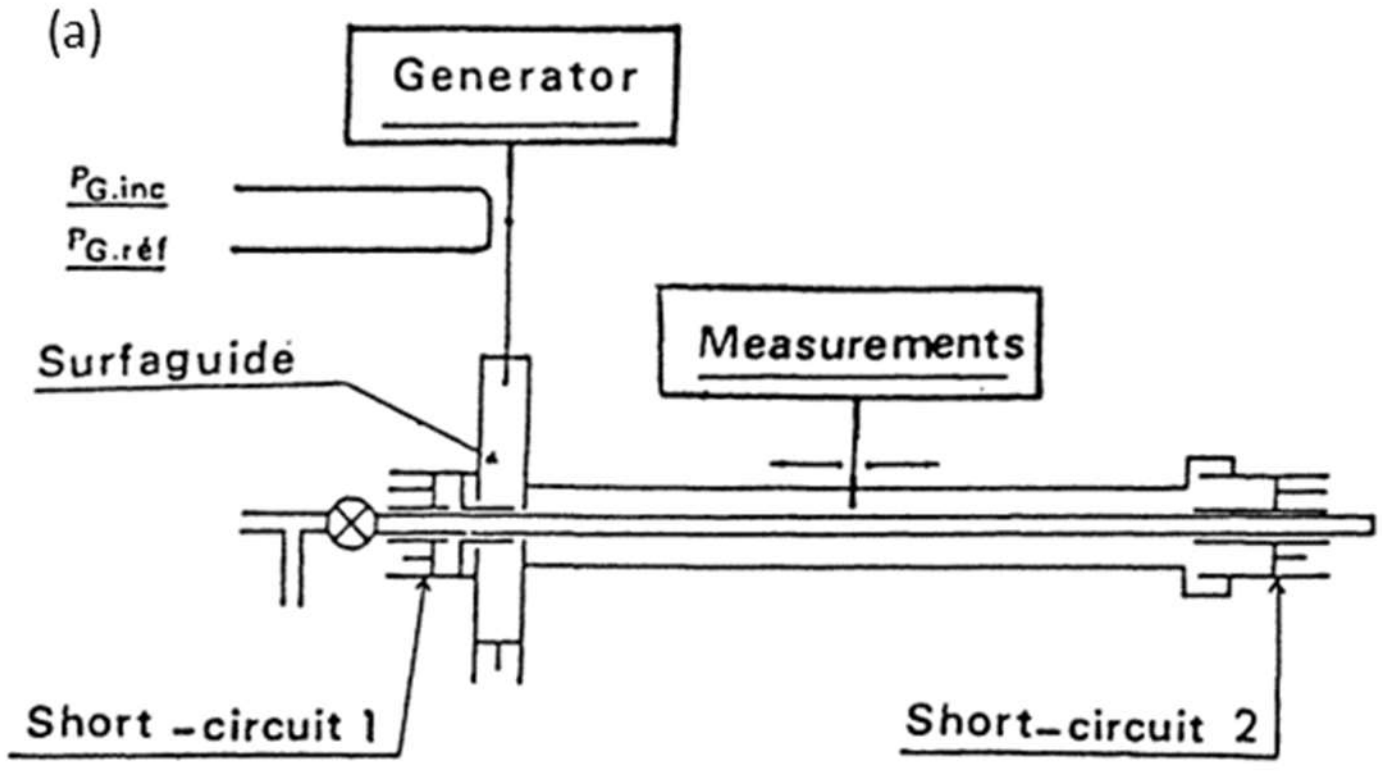

FIG. A6a (foreign laboratory). Schematic setup of a two-wave interference experiment using a Surfaguide, which operates at 2450 MHz with 70 W at 0.5 Torr of argon, in a fused silica tube with inner and outer radii of 2.5 mm and 5 mm, respectively: pure $m = 0$ mode waves are excited (Rakem, Leprince and Marec, 1990).

*b)* Measured axial intensity of $E^2_r$ under standing wave conditions

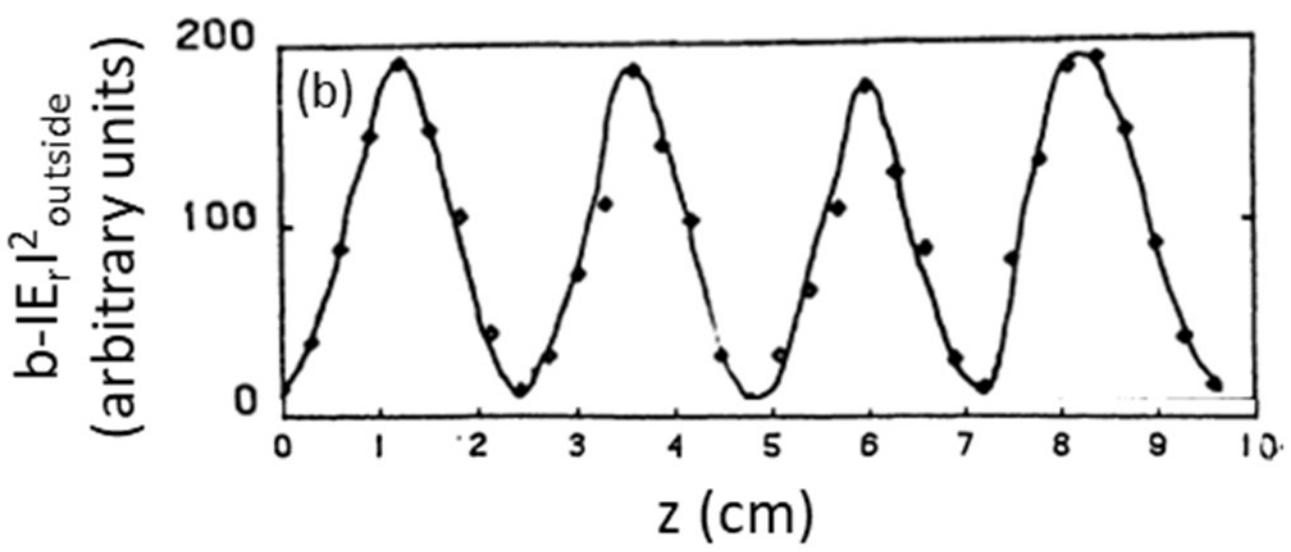

FIG. A6b (foreign laboratory). Measured intensity of $E^2_r$, the radial component of the resulting electric field intensity of two interfering waves, under standing wave conditions (Rakem, Leprince and Marec, 1990).

*c)* Axial variation of radial electron density under standing wave conditions

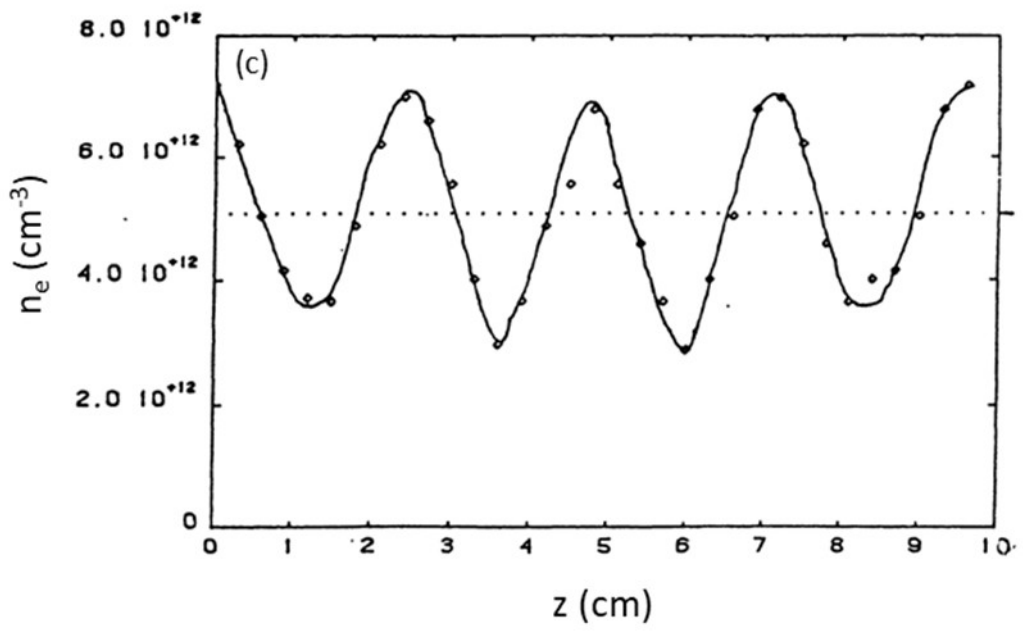

FIG. A6c (foreign laboratory). Axial variation in the average radial electron density under standing-wave conditions, measured using the EM wave phase-shift method (Appendix B in Part I). For an absorbed power of only 70 W, the electron density varies axially around an average radial value of $5.1\times10^{12}$ $cm^{-3}$. In contrast, around 200 W would be required to obtain the same electron density value at the start of a pure travelling wave-sustained plasma column (Rakem, Z., P. Leprince, and J. Marec, 1990).

3- A standing wave configuration has been proposed by Chaker, Moisan and Zakrzewski (1985), which involves two launchers located at opposite ends of the discharge tube that are powered by two separate MW generators, which are therefore not in phase. While this method achieves better axial uniformity than that described in paragraph 2a) above, correctly adjusting the power of each generator to optimise uniformity can be challenging. Another solution for improving the axial uniformity of the electron density, as suggested by the authors, involves gradually reducing the diameter of the plasma tube as one move away from the launcher, in order to compensate for the axial decrease in the wave's power flux. The results obtained with this original configuration, which is more challenging to implement, are nevertheless similar to those described in section 2a) above.

4- Two Surfaguides, each operating at 2450 MHz, are positioned at opposite ends of the discharge tube, and are supplied by the same generator, with a 3 dB power divider sharing the power equally between them (Wolinska-Szatkowska, 1988). This creates a standing wave along the plasma column through the interference of two TM $m = 0$ mode waves launched independently but in phase since the same generator is used. However, the results obtained are again similar to those described in paragraph 2a) above.

5- A single Surfaguide is used to achieve a $m = 1$ mode TWD, setting the standing wave pattern with the help of an axially movable reflecting plane (Zhukov and Karfidov, 2023). The experimental setup is illustrated in Fig. A7 and designated by the authors as an open-air MW interferometer.

It is built around a low-pressure argon gas TWD ($\nu/\omega \ll 1$), which is sustained within a 2 m-long fused silica discharge tube. The tube has inner and outer radii of 21 mm and 27 mm, respectively. This size is sufficient to excite a $m = 1$ propagation mode at 2450 MHz (Margot-Chaker *et al.*, 1989). The configuration consists of two conducting copper mirrors (M1 and M2) that reflect microwaves. The mirrors are perpendicular to the discharge tube, and the distance between them can be adjusted by moving the mobile M2 mirror. What makes this assembly unique is the presence of a copper reflector mesh with the same internal diameter as the discharge tube. This mesh is inserted transversely into the tube to ensure that the reflective plane of mirror M2 is completely uniform (with no radial discontinuities) given the orifice through which the tube passes. The reflector mesh and mirror M2 can also be moved independently of each other.

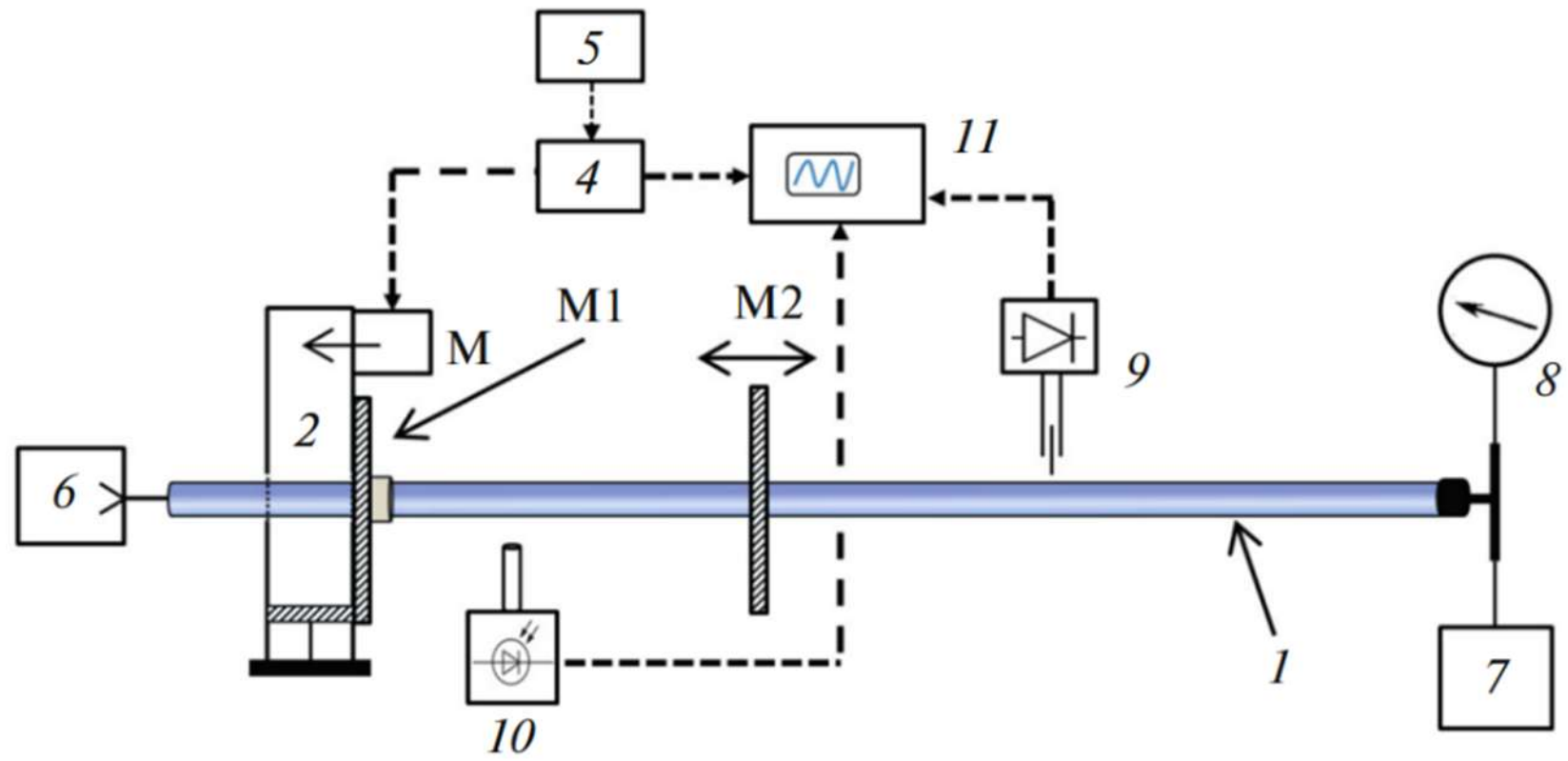

FIG. A7 (foreign laboratory). Schematic diagram of the experimental setup of a so-called open-air MW interferometer (Zhukov and Karfidov, 2023): (1) A TWD is achieved in argon gas inside a 2 m long fused silica discharge tube with an inner radius of 21 mm and an outer radius of 27 mm. This excites a $m = 1$ mode wave at 2450 MHz (Margot-Chaker *et al.*, 1989). (2) A waveguide EM field applicator. (3) A 2450 MHz magnetron generator for MW power. (4) A square wave modulator with a 50 ms duration. (5) A delayed pulse generator. (6) A vacuum pump. (7) A gas inlet valve. (8) A vacuum gauge. (9) An MW electric field probe. (10) A collimated radially oriented photodetector. (11) An oscilloscope connected to the electric field antenna (9) for the EM wave and to the photodiode (10). Mirror M1 is fixed, while mirror M2 is axially movable.

This arrangement is designed to determine the axial intensity distribution of the $E_z$ and $E_r$ components of the electric wave field. Compared to a purely axial linear decrease of electron density, this technique also allows for a more uniform distribution of electron density throughout the plasma column length (see Fig. A6c).

The distance between the mirrors is divisible by an integer number of TW half-wavelengths. As illustrated in Fig. A8, the intensity $E_r^2$of the electric field component along the axis is greater than that of $E_z^2$.

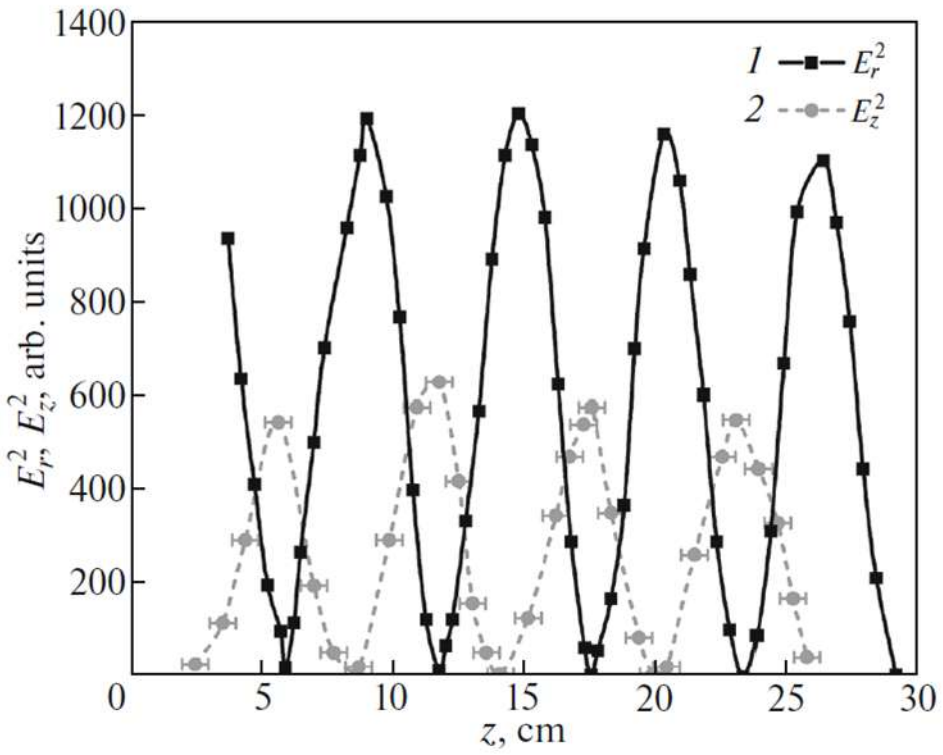

FIG. A8 (foreign laboratory). Measured axial intensity distribution of the electric field components $E^2_z$ and $E^2_r$ of the travelling waves. These waves are confined under a standing wave mode between the two mirrors. The distance between the mirrors is $L = 5/2\lambda = 29$ cm, and the radial distance between the MW probes and the tube outer wall is $r = 1.5$ cm (Zhukov and Karfidov, 2023).

6- Unlike the other experimental situations discussed in this appendix, in this case, a single wave propagates between a reflector plane that is located at a variable distance from the EM field applicator. The front plate of the wave launcher can also act as a second reflector plane (Rakem, Leprince & Marec, 1992). The aim is to achieve a constant axial distribution of excited argon ions, tracking them by using the intensity of the axial emission of excited argon atoms.

Fig. A9 shows the reflector plane in three different axial positions:

(a) Position A: located beyond the end of the plasma column, reflection is avoided, resulting in a linearly decreasing axial distribution of the intensity $I_{ArI}$ of the excited argon atom emission line, as expected when the atom excitation rate follows the axial distribution of electron density;

(b) Position B: the reflected power of the wave reaches the EM field applicator, but with minimal power, resulting in a gentler axial slope of $I_{ArI}$, which is additionally slightly modulated;

(c) Position C: the reflected power level is sufficiently high between the faceplate of the EM field applicator and the reflector plane to ensure an axially constant $I_{ArI}$ intensity characterized by deeper modulation. This modulation reflects the standing wave pattern resulting from multiples of the wave's half-wavelengths (at 2450 MHz). The authors consider this situation to be like the modulation of the electron density observed inside a resonant cavity. The high Voltage Standing Wave Ratio (VSWR) level achieved enabled wave power densities of up to 2 kW/cm$^3$ to be reached. An argon ion laser operating with a 1:1 argon: helium mixture has produced a laser beam of a few tens of milliwatts, based on this structure (Rakem *et al.*, 1986).

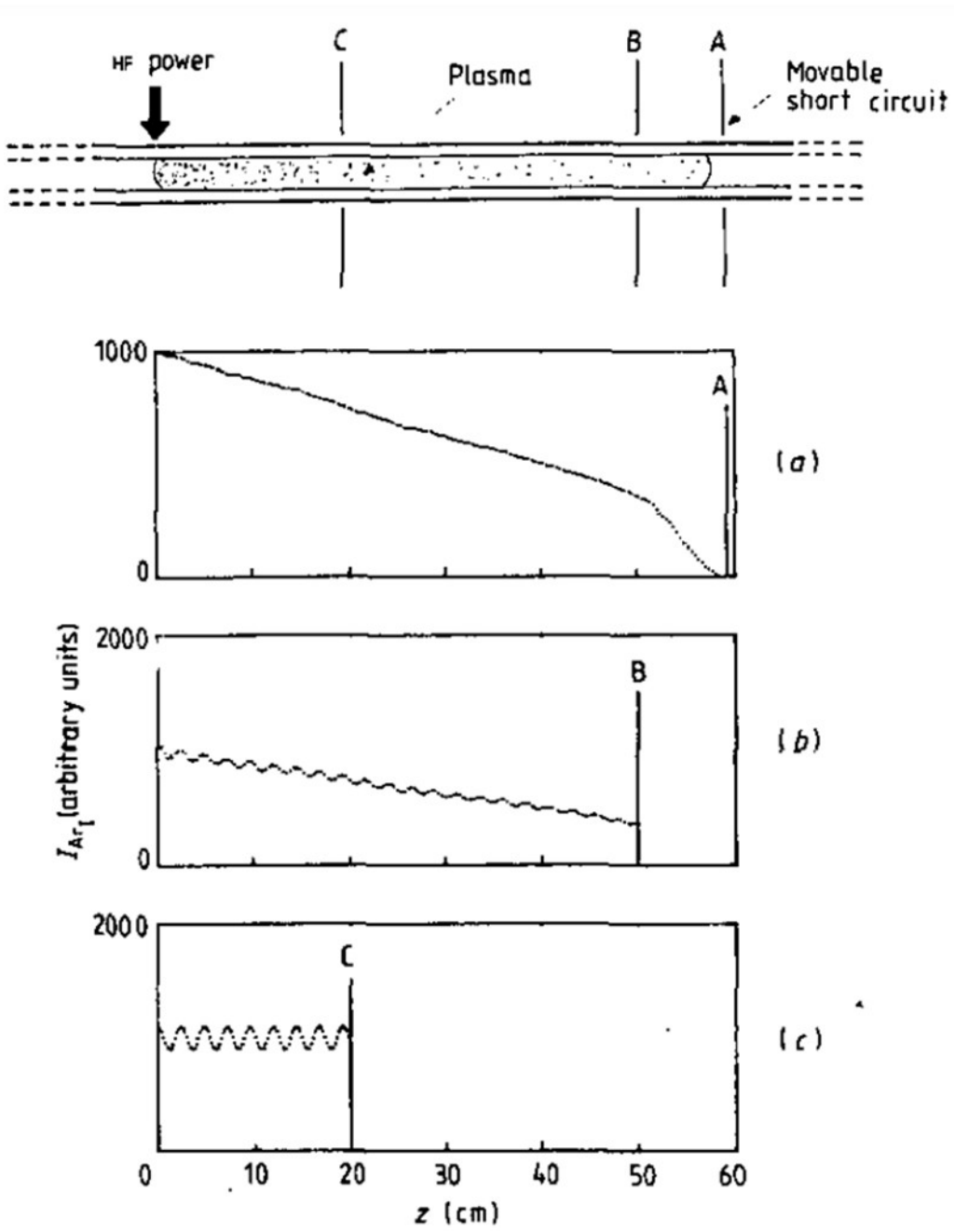

FIG. A9. Axial variation of the intensity of the emission line from given excited argon atoms plotted against the axial position of the reflector plane (Rakem, Leprince and Marec, 1992).

In summary, using standing waves makes the axial electron density of the TW plasma column more uniform than if it were supported by travelling waves alone, in which case the electron density decreases linearly along the axis. Improving the uniformity of the TW plasma column's characteristic quantities can be useful in applications such as lasers and lighting.

Rogers and Asmussen (1982) were the first to suggest using a plasma column supported by surface waves in the standing wave mode.

## APPENDIX D: THE TIAGO PLASMA TORCH, A FILAMENT-LIKE HIGH ELECTRON DENSITY PLASMA SUSTAINED BY A TRAVELLING WAVE PROPAGATING ALONG ITS GAS FLOW AT ATMOSPHERIC PRESSURE

Figure A10a shows a schematic cross-section of the TIAGO plasma torch device. It consists of a hollow conductive rod through which the discharge gas flows, terminating in a conical nozzle with a 1 mm diameter opening. The rod passes through the wide front and rear walls of a modified standard rectangular waveguide section, the height *h* of which has been reduced along the narrow walls to achieve optimal impedance matching: this assembly emerges from a Surfaguide (see section 2.2.3.1). The Surfaguide adjustment plunger can, optionally, be held in a fixed position for specific operating conditions (see section 2.2.3.2).

In the present example, a mixture of helium and hydrogen ($He/H_2$) at a ratio of 9:1 flows at a rate of 10 slm through the copper rod conduit. It exits as a filamentary plasma through the nozzle, dispersing into its own ambient gas before at the column end being partially penetrated mainly by the nitrogen in the air. The discharge is maintained at 2450 MHz with a power of 700 W (Ricard *et al.*, 1995). As shown in Figure A10a, the plasma consists of a bright filament, the dart, surrounded by a diffuse plume that extends beyond it. The dart constitutes the TWD plasma, whilst the plume corresponds to its flowing afterglow.

Many small-size TIAGO plasma torches can be assembled on a single waveguide ensuring non-interference and equal sharing of MW power among the various torches (Pelletier *et al.*, 1999). Further information on the TIAGO concept can be found in Moisan, Zakrzewski and Rostaing (2001).

Figure A10b shows the average electron density along the filament-like plasma as a function of the axial distance from the nozzle tip, as obtained from the Stark broadening of the $H_\beta$ line (Ricard et al., 1995). The first two points, which do not lie on the least-squares regression line, are in fact not part of the TWD plasma, as they belong to the antenna-type radiation region (Moisan, Levif and Nowakowska, 2019). The last two points must also be excluded, as they are believed to result from the gradual penetration of ambient air into the plasma filament. Least-squares regression analysis of the remaining data points reveals an almost perfectly linear decrease in axial electron density distribution along the filamentary plasma, starting 3.5 mm from the nozzle tip ($r^2 = 0.999$). For further details on the plasma plume and jet composition, see Rincón *et al.* (2013, 2014a, 2014b) and Toman *et al.* (2023).

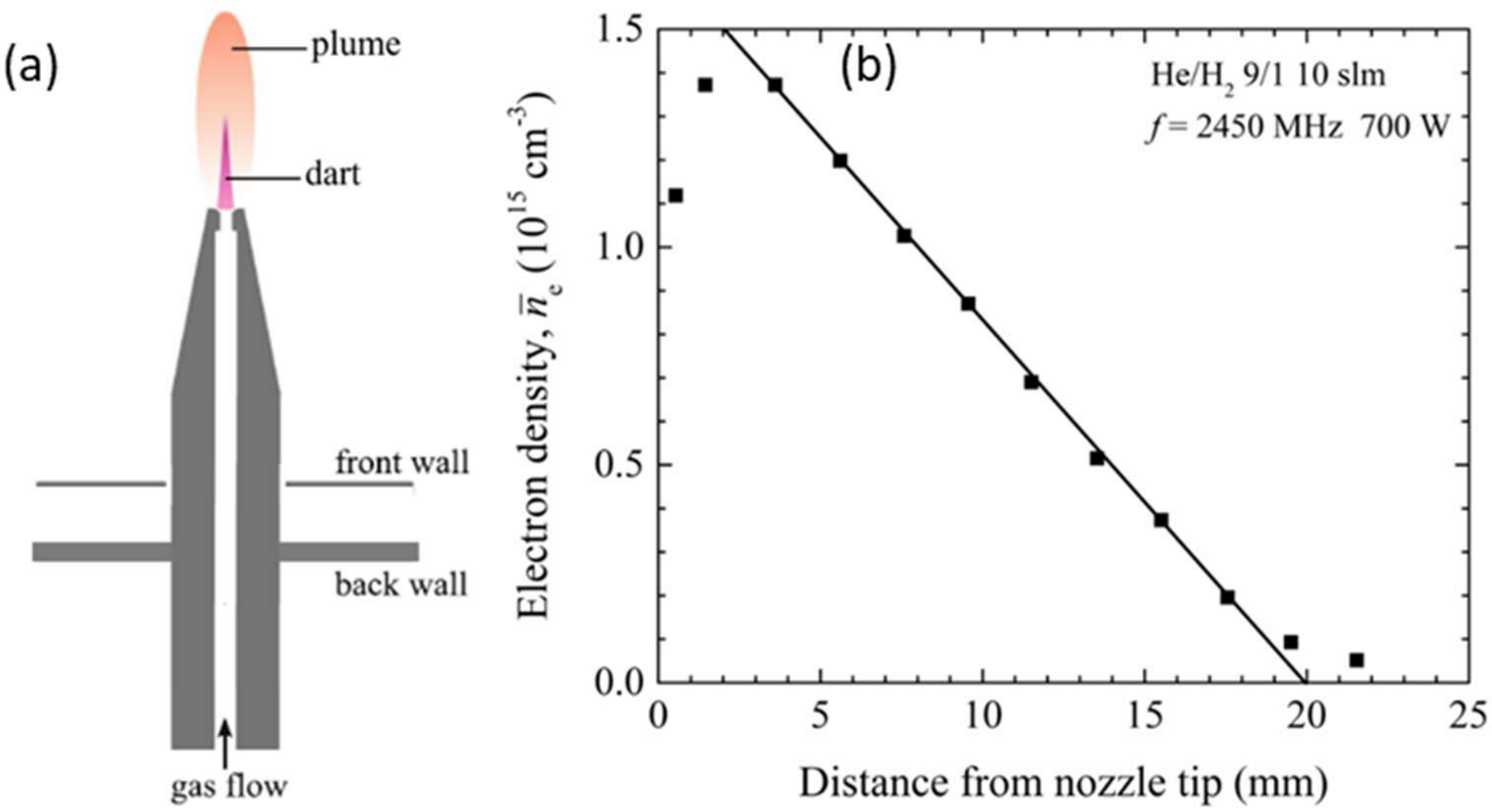

FIG. A10: a) shows a schematic diagram of the TIAGO MW plasma torch. It consists of a hollow conductive rod through which the discharge gas flows to the nozzle through a 1 mm-diameter hole (Ricard *et al.*, 1995). This self-standing plasma is supported by an EM travelling wave launched from the Surfaguide gap. The travelling wave gives rise to a dense filamentary plasma (the *dart* : the TWD segment) and a surrounding diffuse plasma (the *plume*, the flowing afterglow); b) the figure shows the average electron density obtained from the Stark broadening of the $H_\beta$ line along the filamentary plasma (Christova *et al.,* 2004, Moisan, Kéroack and Stafford, 2016). The plasma is generated in a premixed $He/H_2$ 9/1 gas flowing at a speed of 10 standard litres per minute (slm) and sustained at 2450 MHz with a power of 700 W (Ricard *et al.*, 1995).

The fact that the electron density distribution decreases linearly along the plasma filament held by an EM travelling wave indicates that the discharge is controlled by Zakrzewski's stability criterion (see Section 4.1). Under these conditions, the EM travelling wave, whose electric field heats the electrons in the carrier gas, propagates along the exterior of the plasma filament embedded in the ambient gas.

In terms of applications. the TIAGO has a proven track record in highly efficient graphene production (Melero *et al.*, 2023). This process was recently improved by over 20% (Morales-Calero *et al.*, 2024) using a Faraday cage around the TIAGO torch. Further improvements could be achieved by reducing the outlet orifice diameter thereby increasing the carrier gas temperature.

## APPENDIX E: THE EM TRAVELING WAVE THAT SUSTAINS TWDs PROPAGATES PRIMARILY ON THE SURFACE OF THE DISCHARGE TUBE OUTER WALL, WHICH IS IN A VACUUM (AMBIENT AIR)

As we have repeatedly emphasised, the main issue when modelling the axial distribution of electron density in a plasma column generated by a guided EM travelling wave is the, erroneous, assumption that this wave propagates within the plasma column or along the interface between the plasma column and the inner wall of the containing tube. As previously mentioned, and as will now be demonstrated, wave power primarily flows along the interface between the outer wall of the discharge tube and the surrounding ambient air.

To prove this, we will first revisit the fundamentals of non-ionising EM surface waves propagating along an existing plasma column, such as the positive column of a direct current (DC) discharge (Trivelpiece and Gould, 1959). This travelling wave was initially referred to as a *space charge wave*. It was later shown that its fundamental propagating mode is an azimuthally symmetric transverse magnetic (TM) wave (e.g. Margot-Chaker et al., 1989), which has two electric field components, $E_r$ and $E_z$, and a unique magnetic field component, $H_\varphi$. Fig. A11 shows the radial variation of the two electric field components calculated along the discharge tube while the magnetic field component, which is radially constant, is not shown. We observe that the amplitude of the radial component of the wave electric field $E_r$ is a few times larger than that of its axial electric field component $E_z$.

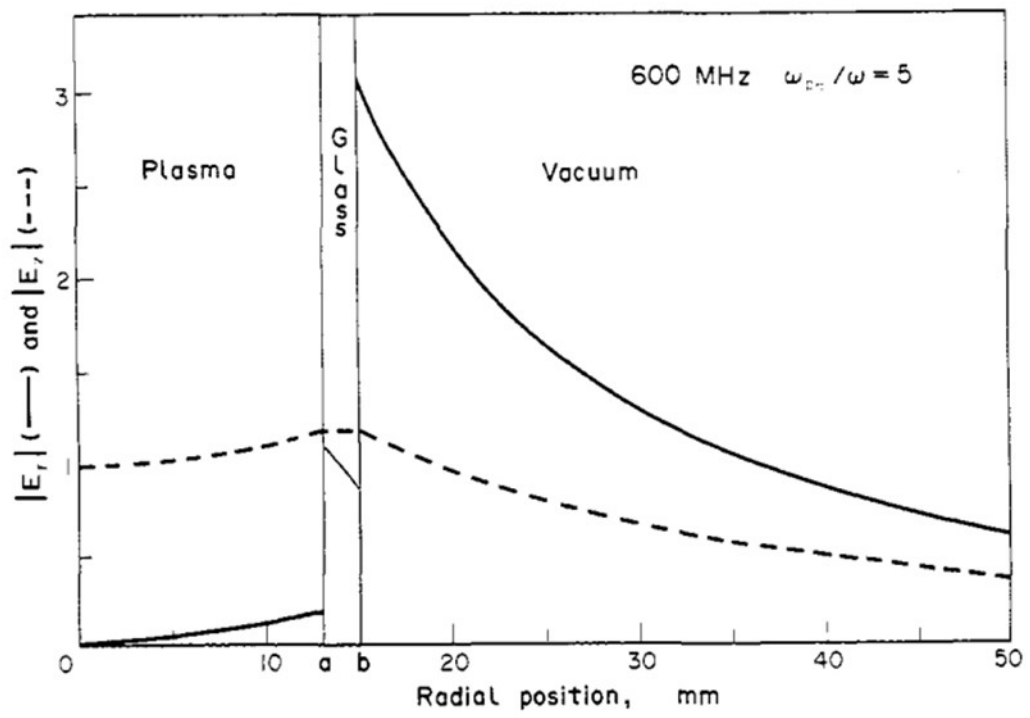

FIG. A11. Calculated radial amplitude of the radial and axial electric field components, $E_r$(r) and $E_z$(r) respectively, of the azimuthally symmetric TM mode wave propagating along a preexisting cylindrical DC plasma column enclosed in a dielectric tube surrounded by vacuum. The relative permittivity of the tube is 3.8, with inside and outside radii respectively *a* and *b*. The intensities are normalized by requiring $E_z(0) = 1$ (Moisan, Shivarova and Trivelpiece, 1982a).

Fig. A12 now shows the measured intensity of the radial electric field component $E_r$ of the travelling wave along this time a TWD. As calculations in Fig. A11 already showed, it decreases radially almost exponentially.

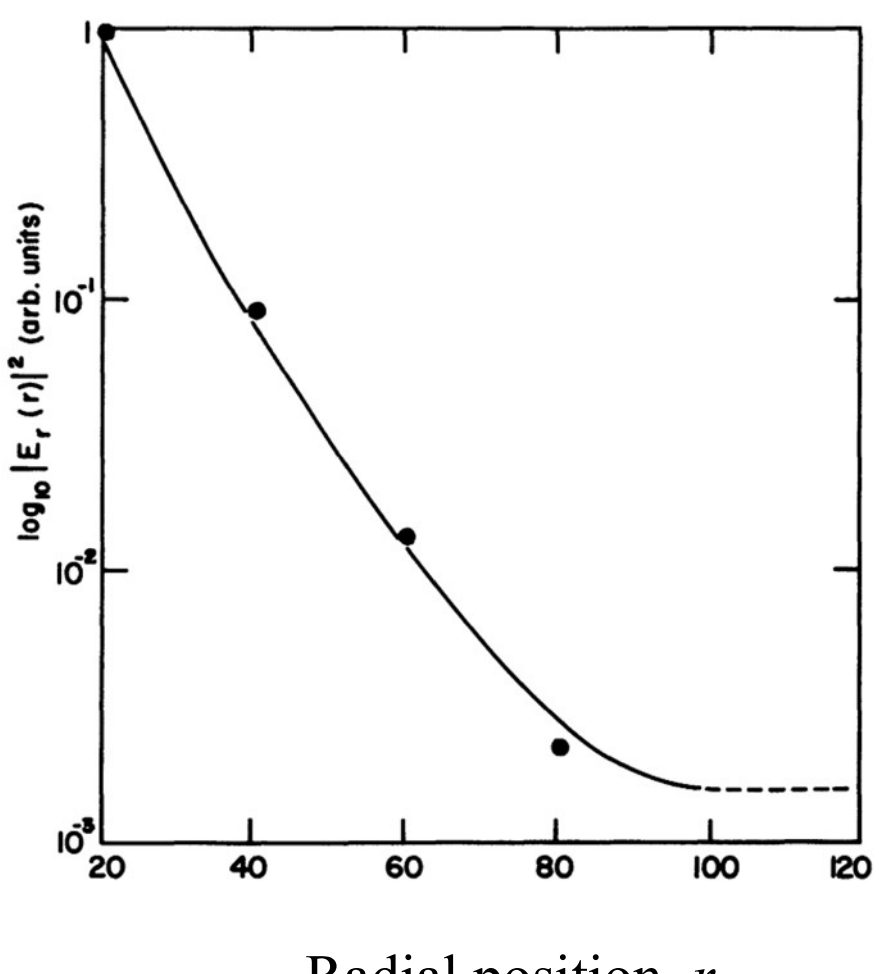

Radial position, *r*

FIG. A12. The solid line shows the $\log_{10} [E_r(r)]^2$ value plotted against radial position, as measured using a radially oriented electric-field type antenna (see footnote 9) located on the outer wall of the discharge tube. The $r$ = 20 mm position corresponds to the antenna tip almost touching the external tube wall (Moisan i, 1982b). The points have been fitted with a modified Bessel function of the first kind, $K_1(r)$, which is required for the *surface wave* field intensity distribution (Moisan and Nowakowska, 2018). As the *fR* product (0.5 GHz × 1.15 cm = 0.575 GHz cm) is much less than 2 GHz-cm, propagation of an azimuthally symmetric TM travelling wave with three field components $E_z$, $E_r$ and $H_\varphi$ is ensured (Margot-Chaker et al., 1989).

The wave Poynting vector which is given by $\boldsymbol{S} = \boldsymbol{E} \times \boldsymbol{H}$ (Lorrain and Corson, 1972) has thus two components due to the $E_z$ and $E_r$ field values, respectively:

$$\boldsymbol{S}_z = \boldsymbol{E}_r \times \boldsymbol{H}_\varphi \tag{A1}$$

and:

$$\boldsymbol{S}_r = - \boldsymbol{E}_z \times \boldsymbol{H}_\varphi. \tag{A2}$$

The measured values of the field intensity components $E_z^2$ and $E_r^2$ are easily obtained under standing wave conditions. As shown in Appendix C, the $E_r^2$ intensity of the radial electric field component is always larger than $E_z^2$: consequently, $\boldsymbol{S}_z$ is greater than $\boldsymbol{S}_r$ meaning that the wave power primarily flows axially. The remainder of the total wave power flows radially into the plasma column to sustain it. This should put an end to the debate about how the wave propagates along the TWD and how it sustains it. Aliev, Boev and Shivarova first raised theoretically the possibility that most of the wave power circulates along the outer wall of the discharge tube in a vacuum (see Sec. 5.1.3 and 5.1.4), but the idea never really took hold.

**Acknowledgments**

We gratefully acknowledge the contribution of Dr. Helena Nowakowska (Institute of Flow Machinery, Polish Academy of Sciences, Gdansk, Poland) to the equations used in Section 4. Dr. Kinga Kutasi (Wigner Institute, Budapest, Hungary) is thanked for her relevant reading of Section 2. We would also like to thank Professors Émile Carbone (INRS-Énergie, Matériaux, Télécommunications, Varennes, Québec), Vasco Guerra (Instituto Superior Técnico, Lisbon, Portugal) and J. J. A. M. Van der Mullen (Eindhoven University of Technology, the Netherlands and Université Libre de Bruxelles, Belgium) for their comments on the manuscript during its preparation. We are also indebted to Dr Mariam Rachidi (Montréal, Québec) for her help with data processing. The Natural Sciences and Engineering Research Council of Canada (NSERC) covered the costs of editing and publishing this manuscript.

## Section 7: Sterilising dielectric medical devices at low temperatures without causing damage or requiring subsequent ventilation is a prime example of TWD application

The advent of the Surfatron and Surfaguide travelling wave launchers for generating TWDs (1972-1978, see Section 2) quickly sparked considerable interest in studying these discharges and applying for inventive patents thanks to their stability, reproducibility and unrivalled range of operating conditions. According to Mendeley (IOP), as of February 2026, the article by Moisan and Zakrzewski (1991) on the family of travelling wave launchers had resulted in 240 individual patents. Of the many applications cited in the literature, the Université de Montréal (UdeM) has focused specifically on sterilising medical devices (MDs), achieving notable success with dielectric objects.

This method inactivates microorganisms, including bacteria, spores, viruses and prions (non-conventionally transmissible agents), in tens of minute by applying an EM field to a gas. This ionised gas emits particles and photons, the combined physicochemical action of which causes the cellular death of these germs (e.g., Moisan *et al.*, 2013). Although this technique received financial support from the companies Air Liquide and Getinge for its development, it was not commercialised in the end due to its lower economic viability compared to gaseous sterilisation using ethylene oxide (EtO), which is used in most central sterilisation facilities. Apart from the low cost of using EtO, it must be acknowledged that this process could otherwise be described as disastrous: EtO is a toxic gas that is lethal to humans at concentrations of just a few parts per billion (ppb). It is also a greenhouse gas and is flammable. Furthermore, following sterilisation, a ventilation period of typically more than ten hours is required. This is because EtO penetrates deep into the treated material, making it toxic if not allowed to desorb completely beforehand. Objects sterilised using plasma sterilisation by TWD, on the other hand, are immediately available, which is particularly useful in emergency situations.

### 7.1 The TWD sterilisation device developed at UdeM

Figure 22 shows a diagram of the device that our work ultimately led to. The control substrate to be sterilised is placed in a 50-litre (61 × 26 × 30 cm³) chamber preferably filled with the late afterglow of a $N_2$–$O_2$ TWD[19], with the percentage of $O_2$ adjusted to maximise UV emission from the $NO_\beta$ band (Patent by Moisan *et al.* 2004). Some earlier TWD sterilisation chambers, which were designed to examine specific operating conditions, can be found in Moreau *et al.* (2000).

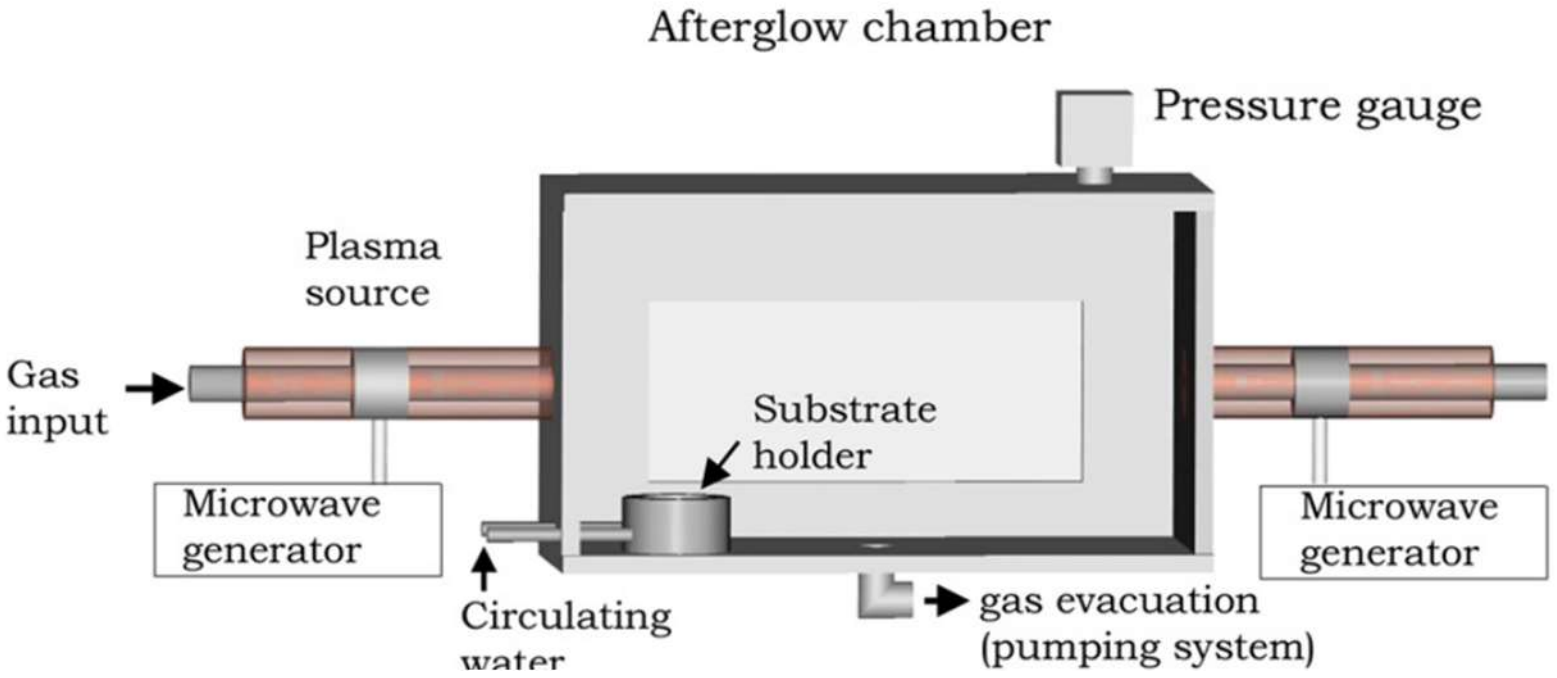

FIG. 22 shows a side view of an advanced plasma sterilisation chamber developed at UdeM. Gas discharge is sustained by a TWD at either 915 MHz (Surfatron) or 2450 MHz (Surfaguide), either in a single discharge tube or in two tubes simultaneously inputting the chamber, with inner and outer diameters of 26 mm and 30 mm, respectively. The aluminium enclosure has a volume of 50 litres and inner dimensions of 61 × 26 × 30 cm³. The observation windows, which are used for UV spectroscopic studies, are made of fused silica. The temperature-controlled substrate holder, cooled by a circulating water heat exchanger, is located at the base of the chamber (Moisan *et al.*, 2013).

As will be discussed in more detail later, sterilisation is mostly achieved in an $N_2$-$O_2$ discharge in our experiments. To fully benefit from this regime, the distance between the wave launcher slit and the sterilisation chamber entrance must exceed a certain value. In the present case, setting this distance to a value between 150 and 220 mm fills the chamber with early afterglow (no EM wave field, but electrons remain), whereas setting the distance to 820 mm or greater fills the chamber with late afterglow (no EM wave field and few or no electrons), as discussed further in Section 7.4.

7.2 Factors that contribute to the efficiency and safety of the $N_2O_2$-based TWD sterilisation method, which avoids damaging the spore outer structure or the sterilised devices

Such a discharge is produced in a flowing gas, which provides N and O atoms. These long-lived species can be transported by the gas flow downstream to a region beyond the discharge regime, where there is no wave electric field, the 'afterglow' (Mérel *et al.*, 1998). Under collision processes involving a third body, such as $O_2$ or $N_2$, these atoms form excited NO molecules. Figure 23 shows the particularly dense UV spectrum of $NO_\beta$ and $NO_\gamma$ molecular emission obtained under reduced gas pressure (8 Torr). This emission extends down to 180 nm, providing photons with very high energy. At atmospheric gas pressure, only the $NO_\gamma$ molecular spectrum is observable (Boudam *et al.*, 2006).

[19] Although the sterilisation process is slower, the flowing late afterglow is considerably less damaging to MDs than the discharge itself and the early afterglow (see Sec. 7.4 for details).

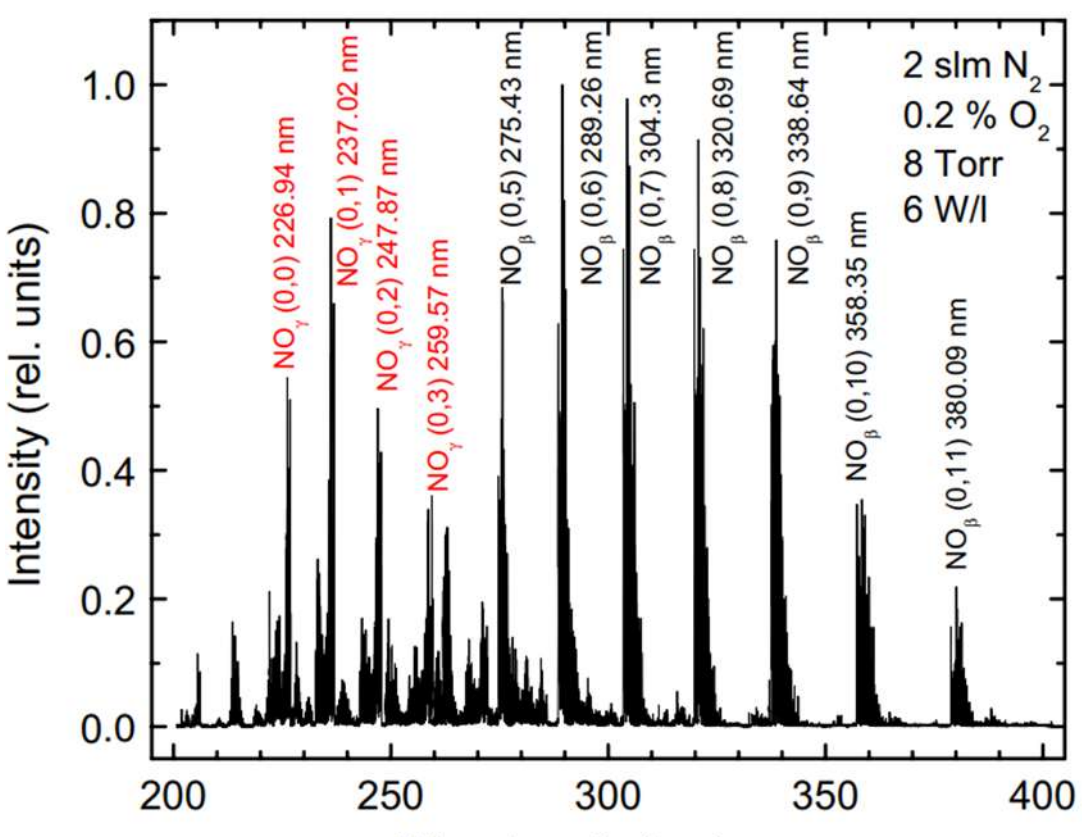

FIG. 23. Emission spectrum recorded in the flowing afterglow of a reduced-pressure (8 Torr) $N_2$–$O_2$ 2450 MHz TW discharge with 0.2% $O_2$ added to $N_2$ (flow rate 2 slm) to maximize the intensity of this UV spectrum (Patent by Moisan *et al.*, 2004). Line identification shows $NO_\beta$ and $NO_\gamma$ molecular bands (Boudam *et al.*, 2006).

As shown in Fig. 24, tuning the percentage of $O_2$ in the $N_2$-$O_2$ mixture maximises the UV emission. UV photons efficiently inactivate microorganisms by damaging their DNA without affecting their envelope (see Sec. 7.4). Unlike in the discharge itself, in the afterglow region there are minimal heating and ion bombardment effects on the processed material. Furthermore, since a flowing afterglow can expand spatially, it therefore fills volumes much larger than the discharge vessel itself. All of these clearly benefit our method.

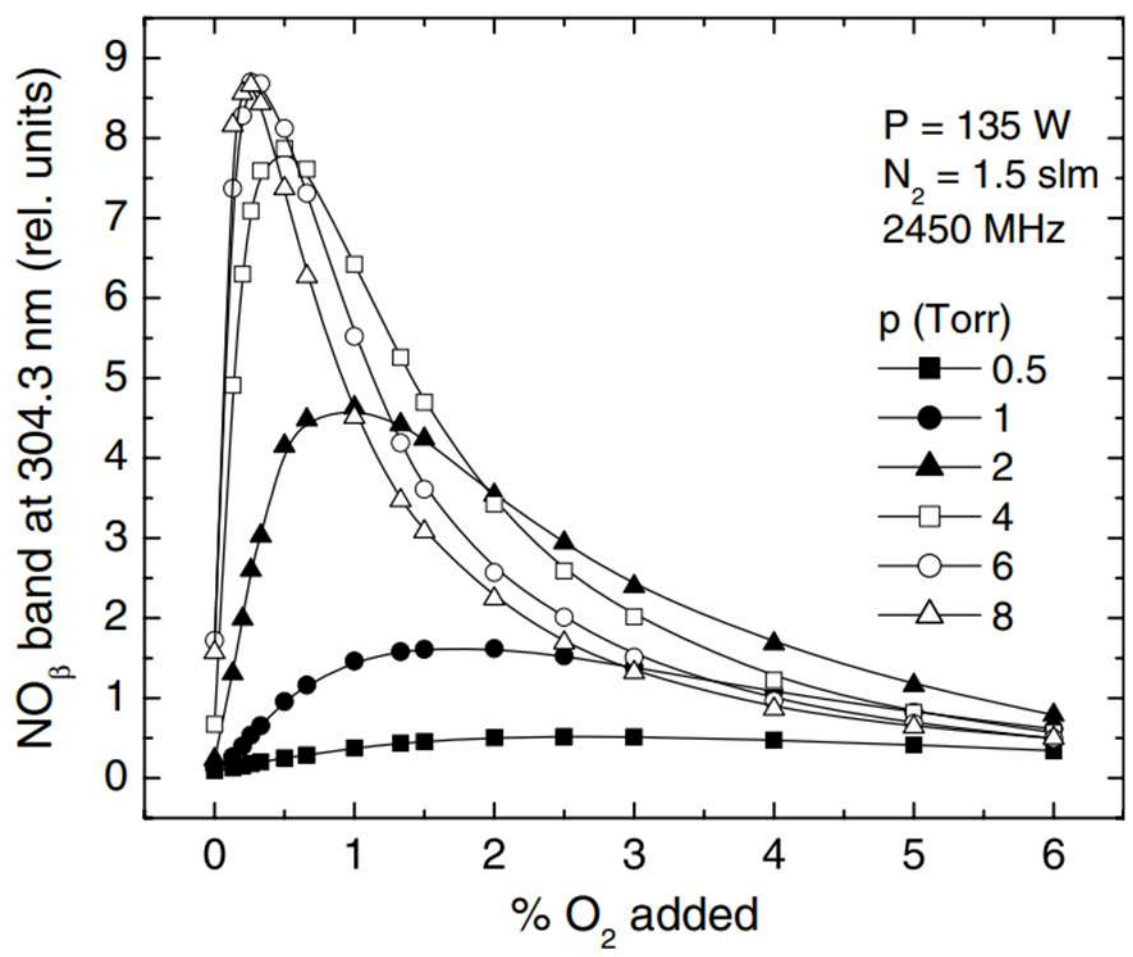

Fig. 24. Emission intensity of the 304.3 nm line, a band head of the $NO_\beta$ system, recorded in the flowing afterglow of an $N_2$–$O_2$ TWD operated at 2450 MHz[20] as a function of the $O_2$% added to $N_2$, at various total gas pressures (Boudam *et al.*, 2006).

7.3 A further advantage of operating with the TWD afterglow is that UV photons can also be emitted inside spore clusters

[20] The higher the TWD frequency, the higher the maximum (saturation) N-atom density, hence the eventual maximum NO UV emission (Mérel *et al.*, 1998).

UV radiation used for sterilisation is sometimes provided by a UV lamp. However, the lamp's directed emission can prevent spores located inside clusters from being reached. A similar situation occurs when photons are emitted from a plasma column, which acts as a partially directional UV emitter. In our case, the picture is different since some N and O atoms from the TWD afterglows can circulate freely within the clusters before forming NO molecules that emit UV photons targeting a spore, as shown in Fig. 25.

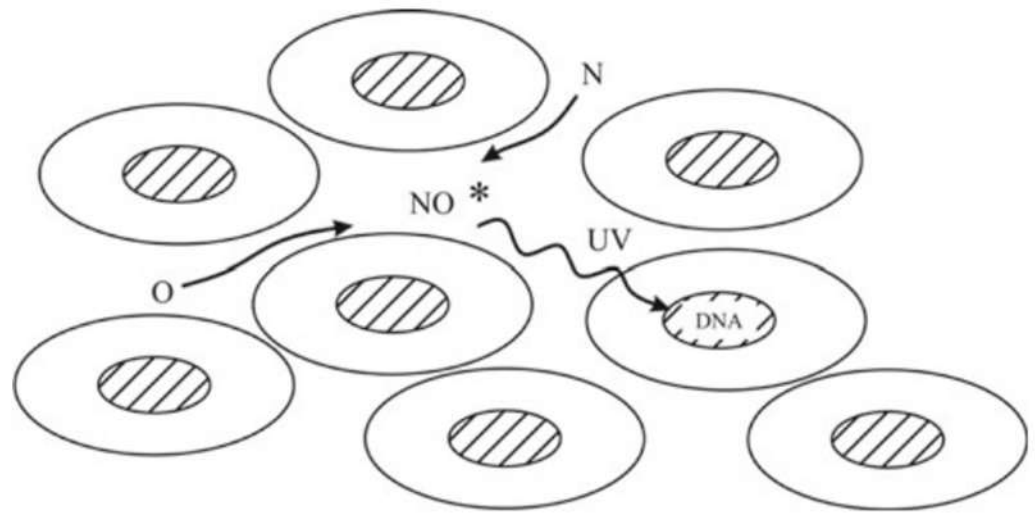

FIG. 25. Schematic representation showing how N and O atoms can diffuse between spores and eventually form a NO* excited molecule that emits UV photons which reach the DNA of a hidden spore in a stack (Moisan *et al.,* 2013).

### 7.4 In comparison to the discharge itself and the early afterglow, the late afterglow of the $N_2$–$O_2$ TWD ensures that polymers and bacterial spores remain structurally intact under sterilising conditions

Unlike the early afterglow, there are no electrons in the late afterglow. This prevents the significant damage to exposed spores and their lift off (see Remark) as was observed during the early afterglow (Boudam *et al.*, 2006). These two types of afterglows can be identified using optical emission spectroscopy. The early afterglow is characterised by intense emission from the $N^+_2$ first negative system (band head at 391.4 nm), while the late afterglow is typified by an overpopulation of the v = 11 rotational-vibrational level of the $N_2$(B) state (Mérel *et al.*, 1998, Moisan *et al.*, 2002).

More specifically, the afterglow of the $N_2$–$O_2$ TWD in the 0.6–10 Torr range has been examined in the context of sterilising medical devices (MDs). To minimize damage to polymer-based MDs, the percentage of $O_2$ in the $N_2$–$O_2$ mixture is adjusted to ensure maximum UV intensity (thereby also reducing sterilisation time). The influence of operating conditions such as pressure and, as well as the distance from the discharge to the afterglow chamber entrance, has been examined to determine the optimal conditions for achieving a dominant late afterglow and ensuring maximum, uniform UV intensity in the sterilisation chamber.

Figure 26 introduces the characteristic distance $x_d$, showing that positioning the plasma source at a sufficient distance from the chamber inlet allows the early and late afterglow phases to be reached. In the case of $x_d$ = 100 mm (Fig. 26), to implement late afterglow treatment to minimise damage to MDs[21], the substrate should be placed at the bottom of the chamber, far from the discharge axis (Moisan *et al.,* 2014). Full late afterglow is achieved in the whole chamber when this distance is set to at least 820 mm (Moisan *et al.*, 2013).

---

[21] The structural erosion eventually caused by species specific to the early and late afterglows of $N_2$-$O_2$ discharges has been mapped using polystyrene microspheres deposited on glass and *B. atrophaeus* spores on a Petri dish (Moisan *et al.*, 2013).

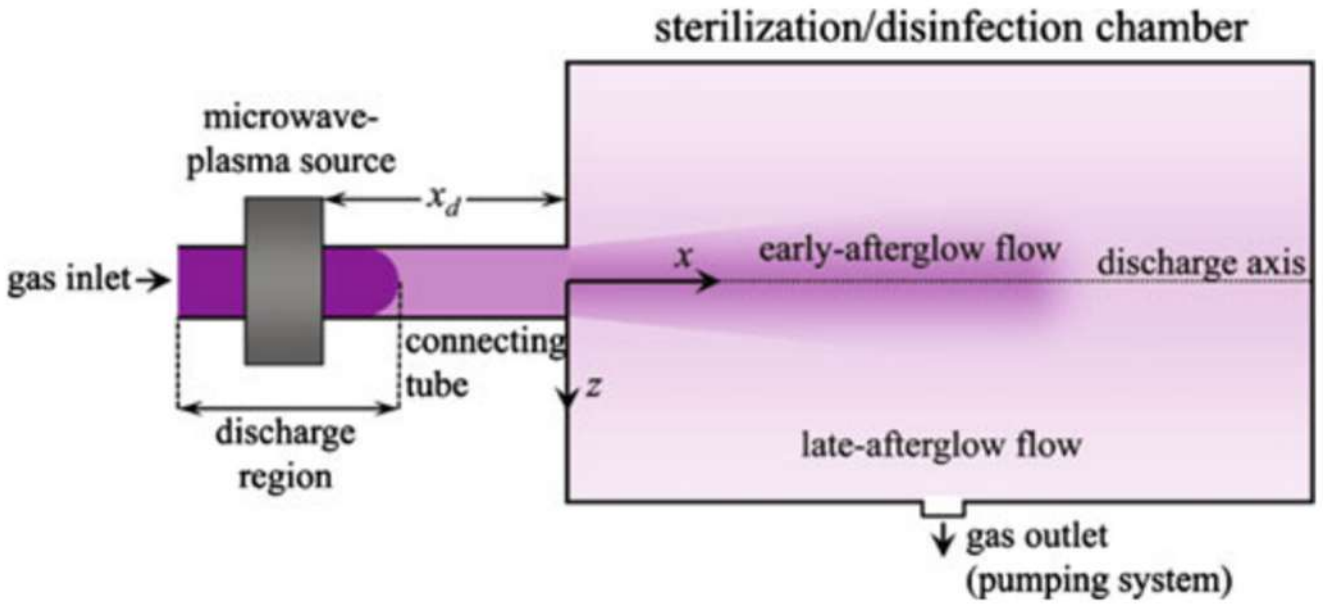

FIG. 26. 'Artist's view' representation of the discharge outflow of long-lived species as they enter our sterilization/disinfection chamber after travelling a distance $x_d$ from the microwave-plasma source (Moisan *et al.,* 2014). The figure illustrates the case where $x_d$ is relatively short (≈100 mm) whereas with $x_d > 820$ mm, the whole chamber is under late afterglow regime (Moisan *et al.*, 2013).

Remark: A key and original issue raised, probably for the first time, by Moisan *et al.* (2014) is that direct exposure to the discharge or its early afterglow (e.g. in $N_2$–$O_2$) can dislodge and release viable microorganisms from their substratum. This is attributed to electrostatic charging by electrons, resulting in an outward electrostatic stress that exceeds the adhesion of the spores to their substrate. This phenomenon could cause the recontamination of sterilised devices, as well as the contamination of the surrounding walls of the chamber, and the *outside world*. It also leads to an erroneous overestimation of the inactivation efficiency when counting the remaining microorganisms after the sterilisation process.

7.5 Number of surviving bacterial spores as a function of exposure time in the late afterglow regime in a 915 and 2450 MHz $N_2$-$O_2$ TWD: a two-stage process

Figure 27 shows that, in the late afterglow regime, there are two sets of inactivation rates: a fast initial rate followed by a much slower second rate. Furthermore, the initial slope at 2450 MHz is steeper than at 915 MHz. Sterility is achieved at 2450 MHz and 400 W in approximately 42 minutes. Clearly, the higher the operating frequency and the power absorbed in the discharge, the faster the inactivation rate.

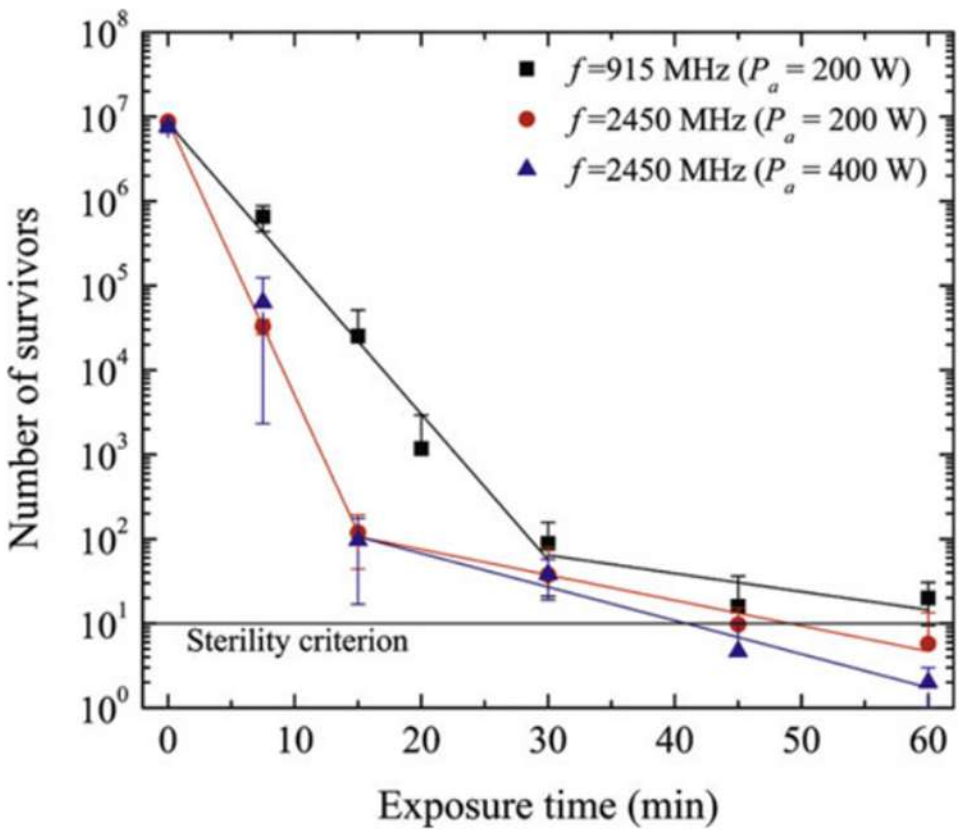

FIG. 27. Survival curves of $10^7$ *B. atrophaeus* spores after exposure to the flowing late afterglow ($x_d$ = 820 mm) from an $N_2$-$O_2$ TWD. The discharge was sustained at either 915 MHz or 2450 MHz, within a 26 mm inner diameter tube connecting to the 50 L chamber. The operating conditions were optimised to achieve maximum UV intensity, with an $O_2$ concentration of 0.2% in $N_2$ at 3.5 torr (≈470 Pa)) and a flow rate of 1.4 slm, to ensure axial uniformity. The absorbed microwave power was set at 200 W at both frequencies and 400 W at 2450 MHz, in all cases providing a fast and a slow inactivation rate. The sterility criterion (a safety assurance level of 6 logs) was reached at 2450 MHz after approximately 50 minutes at 200 W and after 42 minutes at 400 W (Moisan *et al.*, 2013).

Figure 28a illustrates the impact of introducing various amounts of $O_2$ into $N_2$ on the number of survivors in the flowing late afterglow. The inactivation rates are then correlated with the UV irradiation dose (fluence), providing a common unifying parameter, as shown in Figure 28b.

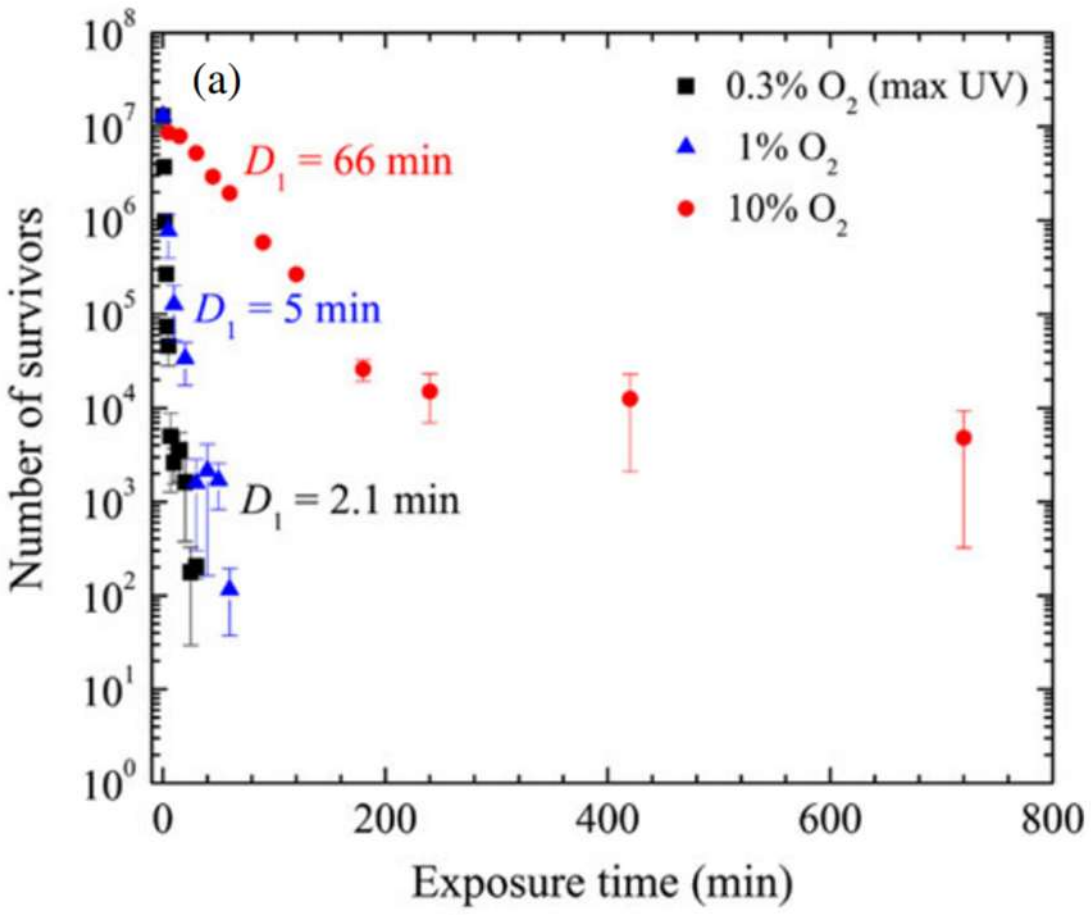

FIG. 28(a). Survival curves of *B. atrophaeus* spores following their exposure to the $N_2$-$O_2$ flowing late afterglow at three different UV intensity levels obtained by setting the $O_2$ admixture percentages to $N_2$ at 0.3%, 1% and 10%. The $D_1$ time values for the fast inactivation rate are shown (Moisan *et al.*, 2013).

Fig. 28b shows the number of surviving *B. atrophaeus* spores as a function of UV fluence (irradiation dose). Fluence is plotted explicitly as $k_iI$ where $I$ is the UV radiation intensity and $k_i$, the reaction coefficient (sensitivity). This coefficient, which depends on the type of exposed microorganism and UV wavelength range considered, can be obtained from the experimental slopes in the figure, following Cerf (1977) who proposed to express the decrease in the number of viable microorganisms with time as a first-order rate, with a dependence in the form:

$$N(t) = N_0\, 10^{-t/D}, \tag{42}$$

which for a bi-phasic survival curve[22] leads to:

$$N(t) = N_{01} \times 10^{-k_1 It} + N_{02} \times 10^{-k_2 It}, \tag{43}$$

where $k_1I = 1/D_1$ and $k_2I = 1/D_2$ correspond to isolated spores and top ones on a stack (readily inactivated), and to stacked spores or hidden in debris (much longer inactivation time), respectively (see also, Rossi *et al*. (2009) and von Keudell *et al*. (2010)). In Fig. 27, $N_{02}$ is approximately 100 spores (i.e., only 0.001% of an initial deposit of $10^7$ spores).

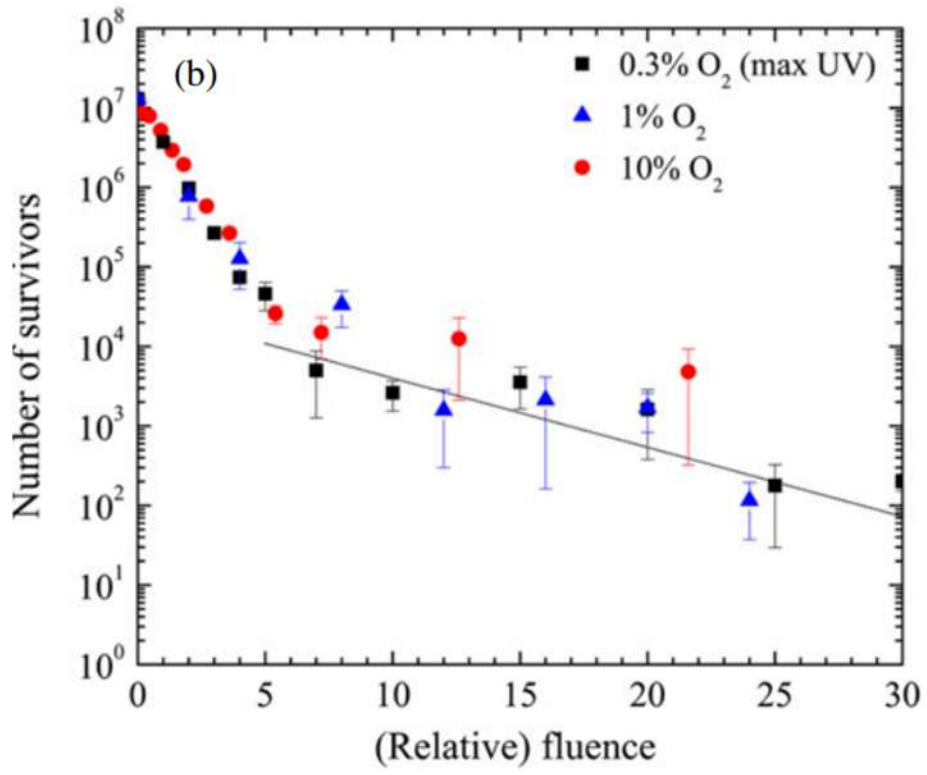

FIG. 28(b). Survival curves of *B. atrophaeus* spores extracted from Fig. 28a when plotted as a function of fluence. Data from Moisan *et al.*, 2013.

The curve representing the first inactivation phase in Figure 28b fits the data points obtained at three different UV intensities particularly well. This strongly suggests that UV photons predominantly control spore inactivation, regardless of the levels of UV intensity considered here. This is noteworthy, given that the density of O atoms increases by approximately a factor of 10 when the percentage of added $O_2$ is raised from 0.3% to 10%. This shows that the inactivation rate in the first phase is unaffected by the erosion (etching) of the spores, despite the increased density of chemically reactive O radicals (but at low gas temperature) and the much longer exposure time to these radicals compared to 0.3% added $O_2$. These features are confirmed by the absence of erosion observed during the flowing late afterglow regime (see footnote 21).

7.6 The nature of MD materials and the average sterilisation time using our method

We noted at the beginning of this chapter that our sterilisation method is primarily intended for dielectric MDs. MDs made of aluminium, or even copper — one of the best conductors — can also be sterilised without sustaining damage when using the flowing late afterglow of an $N_2$-$O_2$ TWD. However, this method takes much longer than the average duration of the various other sterilisation methods used for this type of MD. This is due in our case to the significant recombination of nitrogen atoms, which are essential for forming the NO molecule that emits UV photons. This can be seen in Figure 29. Thus, the intensity of UV emission around a copper MD would be approximately three times weaker than around a Teflon MD. Consequently, it would take at least two and a half hours to reach sterility.

[22] Cerf (1977) set out in his Review to identify all types of bacterial inactivation rate as a function of time. It follows that there is no single such curve. In our case, the two slopes of inactivation are determined by the flux of UV photons, and when this was maximised by selecting the correct percentage of $O_2$ in the $N_2$–$O_2$ mixture, the observed biphasic curve is indeed uniquely due to the damage caused by UV photons to the DNA of the exposed microorganisms.

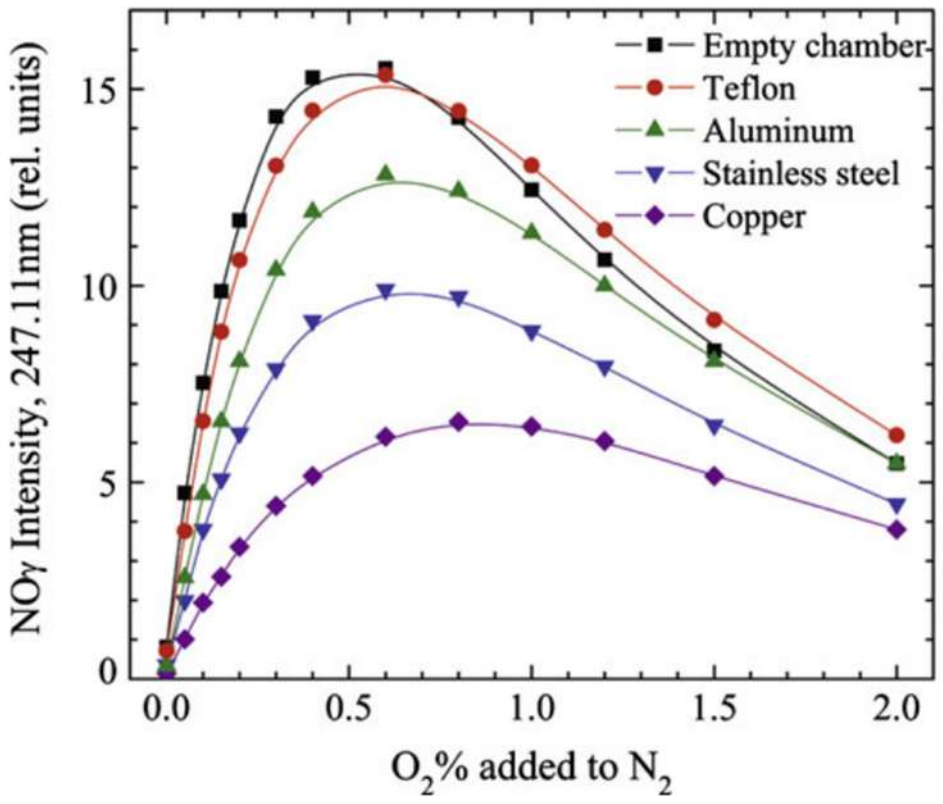

FIG. 29. Variation in the UV intensity (recorded on the 247.11 nm band head of the $NO_\gamma$ system) in the 50 L chamber as a function of the percentage of $O_2$ added to $N_2$ in the presence of plates made from different materials, compared with the empty vessel. Operating conditions: discharge sustained in a 6 mm i.d. Pyrex tube at 2.45 GHz with 100 W absorbed power, at a pressure of 2 torr (≈270 Pa) in the 50 L chamber with a 0.9 L/min of $N_2$ gas (Moisan *et al.,*2013).

7.7 Packaging of MDs that have been sterilised using TWDs

To preserve their on-shelf sterility, MDs are sealed/wrapped in packaging material. Porous packaging materials utilized in conventional sterilization systems (where MDs are packaged before being subjected to sterilization) were tested and found inadequate for the $N_2$-$O_2$ late afterglow system in contrast to a (non-porous) polyolefin polymer (BagLight Polysilk$^{TM}$). As the latter is non-porous, the corresponding pouch must remain unsealed until the end of the process. Even though it is unsealed, but because the opening is very small the $O_2(^1\Delta_g)$ metastable-state molecules are expected to be strongly quenched by the pouch material as they try to enter it and, as a result, only N and O atoms, together with UV photons, are present in significant quantities.

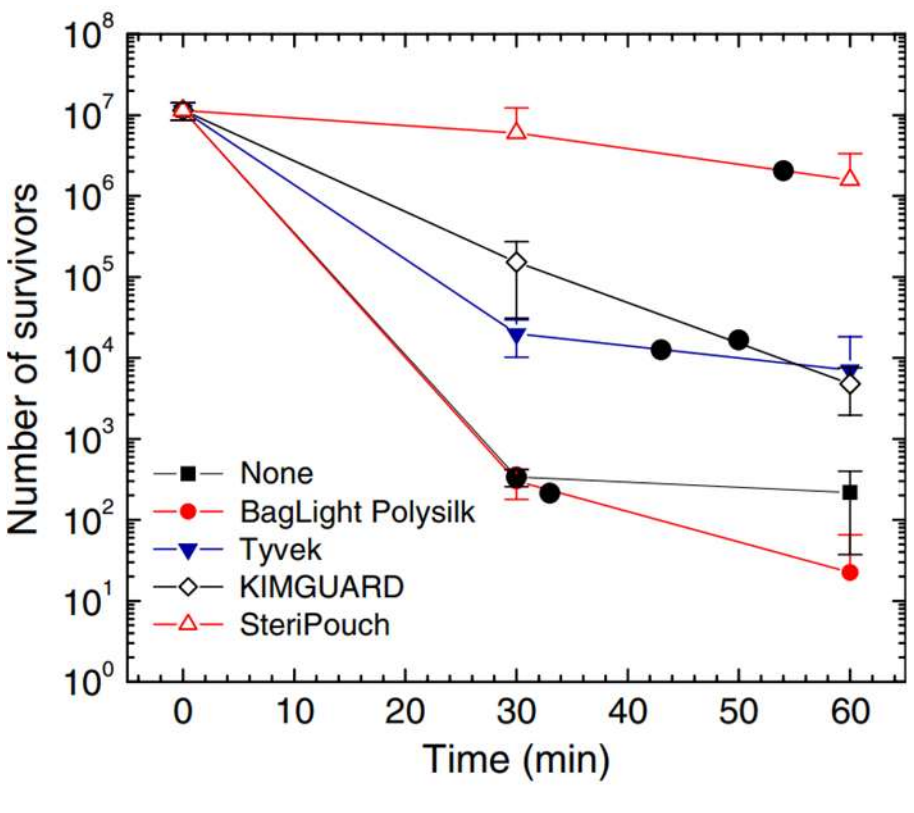

FIG. 30. Survival curves of dried deposits of *B. atrophaeus* spores laid on a PS Petri dish inserted in different packaging materials and then exposed to the $N_2$–$O_2$ late afterglow compared with an unpackaged Petri dish. The large black dot on each curve corresponds to a reference UV photon dose: it is the number of photons collected in the empty chamber after 30 minutes. The more the packaging material reduces the available UV intensity on the outside of the packaging material (figure 23), the more time it takes to achieve such a dose (Moisan *et al.*, 2013).

## 7.8 Advantages and limitations of the TWD late afterglow technique

Besides its more or less conventional use in medicine, dentistry and pharmaceutics — for example, for sterilising implants (including casts, impressions and dentures in dentistry), tubing of all kinds and vials — the system could be used for applications that cannot be carried out using autoclaving (due to the risk of high-temperature damage) or chemical processing (due to the risk of toxic chemical impregnation of the objects). This could include sterilising toys for young patients with weakened immune systems, sterilising dehydrated food for space and military needs and for newborn babies, and sterilising synthetic food containers. Of particular economic interest would be the treatment of seeds for greenhouse facilities (insect killing, fungi inactivation, etc.) and, more "exotically", the decontamination (odour elimination) of sports equipment (e.g. baseball gloves, hiking boots) in rental shops.

Economically viable applications for scaling up this afterglow steriliser to 200 or 500 litres must be identified[23]. The maximum volume tested so far is 60 litres. Nevertheless, we believe that the system is ready for alpha testing on a larger scale. As far as we are aware, our system is one of the most advanced plasma systems reported in the literature, and probably the best characterised in terms of its fundamental mechanisms.

## 7.9 Comparing different plasma sterilisation methods

As a general rule, it is well known that spores can be inactivated by plasma, whether they are exposed to the discharge itself or its flowing afterglow. Inactivation by direct exposure to the discharge is much faster than exposure to a remote plasma. However, exposure of spores and MDs to a discharge can be more aggressive than when they are subjected to a remote plasma, mainly due to erosion (assisted by a higher gas temperature). Direct exposure can also cause interaction with some polymers, which can then interfere with the efficiency of the sterilization/decontamination process, whereas no such limitation was found with the same polymers immersed in the $N_2$-$O_2$ flowing afterglow, however. To the best of our knowledge, we have demonstrated for the first time that direct exposure to argon and $N_2$–$O_2$ discharges, or to the $N_2$–$O_2$ discharge early afterglow, leads to the dislodgement and release of spores (Moisan *et al.,* 2014) This does not occur when they are subjected to the late afterglow. Spore release due to direct exposure to a discharge has been shown to occur with three different types of plasma : argon and $N_2$–$O_2$ discharges at reduced pressure and an argon DBD at atmospheric pressure. Based on the literature reviewed, we can propose that the dislodgement of spores from their substrate results from their charging, which leads to an (outward) electrostatic stress that exceeds their adhesion to the substrate.

[23] At a conference, a DARPA official (US Defence research program) asked about the feasibility of sterilising a combat tank. I pointed out that this would be too costly due to the tank's large volume. He replied that this would not be a problem.